\documentclass{cicc}
\usepackage[version=3]{mhchem} 
\usepackage{physics}
\usepackage{subcaption}
\usepackage[pdftex]{hyperref}
\usepackage{color}

\usepackage{amsthm}
\usepackage{amsmath}
\usepackage{graphicx}
\usepackage{mhchem}
\usepackage{amsfonts} 

\usepackage{amssymb} 
\usepackage[shortlabels]{enumitem} 
\usepackage[utf8]{inputenc}
\firstpage{1}
\articletype{Perspective}

\doi{doi: 10.4208/cicc.2026.xxx.xx}
\publishedyear{2026}
\volume{1}
\issue{1}

\receiveddate{20 Oct. 2026}
\revisiondate{25 Oct. 2026}
\accepteddate{25 Dec. 2026}
\onlinedate{28 Dec. 2026}
\publisheddate{30 Dec. 2026}

\title[Off-diagonal low rank]{The off-diagonal low rank property: new opportunities for low-scaling computational chemistry methods}
\author[1*]{Zikuan Wang}

\affil[1]{\textit{Qingdao Institute for Theoretical and Computational Sciences, Center for Optics Research and Engineering, Shandong University, Qingdao, Shandong 266237, P. R. China} }
\protect\vspace{1em}

\affil[*]{Corresponding author: wzkchem5@sdu.edu.cn}

\begin{document}
	
	\twocolumn[{
		\vspace*{0.5em}
		\maketitle
		\thispagestyle{firstpage}
		\label{firstpage}
		
		\begin{abstract}
			Many matrices in computational chemistry are neither sparse nor low-rank, making the design of low-scaling algorithms difficult. In this Perspective, we point out that many important matrices in computational chemistry, such as the Coulomb matrix, the electronic repulsion integral tensor, the density matrix, the localized molecular orbital (LMO) coefficient matrix, the Fock matrix, and the nuclear Hessian matrix share the same property: when their basis functions are suitably ordered, their off-diagonal blocks have low numerical ranks (despite that they as a whole have high numerical ranks). This property, termed off-diagonal low rank (ODLR), has been extensively studied in the mathematics community, but has surprisingly found very little use in computational chemistry. This Perspective reviews the existing mathematical literature on how to use the ODLR property of matrices to compactly store, as well as efficiently calculate or use them. Subsequently, we review the use of the ODLR property in computational chemistry, and point out possible future opportunities of devising new low-scaling methods for dense, full-rank matrices, exploiting the ODLR property. In particular, we prove for the first time that Fock matrices and LMO coefficient matrices satisfy the ODLR property, even if the system is gapless (in which case the matrices are dense). This paves the way to linear scaling electronic structure calculations of gapless systems at zero electronic temperature.
			
		\end{abstract}
		
	}] 
	
	\section{Introduction} \label{sec:introduction}
	
	The recent decades have seen a constant increase in the size of systems that can be studied by computational chemistry. On one hand, massively parallel calculations have pushed the frontiers of molecular mechanics (MM, including machine learning potentials), density functional theory (DFT) and correlated calculations to billions\cite{DeePMD-kit,DeePMD-kit2,carbonMD2021}, millions\cite{DGDFT,DG-TDDFT,LS3DF,CONQUEST,NOLSM} and hundreds of thousands of atoms\cite{Barca2024MP2,Gordon2022MP2,Mochizuki2025MP2}, respectively; on the other hand, the accessible system sizes of routine calculations using up to a few hundreds of processors have increased dramatically as well\cite{orca-5.0,BDF2020,CP2K,QuantumESPRESSO,Molpro,PySCF,TURBOMOLE,Q-Chem}. Apart from hardware advances, much of these improvements can be attributed to the development of more and more efficient algorithms; it has even been argued that the contribution of algorithmic advances dominates over that of hardware improvement\cite{Neese2024Faraday}.
	
	Key to the development of any computational chemistry algorithm is the avoidance of redundant calculation. In its most preliminary form, this consists of using the sparsity of the involved quantities, such that near-zero quantities are identified before they are calculated, and their calculation can be sidestepped. Taking self-consistent field (SCF) calculations as an example, the sparsity of the one-electron Hamiltonian $\mathbf{h}$ and exchange-correlation (XC) potential $\mathbf{v^{\mathrm{xc}}}$ have enabled the linear scaling calculation of the one-electron and XC contributions to the Fock matrix, while the sparsity of the density matrix $\mathbf{D}$ and electron repulsion integral (ERI) tensor $(\mu\nu|\kappa\lambda)$ (where Greek letters indicate local basis functions, including but not limited to atomic orbitals (AOs)) are a prerequisite for linear-scaling exact exchange algorithms\cite{LinK,COSX,sn-LinK,OchsenfeldReview2025,pseudospectral}. The sparsity of the ERI tensor also reduces the scaling of the Coulomb term from quartic to quadratic scaling\cite{prescreening}, albeit not to linear scaling. By noting that the diagonalization of the Fock matrix $\mathbf{F}$ can be replaced by the computation of Chebyshev polynomials of $\mathbf{D}$\cite{shang2010review}, direct minimization (where the energy is minimized as a function of $\mathbf{D}$ subject to idempotency constraints)\cite{DM,DM2}, or the block-diagonalization of localized molecular orbital (LMO) basis Fock matrices\cite{stewart1996LMO,FLMO,FLMO3,ACR-FLMO,Triad,iOI}, the update of the molecular orbital (MO) coeffcients $\mathbf{C}$ or density matrix can also be made linear scaling. Post-SCF calculations benefit even more from sparsity as long as they are formulated under the LMO basis, as evidenced from the success of linear scaling local M\o{}ller-Plesset perturbation theory\cite{DLPNO-MP2,PNO-LMP2,PNO-MP2}, local coupled cluster\cite{DLPNO-CCSD,DLPNO-CCSD2013,DLPNO-CCSDT,LCCSD,LNO-CCSDT,OSV-LCCSD,OSV-LCCSDT}, and local time-dependent DFT (TDDFT) calculations\cite{FLMO,FLMO3,ACR-FLMO,Triad,iOI,NOLMO-TDDFT,TDDFTMI}, to name a few.
	
	In some cases, the matrix or tensor to be approximated is not sufficiently sparse, but has low rank; a low-rank decomposition (including but not limited to Cholesky decomposition, truncated QR factorization, singular value decomposition (SVD) and diagonalization) can then compress the matrix or tensor efficiently. This is best illustrated by the resolution of the identity (RI) method\cite{RIreview}, where the low-rank property of the ERI tensor is used to reduce the formally quartic scaling Coulomb building algorithm to cubic scaling:
	\begin{equation}
	(\mu\nu|\kappa\lambda) = \sum_{PQ}(\mu\nu|P)(\mathbf{V}^{-1})_{PQ}(Q|\kappa\lambda), \quad V_{PQ} = (P|Q), \label{eq:RI}
	\end{equation}
	where $P$ and $Q$ are RI auxiliary basis functions. Though RI has a higher scaling than pure integral prescreening due to the Cholesky decomposition of the Coulomb metric $V_{PQ}$ (which scales cubically), it is usually faster than non-RI algorithms even for systems with hundreds of atoms. Another, more contrived example is the density matrix renormalization group (DMRG) method\cite{DMRGreview2020,DMRGreview2023}, where the high-dimensional tensor of the multielectronic wavefunction coefficients is recursively compressed into the products of many three-dimensional tensors. In some sense, low-rank compression can be regarded as utilizing sparsity in the eigenbasis, where the matrix to be approximated becomes diagonal, and only a small portion of the diagonal elements are non-negligible.
	
\begin{figure*}[htbp]
		\begin{tabular}{ccc}
			\resizebox{0.32\textwidth}{!}{\includegraphics{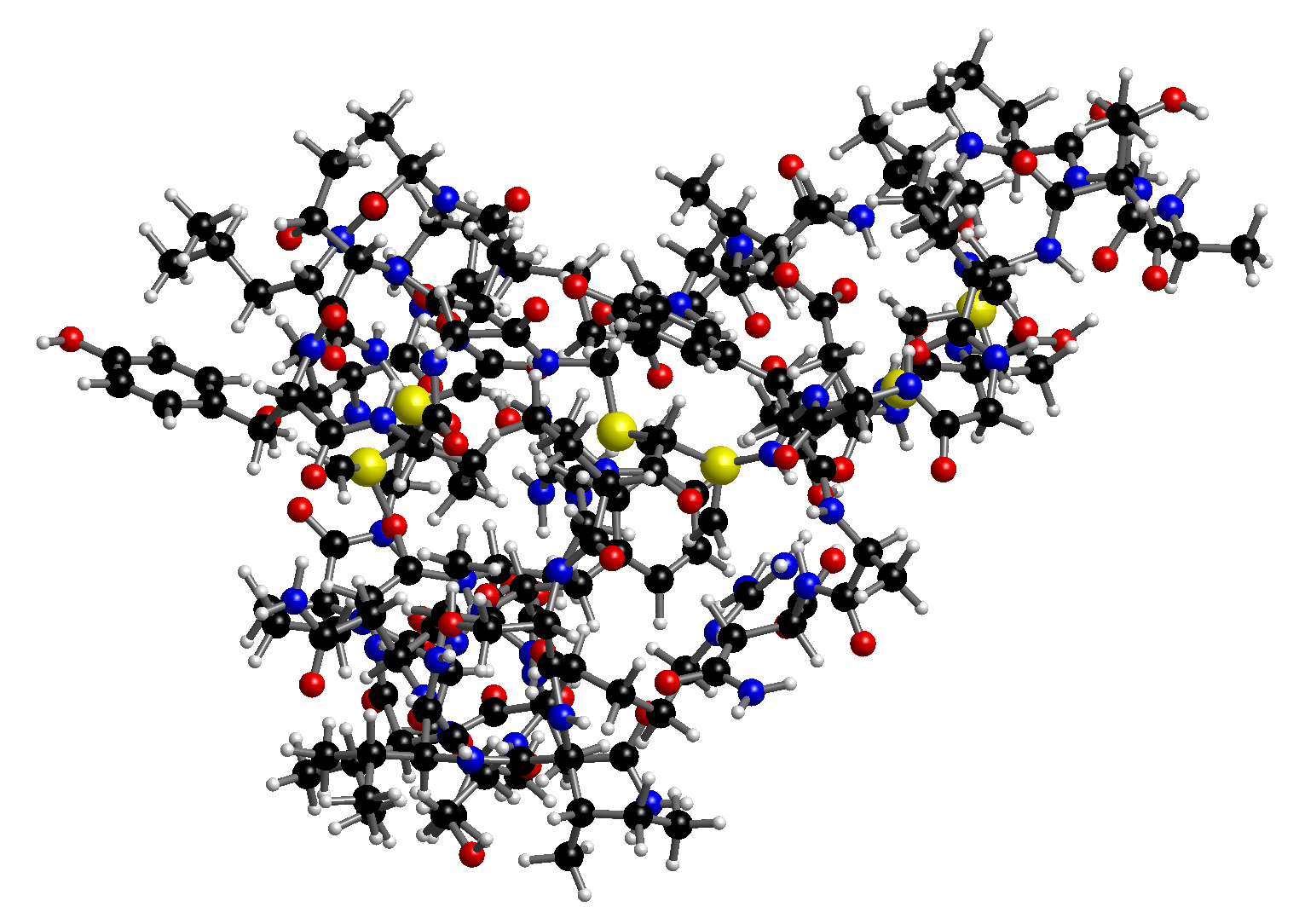}} &
			\resizebox{0.32\textwidth}{!}{\includegraphics{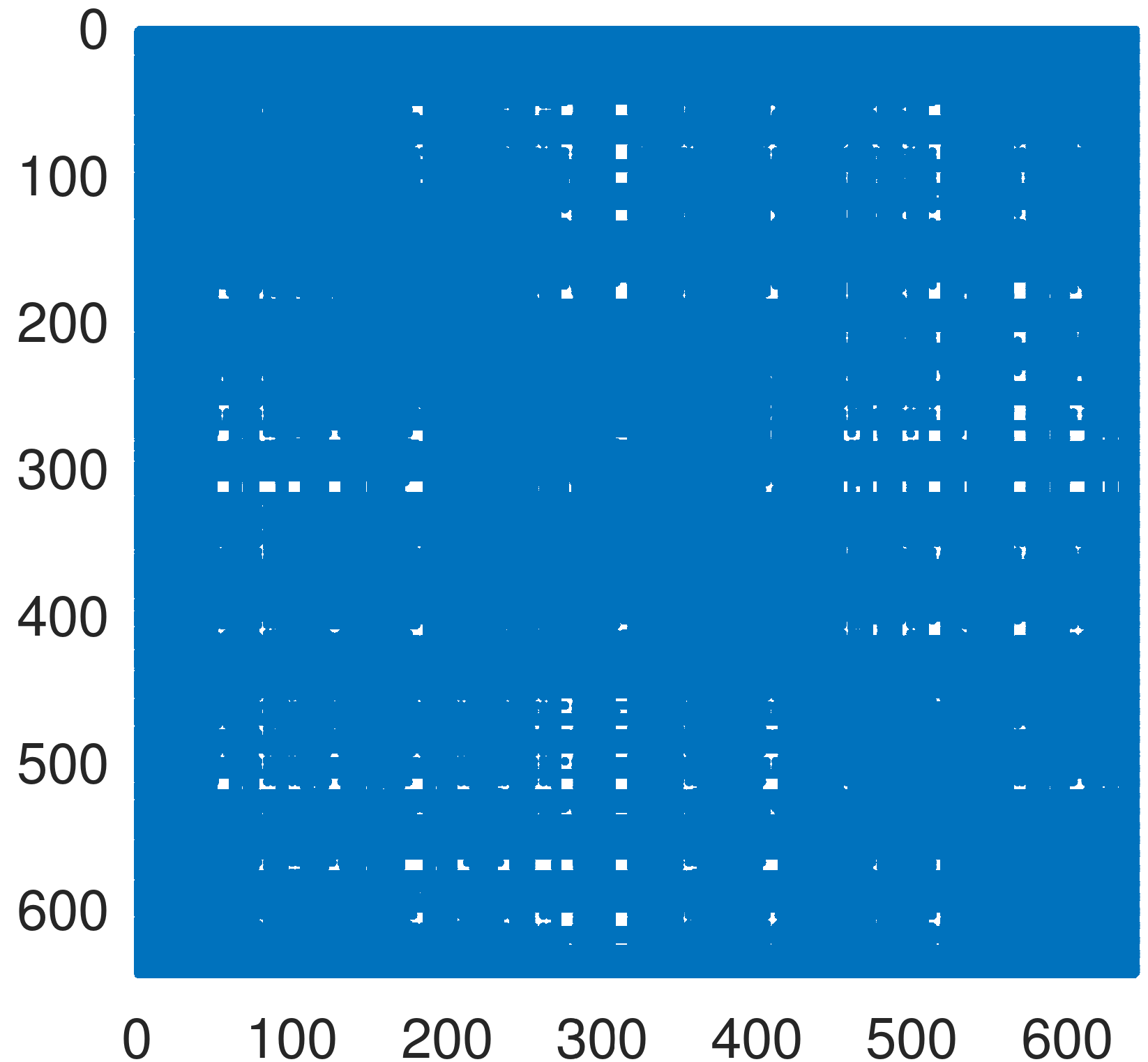}} &
			\resizebox{0.32\textwidth}{!}{\includegraphics{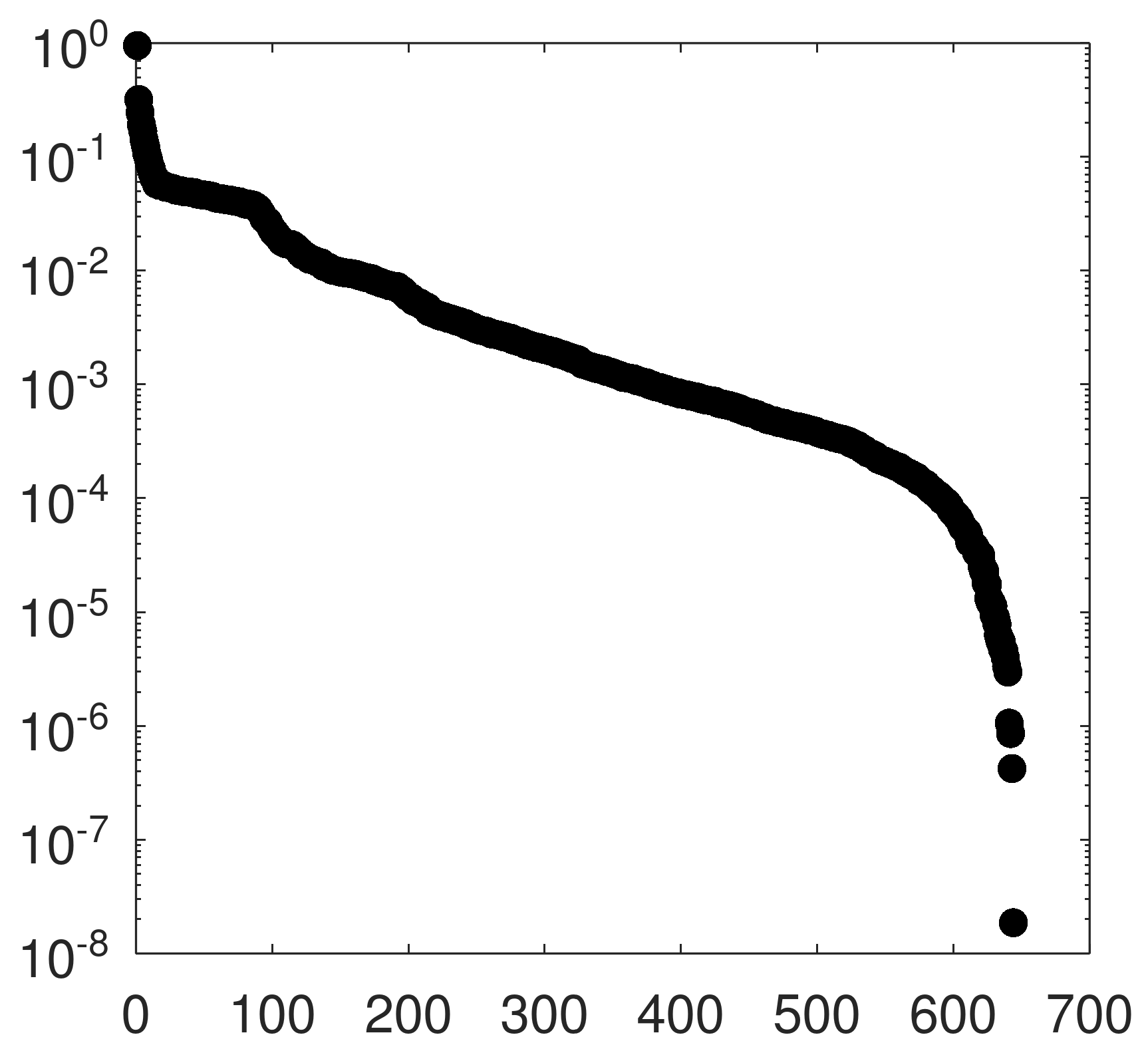}} \\
                    (a) & (b) & (c) \\
		\end{tabular}
		\caption{(a) Molecular structure of a small protein, crambin (geometry taken from Ref.~\citenum{DLPNO-CCSD2013}). (b) Coulomb matrix $\mathbf{J}$ of the GFN2-xTB\cite{GFN2-xTB} L\"owdin charges of crambin calculated using ORCA 6.1.0\cite{ORCA6}, showing elements whose absolute values are greater than $10^{-4}$. (c) Singular values (unit: a.u.) of $\mathbf{J}$, sorted in descending order; note the logarithmic scale.}
		\label{fig:J}
	\end{figure*}

	However, many important matrices and tensors in computational chemistry are neither sparse nor low-rank. A prototypical example is the Coulomb matrix between a set of point charges $\{q_i\}$:
	\begin{equation}
	J_{ij} = \left\{ \begin{array}{cc} \frac{q_iq_j}{R_{ij}}, & i\neq j \\ 0, & i=j \end{array} \right.,
	\end{equation}
	where $R_{ij}$ is the Euclidean distance between the charges $q_i$ and $q_j$. While the denseness of $\mathbf{J}$ is obvious due to the long-range character of the Coulomb interaction, its SVD shows that $\mathbf{J}$ is also almost full-rank, with about 90\% singular values exceeding $10^{-4}$ (Figure~\ref{fig:J}). These explain why prescreening and RI cannot reduce the scaling of the Coulomb building algorithm to below quadratic.
	
\begin{figure}[htbp]
		\begin{tabular}{cc}
			\resizebox{0.22\textwidth}{!}{\includegraphics{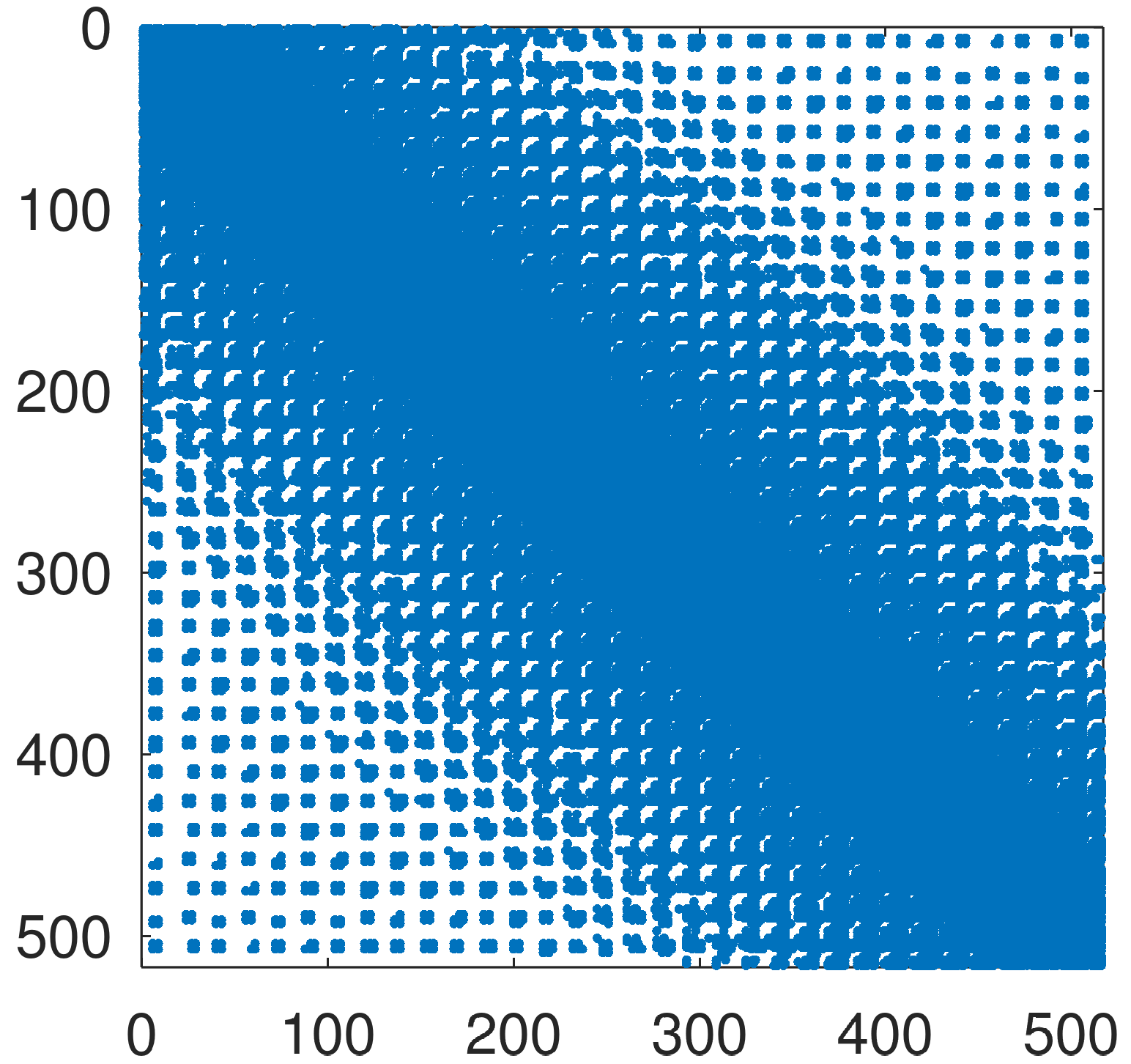}} & \resizebox{0.22\textwidth}{!}{\includegraphics{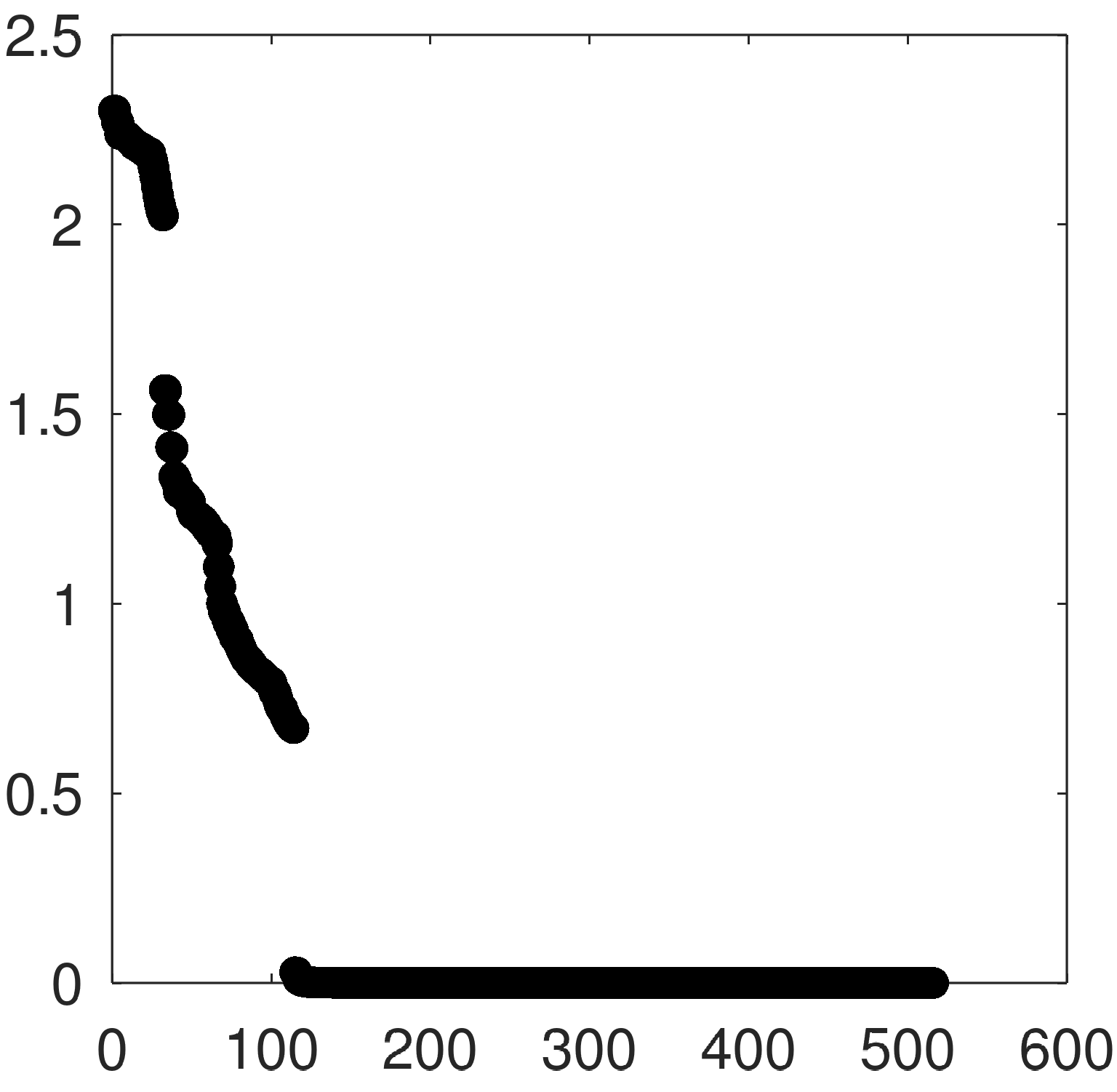}} \\
            (a) & (e) \\
			\resizebox{0.22\textwidth}{!}{\includegraphics{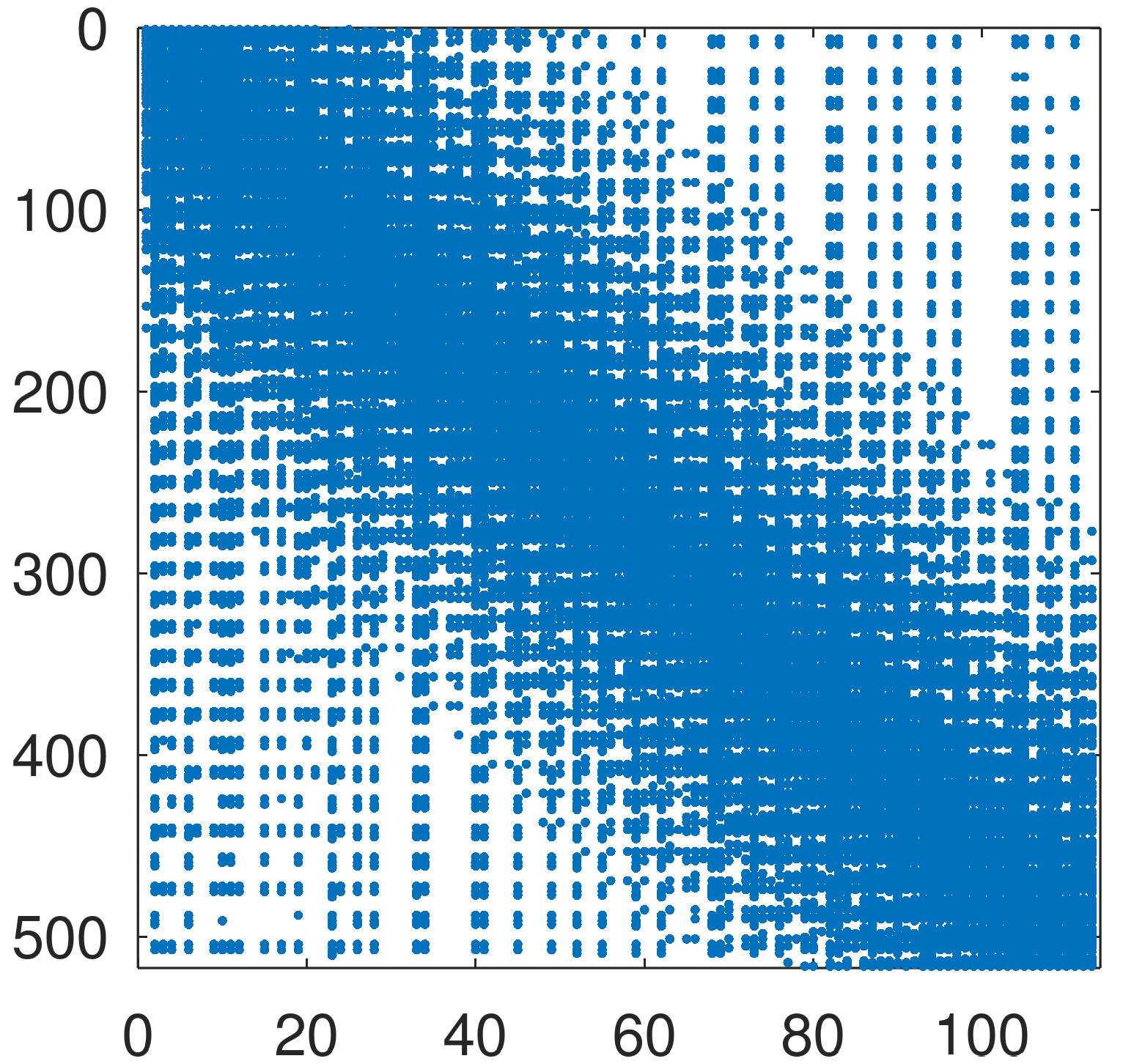}} & \resizebox{0.22\textwidth}{!}{\includegraphics{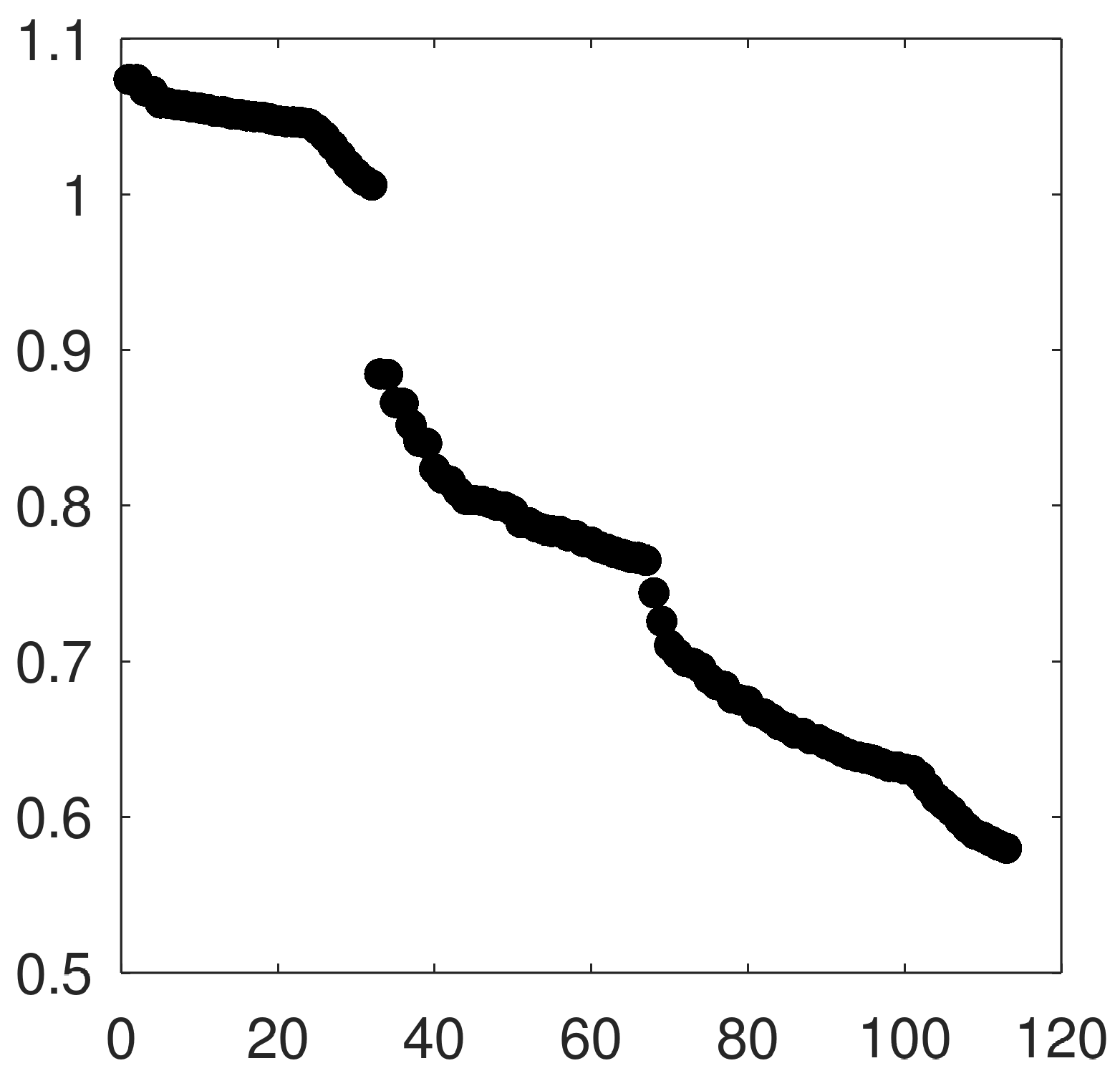}} \\
            (b) & (f) \\
			\resizebox{0.22\textwidth}{!}{\includegraphics{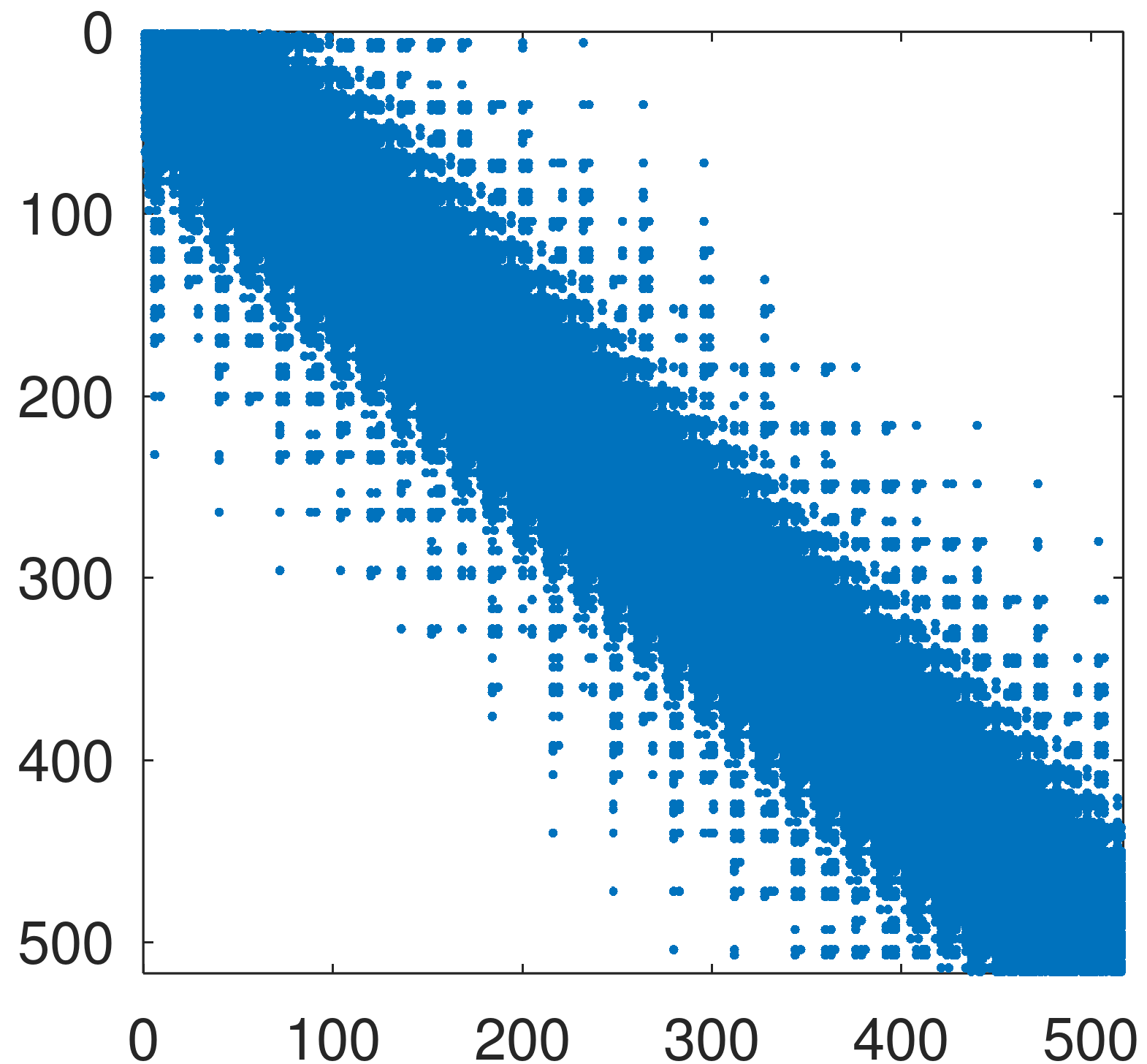}} & \resizebox{0.22\textwidth}{!}{\includegraphics{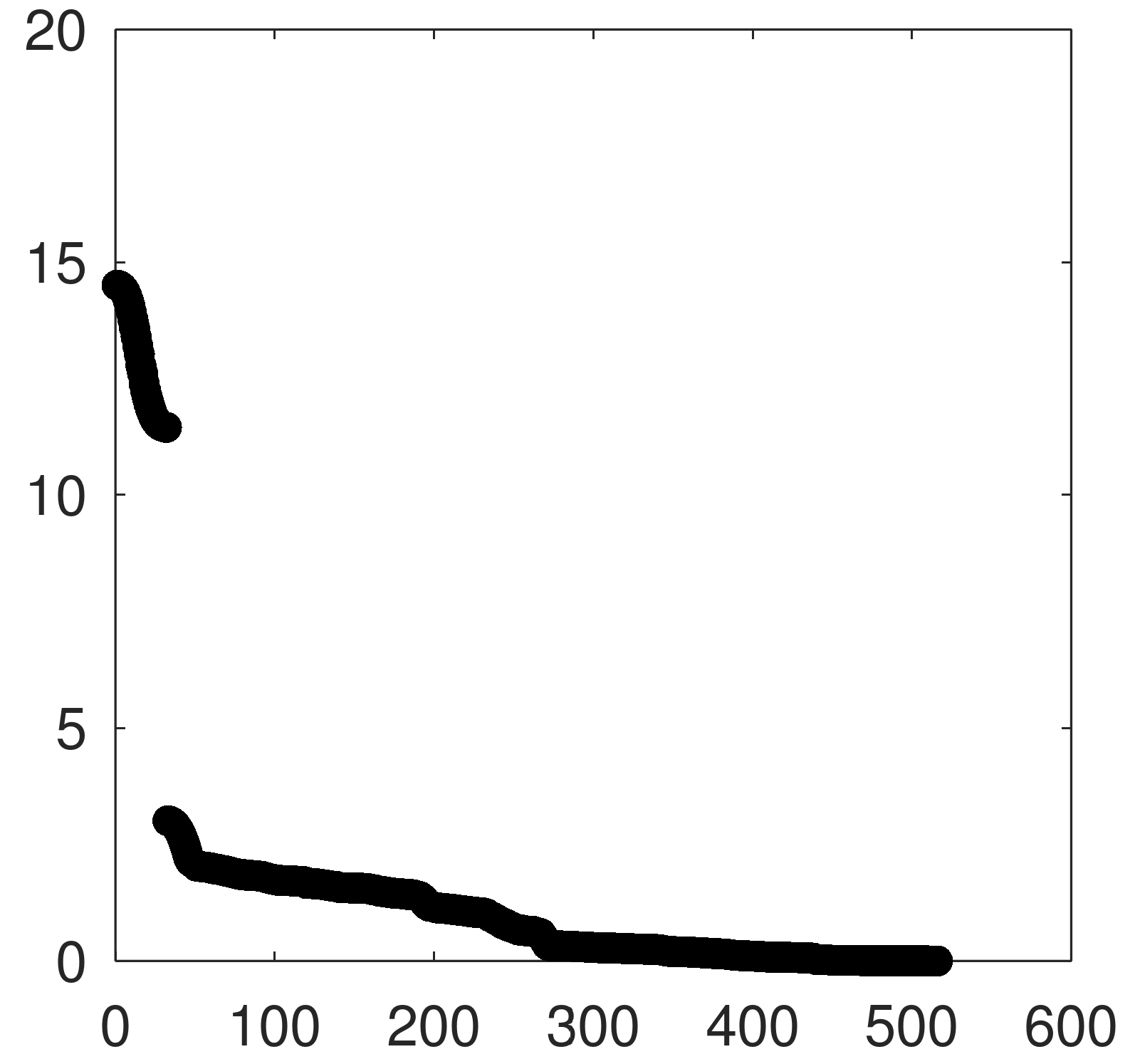}} \\
            (c) & (g) \\
			\resizebox{0.22\textwidth}{!}{\includegraphics{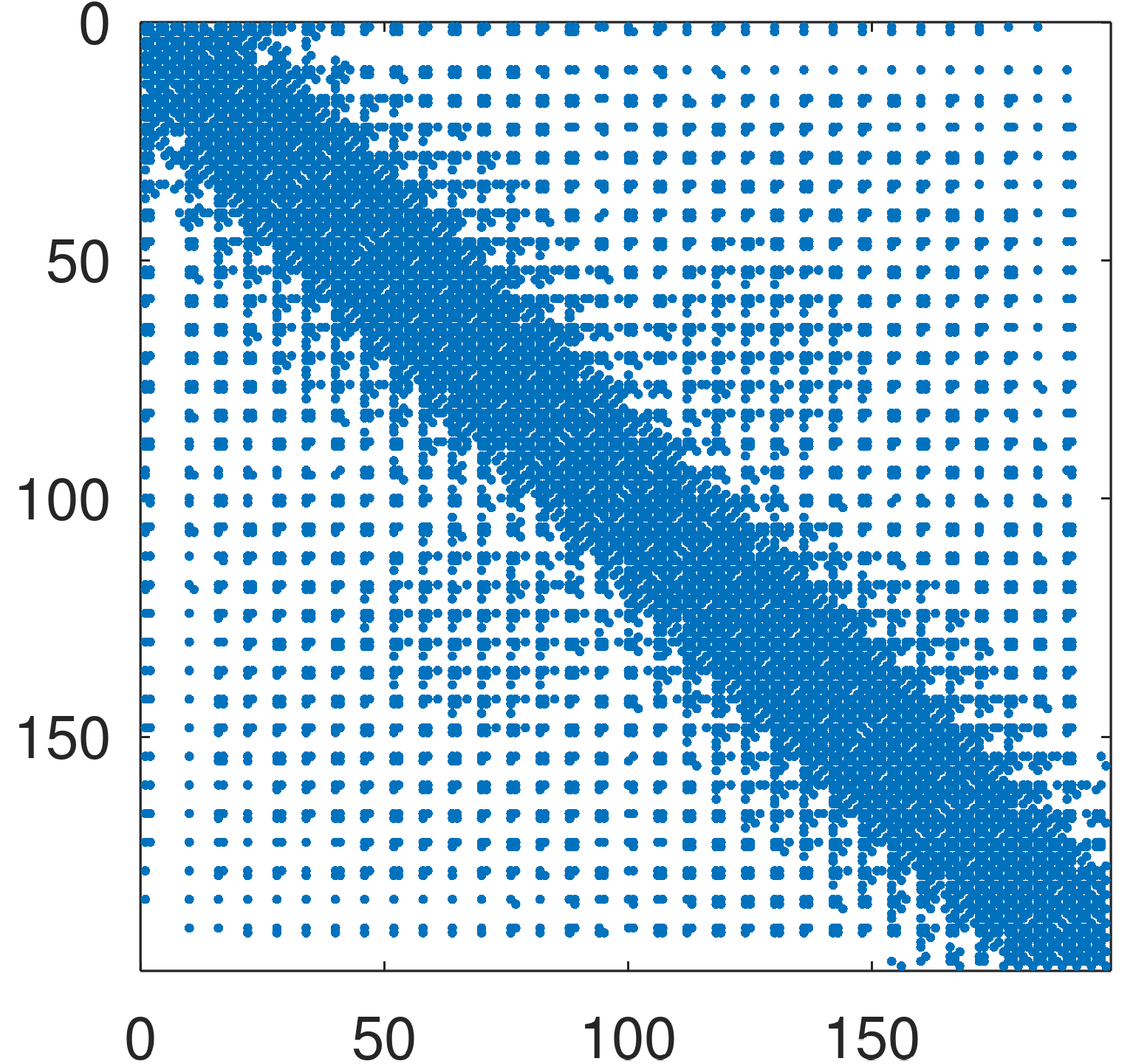}} & \resizebox{0.22\textwidth}{!}{\includegraphics{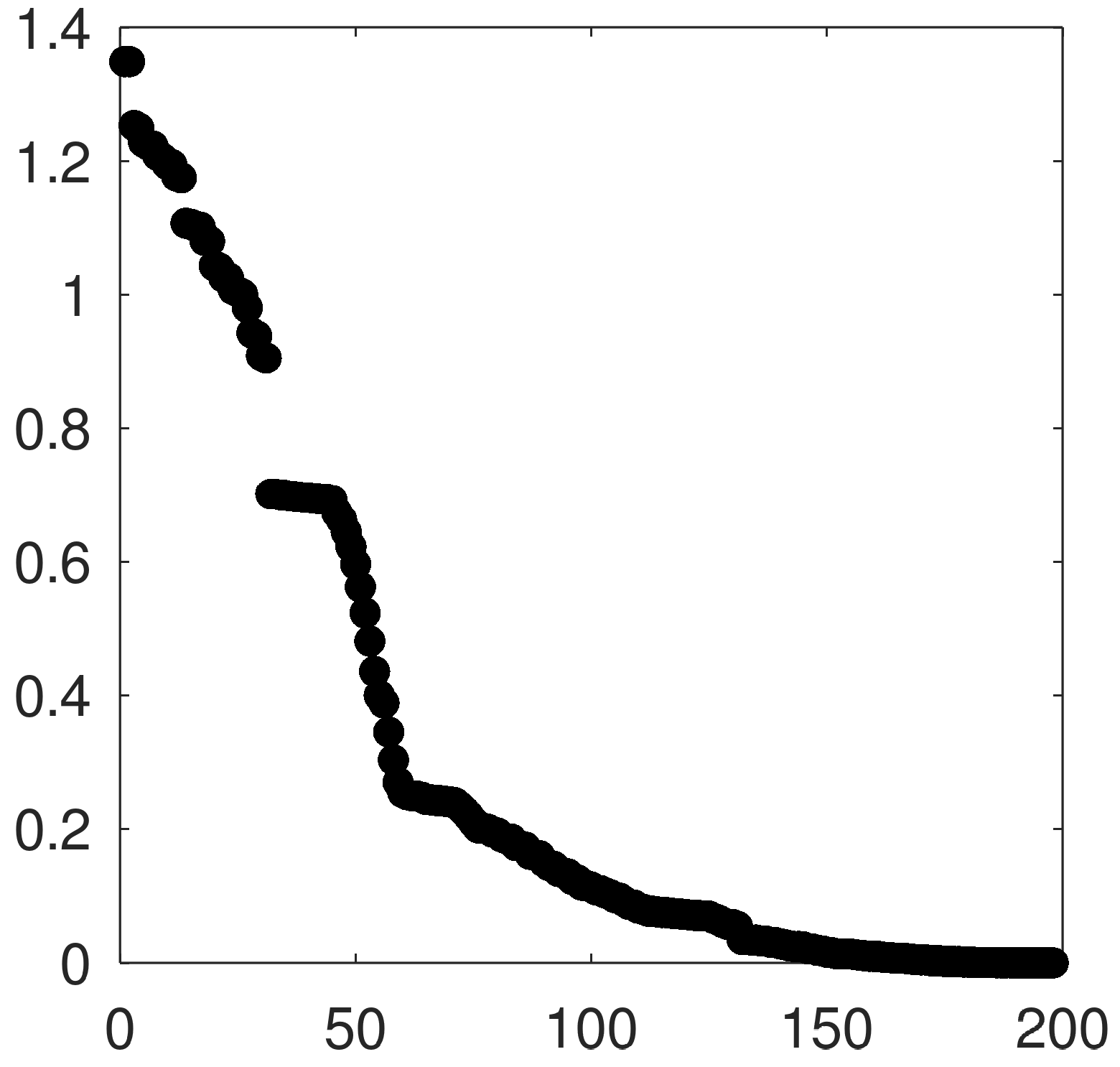}} \\
            (d) & (h) \\
		\end{tabular}
		\caption{(a) Total density matrix, (b) $\alpha$ LMO coefficient matrix calculated using the fragment LMO (FLMO) method\cite{FLMO}, (c) $\alpha$ Fock matrix, and (d) Hessian matrix of the lowest triplet state of the linear polyene \ce{C32H34} (a nearly gapless molecule, HOMO-LUMO gap: 0.88 eV), computed at the B3LYP\cite{B88x,LYP,Becke93,B3LYP}-D3\cite{DFT-D3,DFT-D3BJ}/def2-SVP\cite{DEF2SVP} level of theory, showing elements whose absolute values are greater than $10^{-4}$. (e-h) Singular values of the aforementioned matrices. The atoms, basis functions and LMOs are ordered from one end of the molecule to the other end. All calculations were performed using a development version of BDF\cite{BDF2020}.}
		\label{fig:C32}
	\end{figure}

	Another famous example is the density matrices of systems with near zero highest occupied molecular orbital (HOMO)-lowest unoccupied molecular orbital (LUMO) gaps, such as the triplet linear polyene \ce{C32H34} (Figure~\ref{fig:C32}(a)). They are dense, and while they have low numerical rank, this property cannot be used for developing low-scaling algorithms, since their rank is $O(N)$ (where $N$ is the size of the matrix) owing to the presence of $O(N)$ occupied orbitals (Figure~\ref{fig:C32}(e)). The denseness of these density matrices follows from Kohn's conjecture\cite{kohn1996} (later proved rigorously\cite{benzi2013}), which points out that the density matrix elements $D_{\mu\nu}$ of a system decay exponentially when either the HOMO-LUMO gap or the electronic temperature is non-zero, but only polynomially (for three-dimensional systems, quadratically\cite{goedecker1998}) when both are zero. For brevity, unless otherwise noted, in the following we will assume all calculations are performed at zero electronic temperature. Similarly, the localized molecular orbital (LMO) coefficient matrix also becomes dense for gapless systems (Figure~\ref{fig:C32}(b))\cite{kohn1959}, again reflecting the non-locality of the electronic structure of gapless systems, while it is full rank due to the orthonormality requirement of the LMOs (Figure~\ref{fig:C32}(f)).
	
	Two less recognized examples of (almost) full-rank and dense matrices are the hybrid functional Fock matrices, as well as nuclear Hessians of gapless systems.
	While Fock matrices of (semi)local pure functionals are sparse even for gapless systems, they become dense for hybrid functional calculations of gapless systems due to the nonlocal character of exact exchange, which carries the denseness of the density matrix into the Fock matrix (Figure~\ref{fig:C32}(c)); meanwhile, they are usually full-rank since orbital energies are almost never exactly zero (Figure~\ref{fig:C32}(g)).
	Finally, nuclear Hessians are similar to Fock matrices in that they are sparse for gapped systems but dense for gapless systems\cite{O1NumHess} (Figure~\ref{fig:C32}(d)), and in all cases almost full-rank (Figure~\ref{fig:C32}(h); the null space is usually no more than six-dimensional, spanned by the translational and rotational modes of the molecule). The origin of the denseness of Hessians is however more involved, and we postpone its discussion to Section~\ref{sec:ODLR_Hessian}.
	
\begin{figure}[htbp]
        \begin{tabular}{cc}
			\resizebox{0.22\textwidth}{!}{\includegraphics{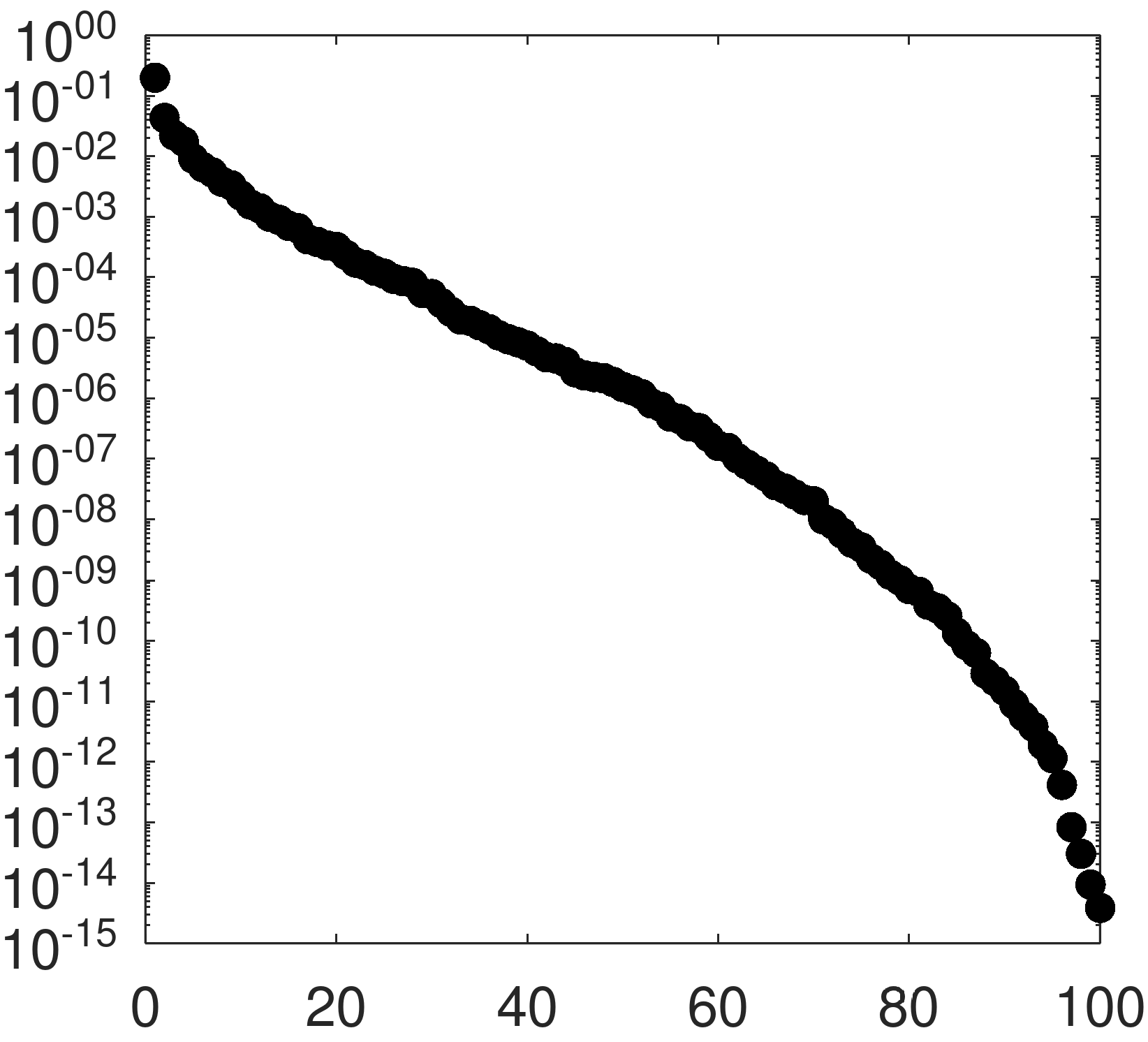}} & \resizebox{0.22\textwidth}{!}{\includegraphics{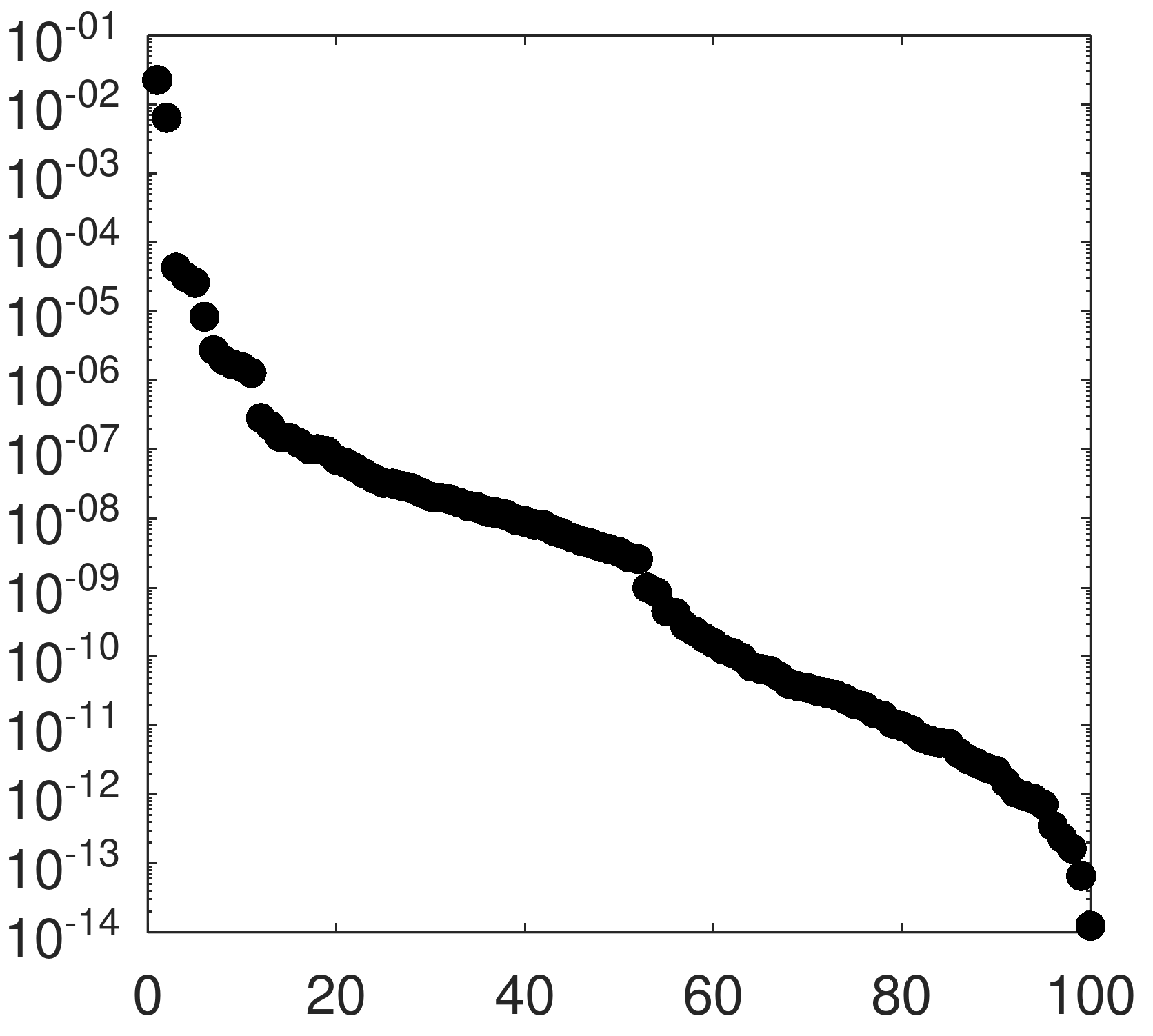}} \\
            (a) & (b) \\
			\resizebox{0.22\textwidth}{!}{\includegraphics{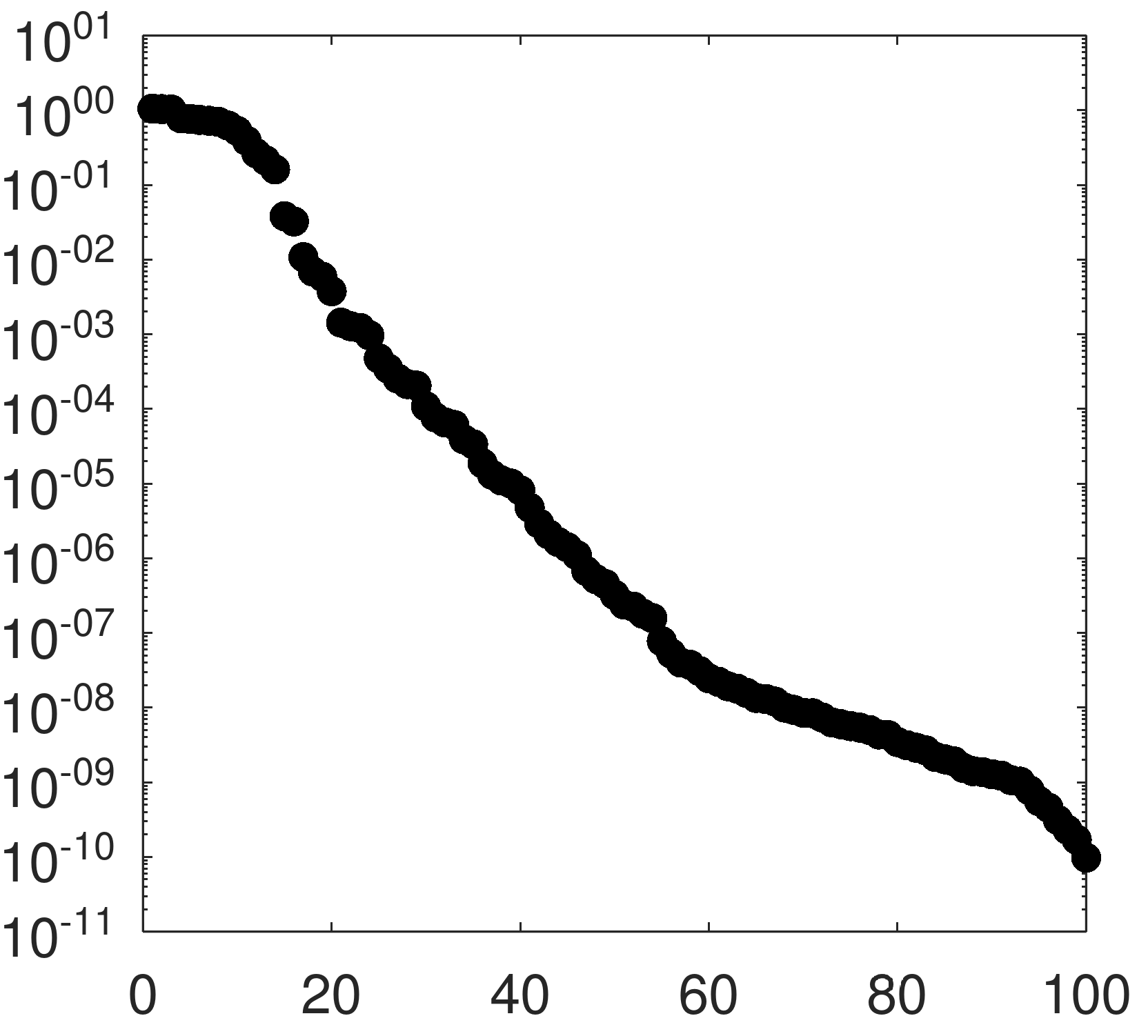}} & \resizebox{0.22\textwidth}{!}{\includegraphics{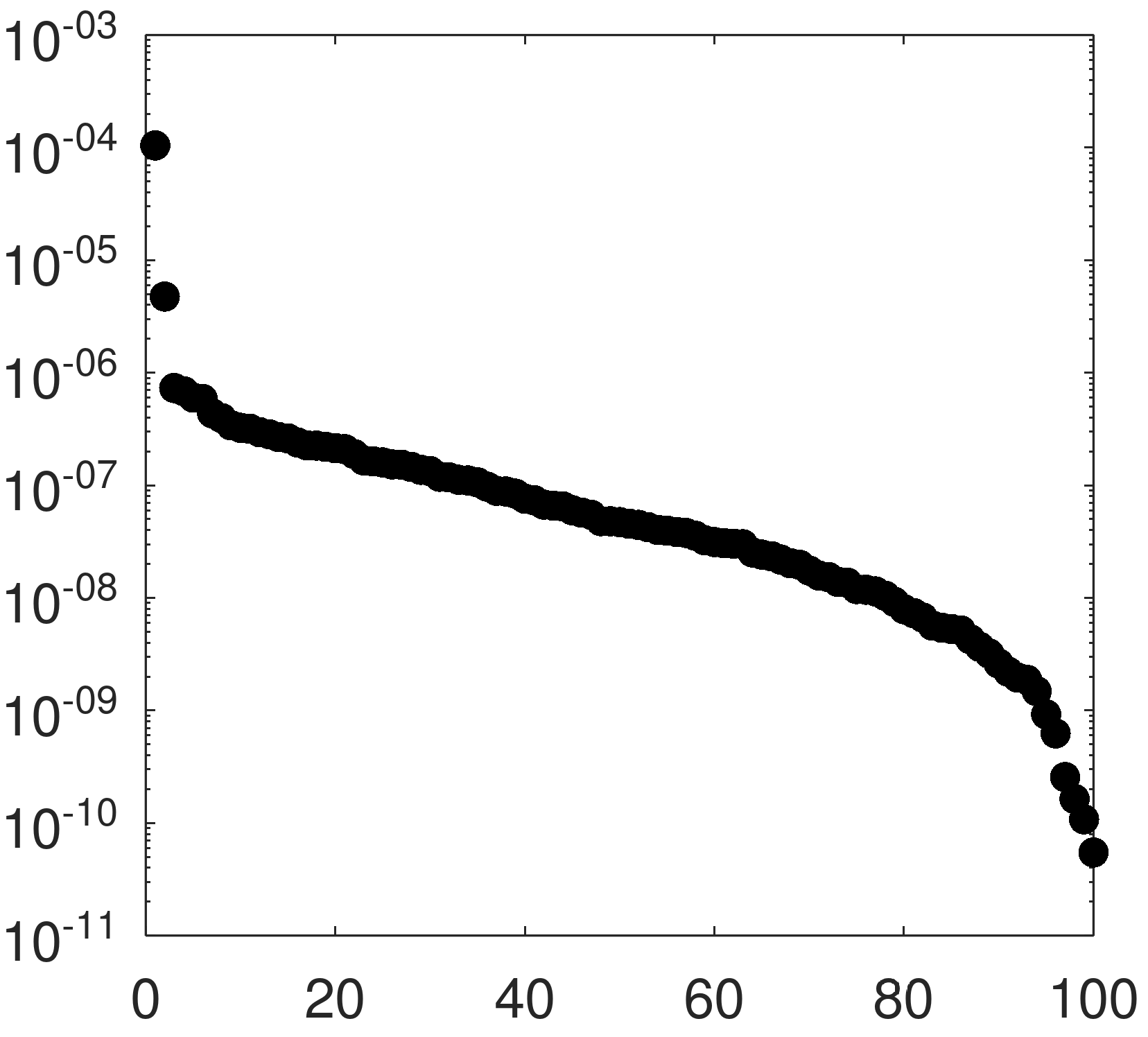}} \\
            (c) & (d) \\
			\multicolumn{2}{c}{\resizebox{0.22\textwidth}{!}{\includegraphics{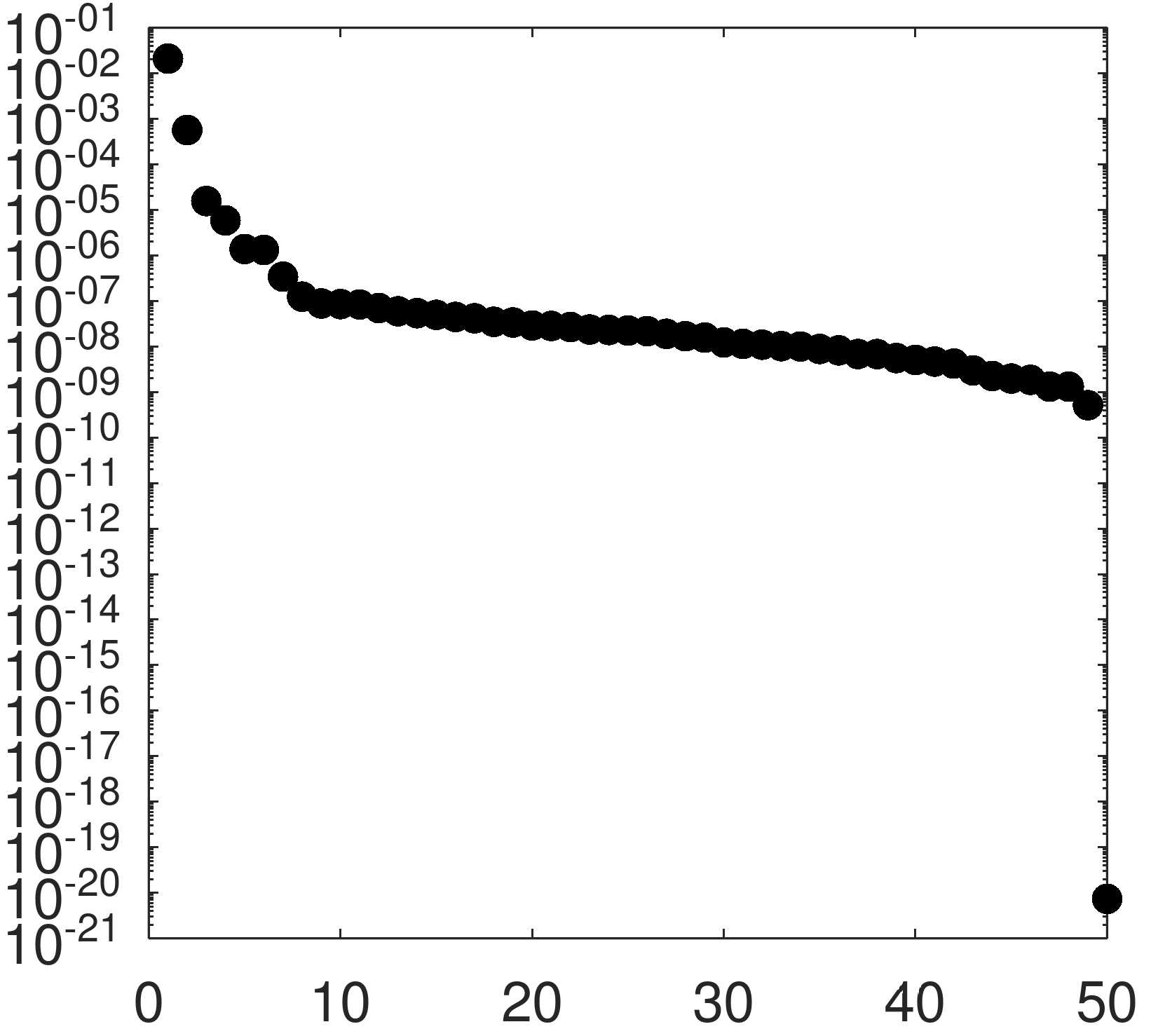}}} \\
            \multicolumn{2}{c}{(e)} \\
		\end{tabular}
		\caption{The singular values of the upper right $100\times 100$ blocks of the (a) Coulomb matrix of the GFN2-xTB point charges of crambin, as well as (b) the total density matrix, (c) $\alpha$ LMO coefficient matrix, and (d) $\alpha$ Fock matrix of the lowest triplet state of the linear polyene \ce{C32H34} at the B3LYP/def2-SVP level; (e) the singular values of the upper right $50\times 50$ block of the Hessian matrix of the lowest triplet state of \ce{C32H34} at the B3LYP-D3/def2-SVP level. Note the logarithmic scale.}
		\label{fig:off_diagonal_SVD}
	\end{figure}

	At this point, one natural question arises: is it possible to consistently and efficiently compress matrices that are both dense and high-rank? We can obtain some useful hints from the matrices in Figure~\ref{fig:C32}(a-d), whose off-diagonal blocks clearly possess some regularity. Indeed, SVD shows that the \emph{far off-diagonal blocks} of the density, LMO coefficient, Fock and Hessian matrices in Figure~\ref{fig:C32}(a-d), as well as the Coulomb matrix in Figure~\ref{fig:J}(b), all have a low numerical rank: the singular values decay exponentially when put in descending order, so that only a handful of singular values are non-negligible (Figure~\ref{fig:off_diagonal_SVD}). Here ``far off-diagonal'' can be loosely understood as blocks that are far from the diagonal line; although this definition only makes sense for one-dimensional systems, we will properly generalize this concept to high-dimensional systems later. Hereafter, we call this property the \emph{``off-diagonal low rank'' (ODLR)} property. The ``ODLR'' acronym has found some popularity in the mathematics community\cite{boulle2024,ODLR_PageRank,xia2010,aminfar2016}, but is yet to be popularized in computational chemistry\cite{O1NumHess}.
	
	The observation that so many matrices encountered in computational chemistry has the ODLR property is surprising, since the Coulomb, density, LMO coefficient, Fock and Hessian matrices contain completely different physics and appear in very different contexts. Meanwhile, the ODLR property offers a new way to compress matrices, and thus may prove useful in developing new low-scaling methods.
	In this Perspective, we will first review the existing mathematical foundations of the ODLR property, and how to best utilize the ODLR property to compress matrices and perform matrix calculus. Subsequently, we derive the ODLR properties of the Coulomb, density, LMO coefficient, Fock and Hessian matrices, and demonstrate how their ODLR properties enable the design of several known low-scaling algorithms. Finally, we conclude this Perspective by possible ways to use the ODLR property to devise novel low-scaling computational methods.
    Note that this Perspective is not limited to a review of known conclusions in the literature; rather, we also present numerous new results, as well as strengthened versions of known results.
    In particular, we demonstrate for the first time that (under mild assumptions) a much larger variety of matrices have the ODLR property than previous known, in particular Fock matrices and LMO coefficient matrices.
	
	\section{ODLR: Mathematical foundations} \label{sec:foundations}
	
	\subsection{Real-space perspective: ODLR property of operators}  \label{sec:foundations_realspace}
	
	Although the concept of ODLR is unfamiliar to the computational chemistry community, it has been known by mathematicians (notably Hackbusch, B\"orm and Grasedyck et al.) since the late 1980s\cite{FMM1987,hackbusch1989,borm2003Hmatrix,hackbusch2000Hmatrix,hackbusch2004,hackbusch2002H2matrix} that the matrix representations of certain linear operators possess the ODLR property. Consider an operator that represents the convolution with a sufficiently smooth kernel $K(\mathbf{r},\mathbf{r}')$, which has no poles except possibly at $\mathbf{r}=\mathbf{r}'$:
	\begin{equation}
	\left(\hat{O}f\right)(\mathbf{r}) = \int K(\mathbf{r},\mathbf{r}')f(\mathbf{r}')d\mathbf{r}'. \label{eq:convolution}
	\end{equation}
	Notably, $K(\mathbf{r},\mathbf{r}')$ can be anisotropic and does not have to decay very fast or possess translational invariance; essentially only smoothness is required. Discretizing the operator on a local basis, one arrives at a matrix representation of $\hat{O}$, $\mathbf{O}$. If $X$ and $Y$ are two ``well-separated'' groups of local basis functions, in the sense that the regions $X$ and $Y$ can be separated by a boundary whose curvature is small everywhere (such as a flat plane or a large sphere),
	then the off-diagonal block of $\mathbf{O}$, $\mathbf{O}^{XY}$ (which is a shorthand notation for the matrix composed of $\{O_{\mu\nu}, \mu\in X, \nu\in Y\}$), is called a far off-diagonal block.
    The \emph{ODLR property} of $\mathbf{O}$ is now defined as the statement that $\mathbf{O}^{XY}$ is numerically rank-deficient as long as it is not too small.
    This constitutes a rigorous definition of the ODLR property for systems with arbitrary dimensionality and basis function ordering; by comparison, the illustrations of Figure~\ref{fig:off_diagonal_SVD} only work for one-dimensional systems and a sequential basis function order. Note that if the basis functions that belong to the region $X$ (or $Y$) do not have contiguous indices, it is understood that a suitable permutation of the basis functions is made, so that $\mathbf{O}^{XY}$ becomes a contiguous block.

\begin{figure}[htbp]
    \resizebox{0.48\textwidth}{!}{\includegraphics{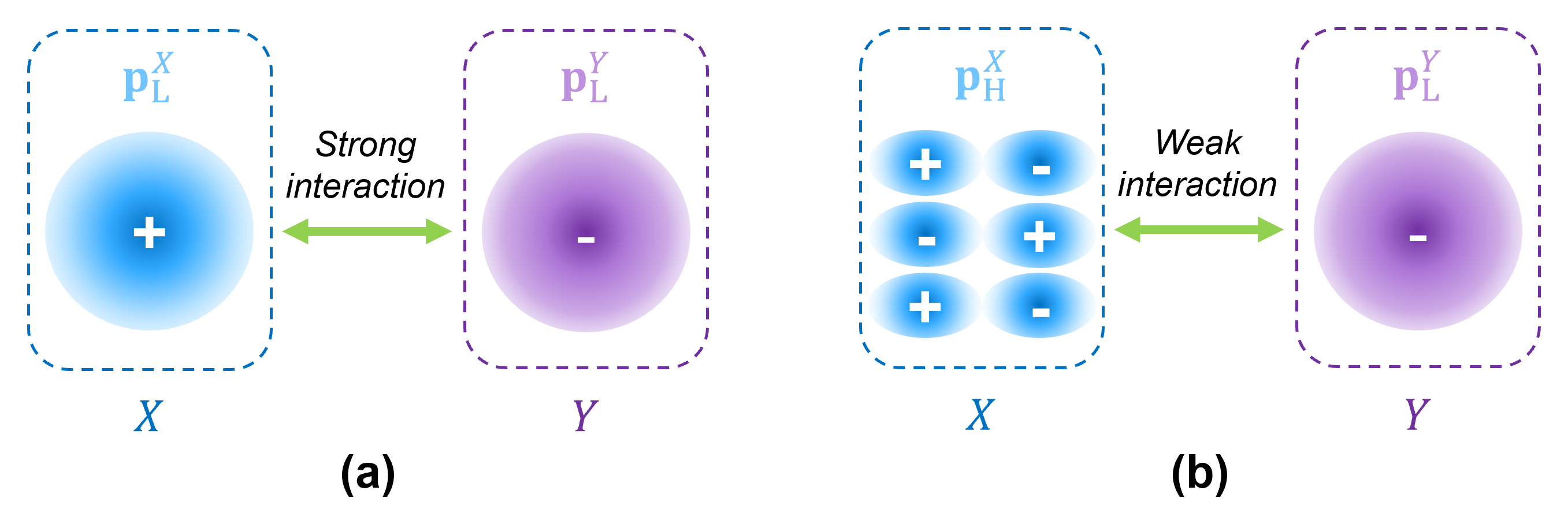}}
		\caption{Schematic illustration of the origin of the ODLR property for the matrix representations of smooth convolution operators. The off-diagonal block between two distant regions has low rank, and the dominant left and right singular vectors tend to be delocalized over the $X$ and $Y$ regions, respectively, with no or few nodes. (a) Interaction between a dominant singular vector in $X$ with a dominant singular vector in $Y$; (b) interaction between a non-dominant singular vector in $X$ with a dominant singular vector in $Y$, where significant cancellation between the opposite phase lobes occur.}
		\label{fig:lobes}
	\end{figure}

    The above results can be understood intuitively as follows. 
	First, construct two orthonormal bases that are localized in the regions $X$ and $Y$, respectively, while being as localized in the momentum space as possible. For example, the bases can be chosen as the eigenvectors $\{p^X_{\mu i}\}, \{p^Y_{\nu i}\}$ of the respective diagonal blocks of the kinetic energy matrix (hereafter called the ``local kinetic eigenbasis'' (LKE), with each eigenvector indexed by $i,j,\ldots$), $\mathbf{T}^{XX}$ and $\mathbf{T}^{YY}$, where
	\begin{equation}
	T_{\mu\nu} = \langle \mu | -\frac{1}{2}\nabla^2 | \nu \rangle.
	\end{equation}
	The basis functions $\{p^X_{\mu i}\}, \{p^Y_{\nu i}\}$ can then be put in ascending order with respect to their kinetic energy expectation values, so that they tend to have an increasing number of nodes. Now, since $K(\mathbf{r},\mathbf{r}')$ is smooth, and the boundary separating $X$ and $Y$ is also smooth, the contributions of positive and negative lobes of an LKE function will tend to cancel out, and this effect becomes more pronounced as the number of nodes increases, where each positive lobe is closer to its neighboring negative lobe(s) and vice versa (Figure~\ref{fig:lobes}). As a result, if we transform $\mathbf{O}$ from the original, localized-in-space basis to the LKE,
	\begin{equation}
	\tilde{\mathbf{O}}^{XY} = (\mathbf{p}^X)^T \mathbf{O}^{XY} \mathbf{p}^Y, \label{eq:pOp}
	\end{equation}
	we will find that only the upper left corner of $\tilde{\mathbf{O}}^{XY}$ is non-negligible (where the subscripts L and H denote low energy and high energy, respectively):
	\begin{equation}
	\tilde{\mathbf{O}}^{XY} \approx \left( \begin{array}{cc} \tilde{\mathbf{O}}^{XY}_{\rm LL} & \mathbf{0}_{\rm LH} \\ \mathbf{0}_{\rm HL} & \mathbf{0}_{\rm HH} \end{array} \right),
	\end{equation}
	as all elements of $\tilde{\mathbf{O}}^{XY}$ involving at least one high-energy LKE function are small.
    $\mathbf{p}^X$ and $\mathbf{p}^Y$ can similarly be split into two rectangular matrices containing the low-energy and high-energy LKE basis functions, e.g.,
    \begin{equation}
    \mathbf{p}^X = \left( \begin{array}{cc}\mathbf{p}^X_{\rm L} & \mathbf{p}^X_{\rm H} \end{array} \right).
    \end{equation}
    It therefore follows that $\tilde{\mathbf{O}}^{XY}$ and thus $\mathbf{O}^{XY}$ has low rank.
	Specifically, given the SVD of $\tilde{\mathbf{O}}^{XY}_{\rm LL}$ (which we assume to have full rank),
	\begin{equation}
	\tilde{\mathbf{O}}^{XY}_{\rm LL} = \mathbf{U}_{\rm L}^{XY}\mathbf{\sigma}^{XY}(\mathbf{V}_{\rm L}^{XY})^T,
	\end{equation}
	we have
	\begin{eqnarray}
	\mathbf{O}^{XY} & \approx & \mathbf{U}^{XY}\mathbf{\sigma}^{XY}(\mathbf{V}^{XY})^T, \label{eq:pUspV} \\
	\mathbf{U}^{XY} & = & \mathbf{p}^X_{\rm L}\mathbf{U}_{\rm L}^{XY}, \\
	\mathbf{V}^{XY} & = & \mathbf{p}^Y_{\rm L}\mathbf{V}_{\rm L}^{XY}, 
	\end{eqnarray}
	where we have used the orthonormality of the LKE functions.
	
    At this point, we emphasize that the use of LKE functions is not essential; the basis functions only need to be localized in the regions $X$ and $Y$, while having different oscillation frequencies. Thus, more generally, one can use a wavelet basis instead of the LKE (recall that wavelets are functions that are mostly localized within a given region, and can oscillate one or more times). For example, Beylkin et al.\cite{beylkin1991fast}~have used Daubechies wavelets as the basis instead of the LKE, giving a compressed representation of $\mathbf{O}$ called the ``nonstandard form''. The compression efficiency of the nonstandard form can be rigorously estimated from wavelet theory. Nonstandard forms of the Coulomb operator have been used in the MADNESS\cite{MADNESS} computational chemistry program, to achieve linear scaling SCF calculations\cite{MADNESS_J,MADNESS_K} with close to basis set limit accuracy. Two similar implementations, also using Daubechies wavelets, are available in the BigDFT\cite{BigDFT} and ABINIT\cite{wavelet_ABINIT} software packages.
	
	We note that weaker claims have been made in the literature as well. For example, Hackbusch et al.~have introduced the concept of admissibility\cite{hackbusch1989,borm2003Hmatrix,hackbusch2004,hackbusch2002H2matrix}, where two regions $X$ and $Y$ are admissible only when the distance between $X$ and $Y$ is larger than or, depending on the context, at least comparable to the dimensions of $X$ and $Y$ themselves. They then concluded that $\mathbf{O}^{XY}$ has low rank when $X$ and $Y$ are admissible. The reason for this additional requirement is that the proofs require a Taylor expansion of the kernel $K(\mathbf{r},\mathbf{r}')$ around a point $\mathbf{r}_0$ in the region $Y$:
	\begin{equation}
	K(\mathbf{r},\mathbf{r}') \approx \sum_{i=1}^{N_{\rm max}} c_i(\mathbf{r}-\mathbf{r}_0)\phi_i(\mathbf{r}'-\mathbf{r}_0), \label{eq:Taylor}
	\end{equation}
	where $c_i(\mathbf{r}-\mathbf{r}_0)$ are the expansion coefficients, and $\phi_i$ are polynomials. Such an expansion separates the variables $\mathbf{r}$ and $\mathbf{r}'$, and naturally yields a rank-$N_{\rm max}$ approximation to $\mathbf{O}^{XY}$:
	\begin{equation}
	O^{XY}_{\mu\nu} \approx \sum_{i=1}^{N_{\rm max}} \left( \int\psi_{\mu}(\mathbf{r})c_i(\mathbf{r}-\mathbf{r}_0) d\mathbf{r} \right) \left( \int\psi_{\nu}(\mathbf{r}') \phi_i(\mathbf{r}'-\mathbf{r}_0) d\mathbf{r}' \right).
	\end{equation}
	This construction requires that Eq.~\eqref{eq:Taylor} converges quickly, so that a decent approximation can be obtained already for small $N_{\rm max}$. However, if $K(\mathbf{r},\mathbf{r}')$ is non-analytic at $\mathbf{r}=\mathbf{r}'$, then Eq.~\eqref{eq:Taylor} diverges when $|\mathbf{r}'-\mathbf{r}_0| > |\mathbf{r}-\mathbf{r}_0|$, i.e.~when the size of $Y$ is large compared to the distance between $X$ and $Y$.
	The admissibility criterion is thus required for the expansion to converge in the whole region $Y$.
	
\begin{figure}[htbp]
        \begin{tabular}{cc}
			\resizebox{0.22\textwidth}{!}{\includegraphics{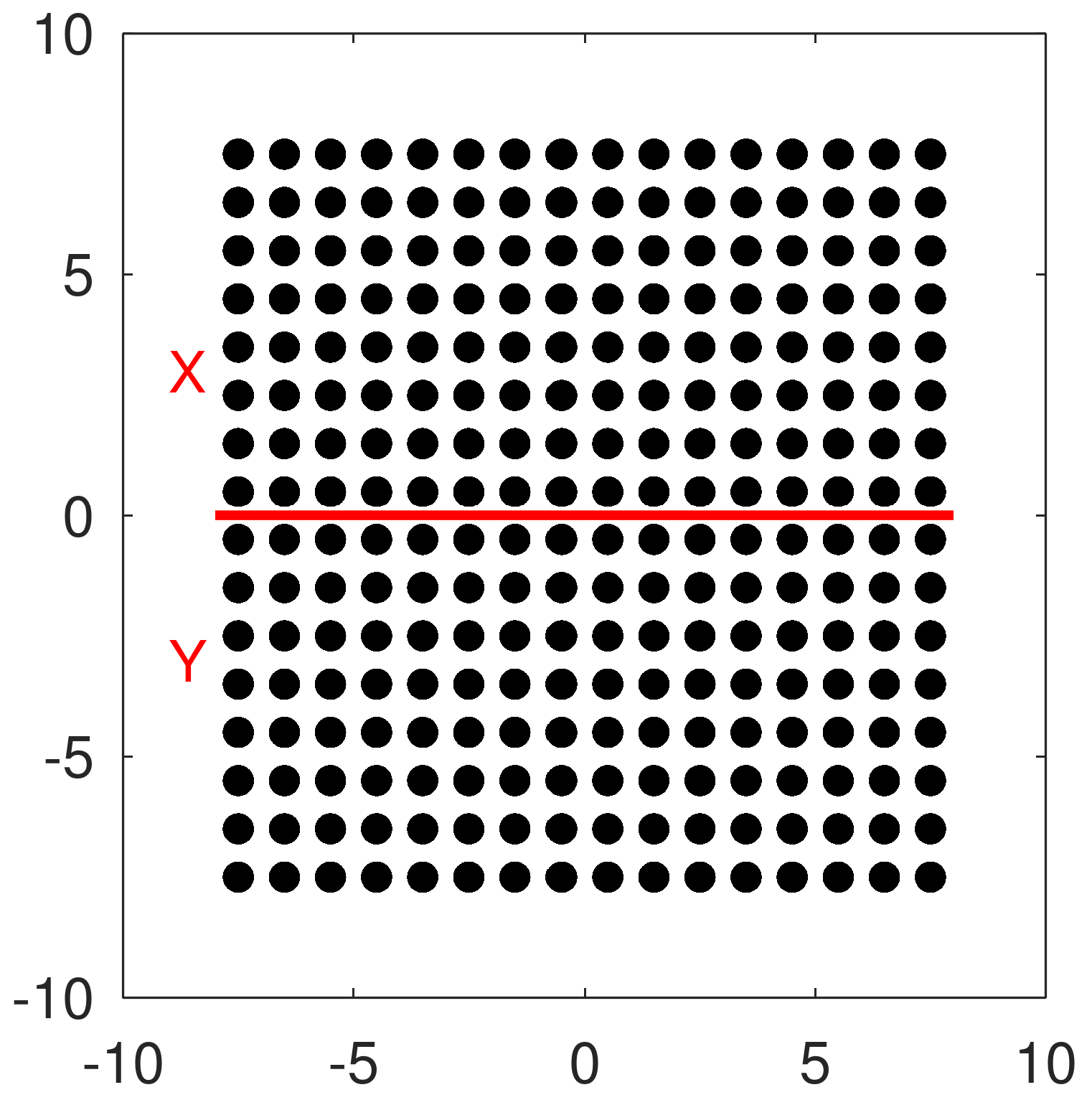}} & \resizebox{0.22\textwidth}{!}{\includegraphics{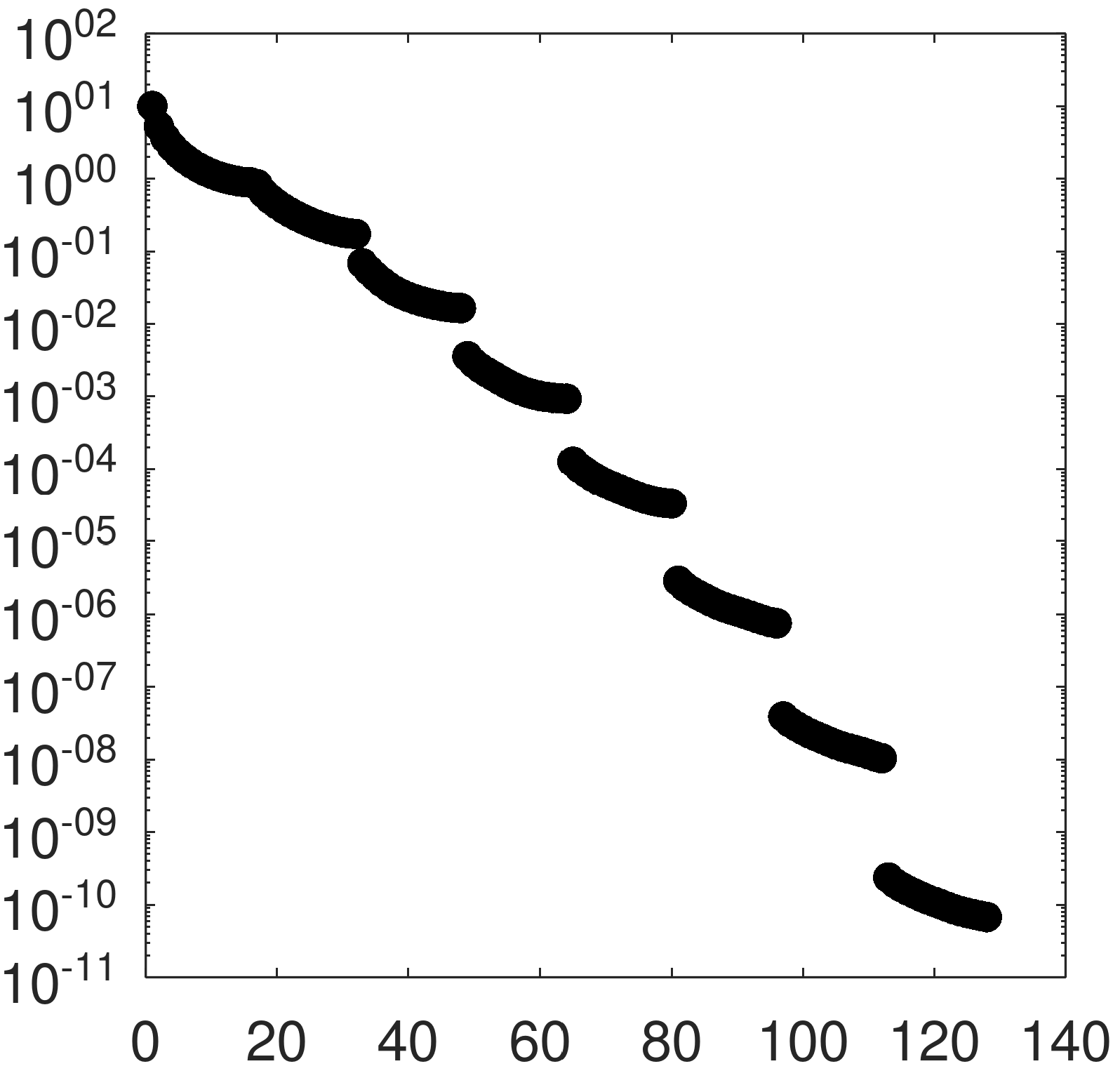}} \\
            \multicolumn{2}{c}{(a)} \\
			\resizebox{0.22\textwidth}{!}{\includegraphics{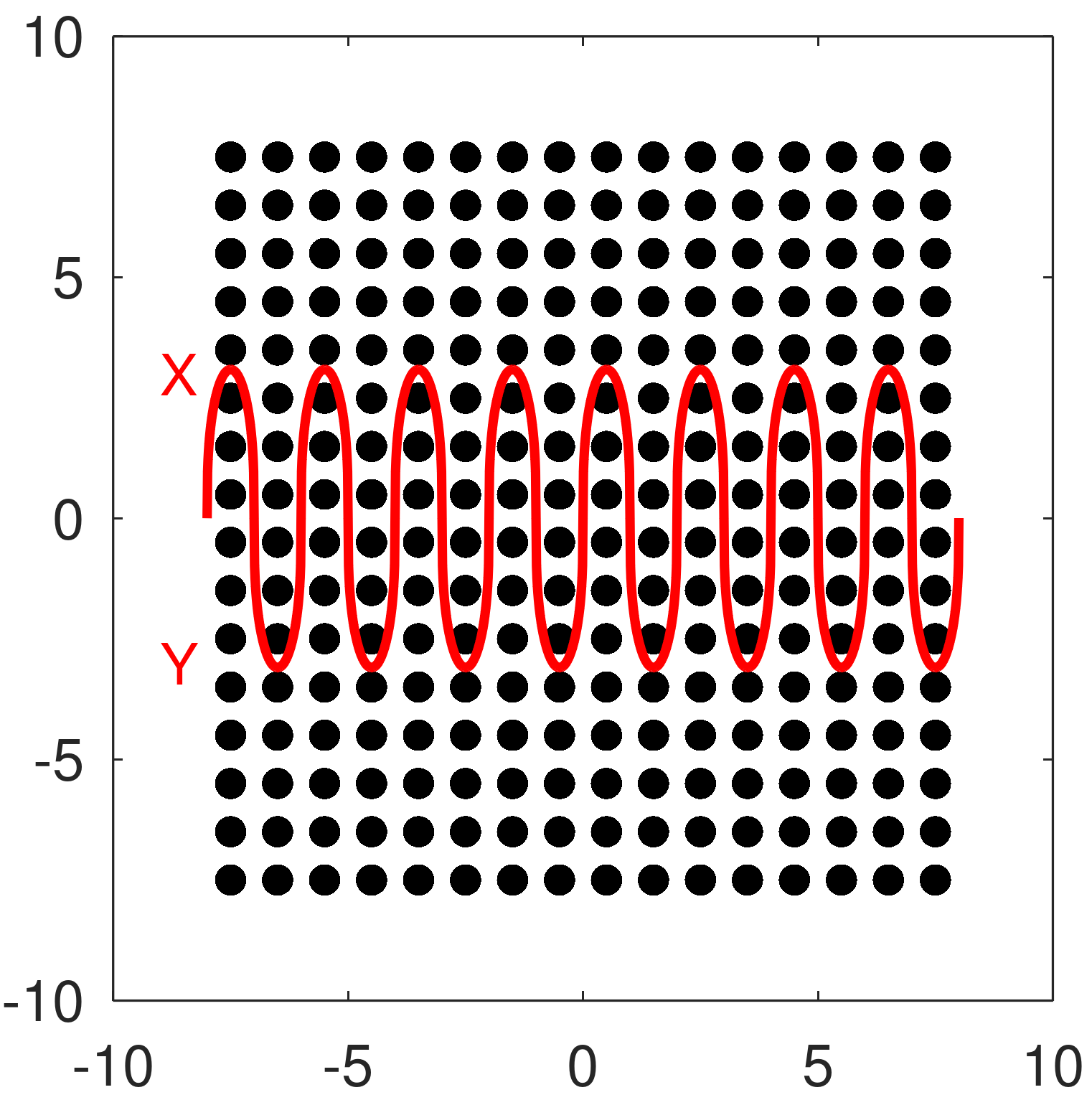}} & \resizebox{0.22\textwidth}{!}{\includegraphics{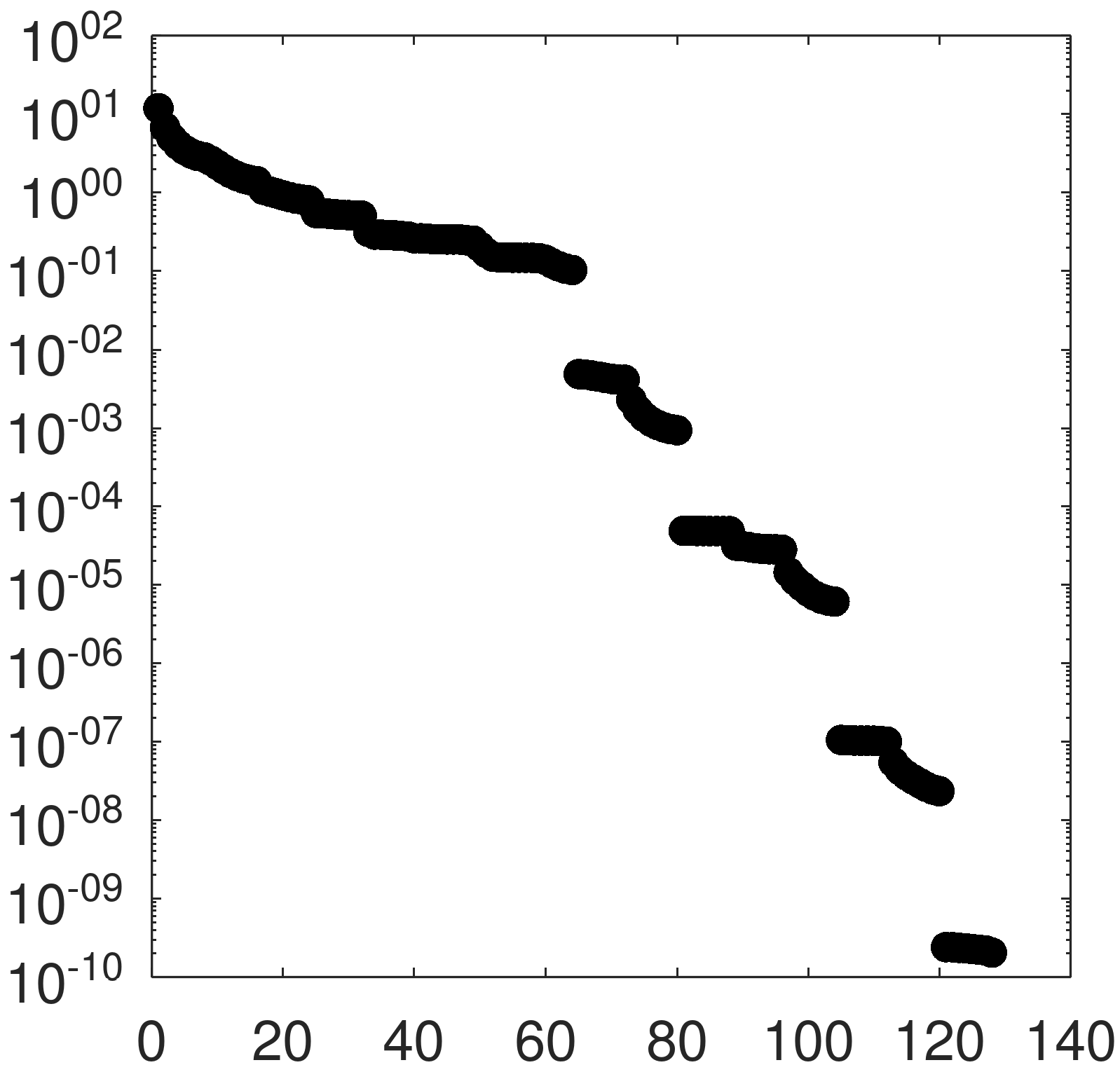}} \\
            \multicolumn{2}{c}{(b)} \\
			\resizebox{0.22\textwidth}{!}{\includegraphics{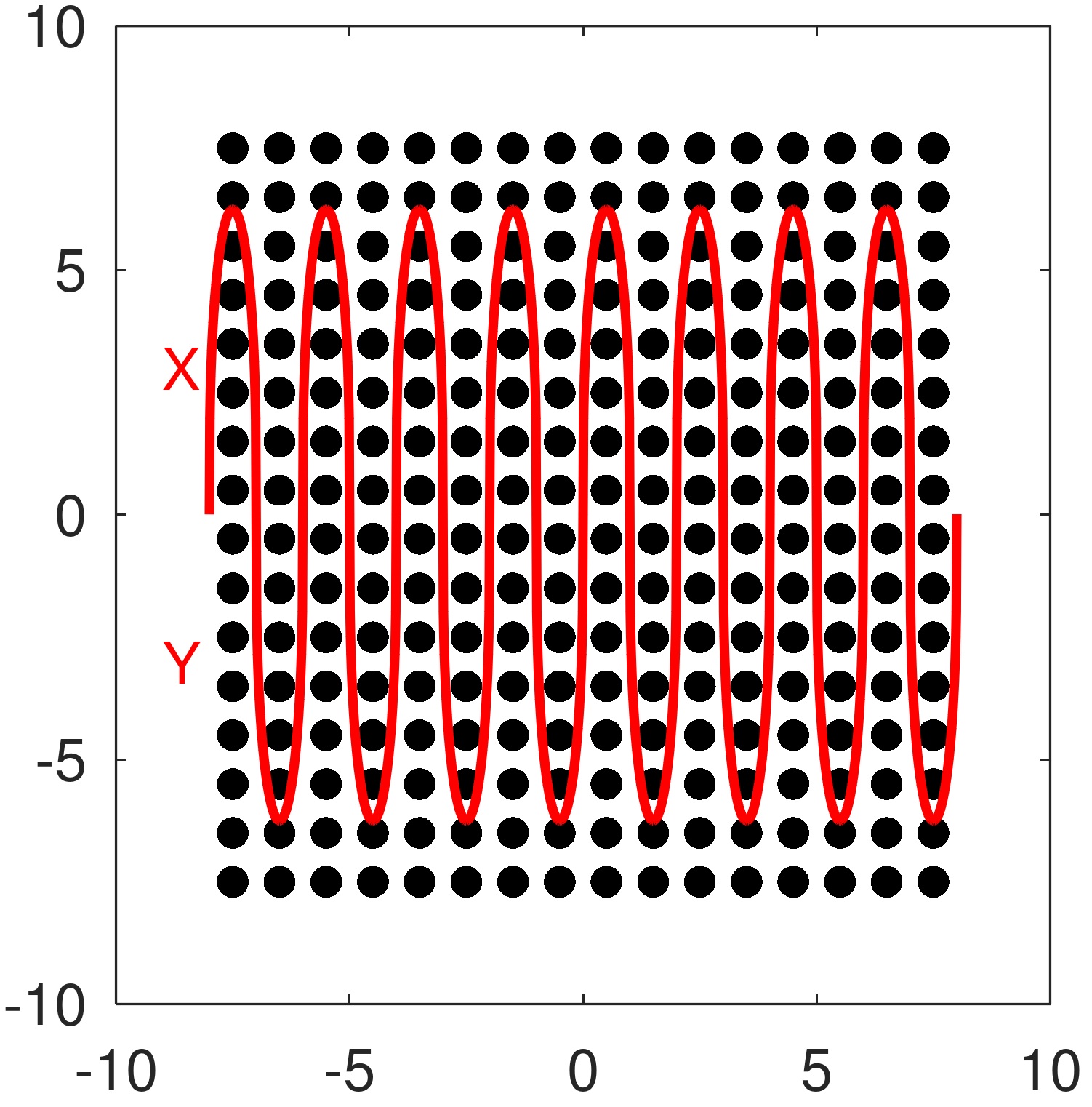}} & \resizebox{0.22\textwidth}{!}{\includegraphics{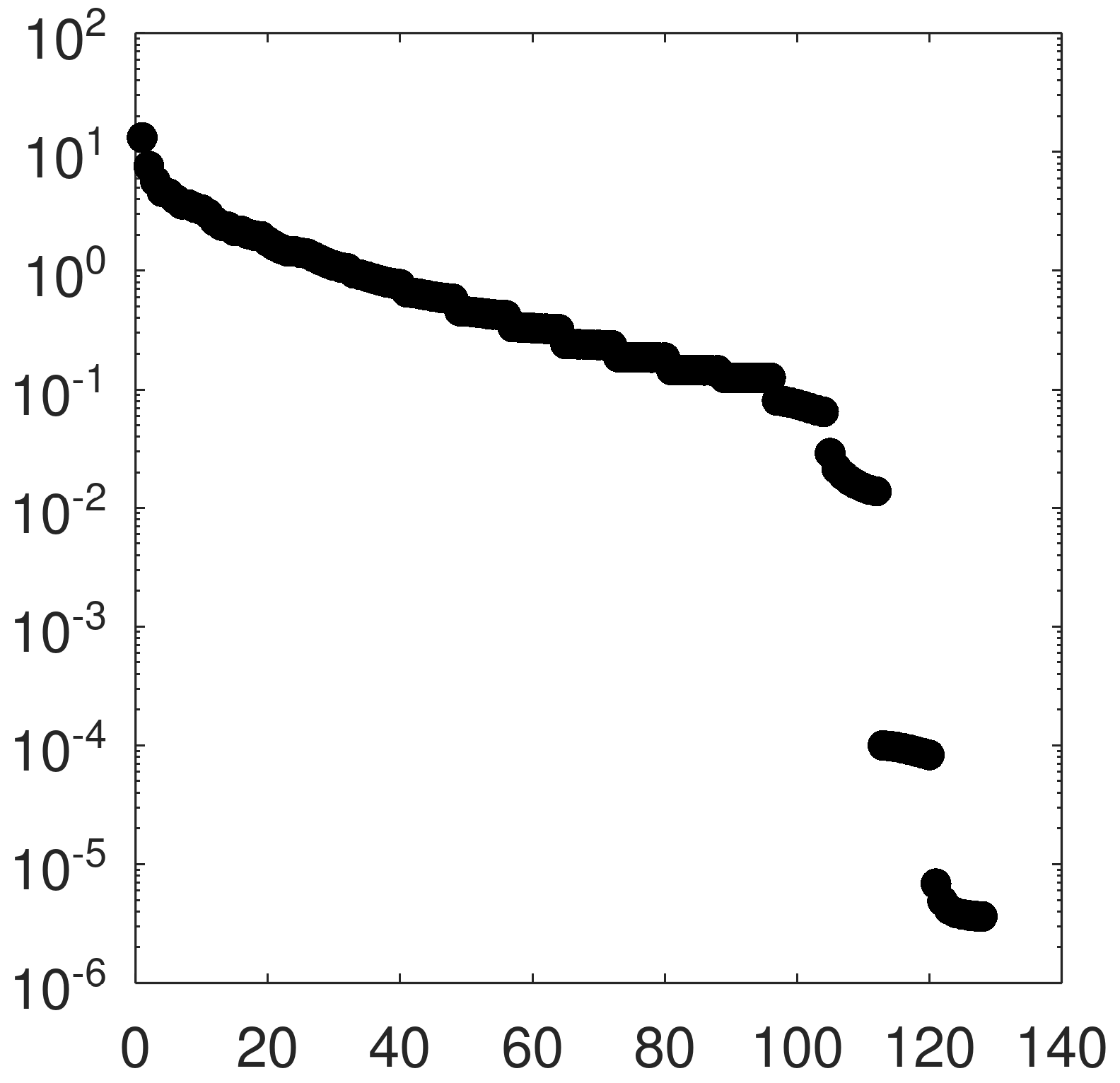}} \\
            \multicolumn{2}{c}{(c)} \\
		\end{tabular}
		\caption{Singular values (unit: a.u.)~of the upper right off-diagonal blocks of three different permuted versions of the Coulomb matrix of a 16$\times$16 grid of unit point charges, with a grid mesh size of 1 a.u. The corresponding $X$ and $Y$ sets are visualized in real space. (a) $X$ and $Y$ are separated by a straight line; (b) $X$ and $Y$ are separated by a wavy line; (c) $X$ and $Y$ are deeply interspersed with each other. }
		\label{fig:coul_grid}
	\end{figure}

\begin{figure}[htbp]
        \begin{tabular}{cc}
			\resizebox{0.1\textwidth}{!}{\includegraphics{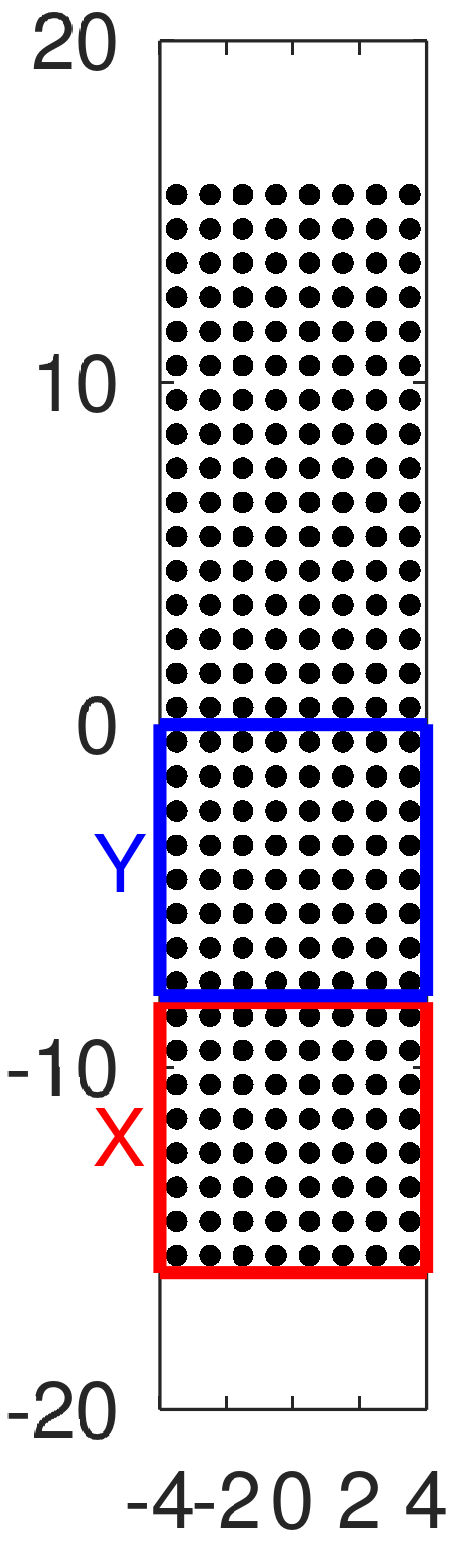}} & \resizebox{0.35\textwidth}{!}{\includegraphics{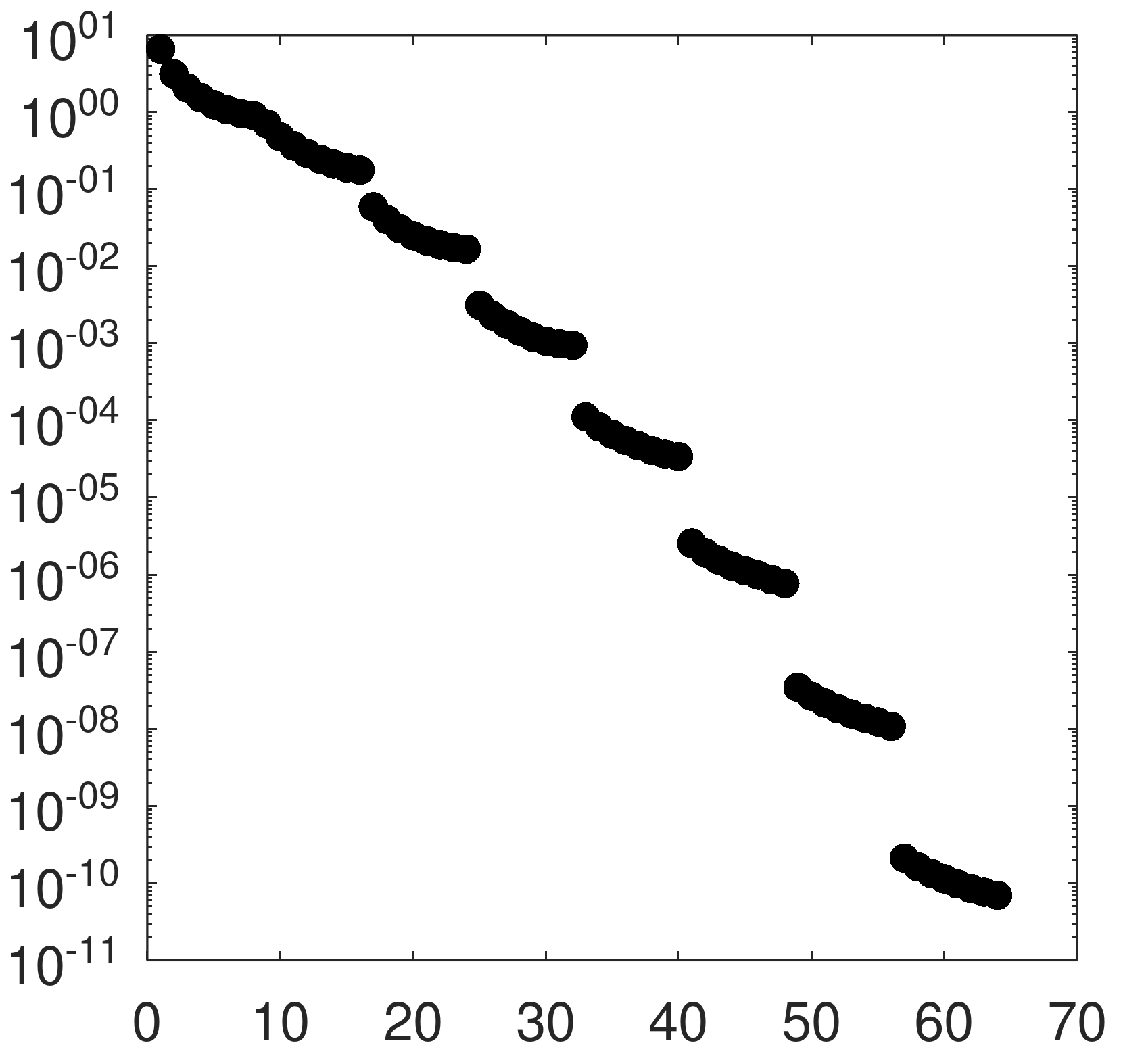}} \\
            \multicolumn{2}{c}{(a)} \\
			\resizebox{0.1\textwidth}{!}{\includegraphics{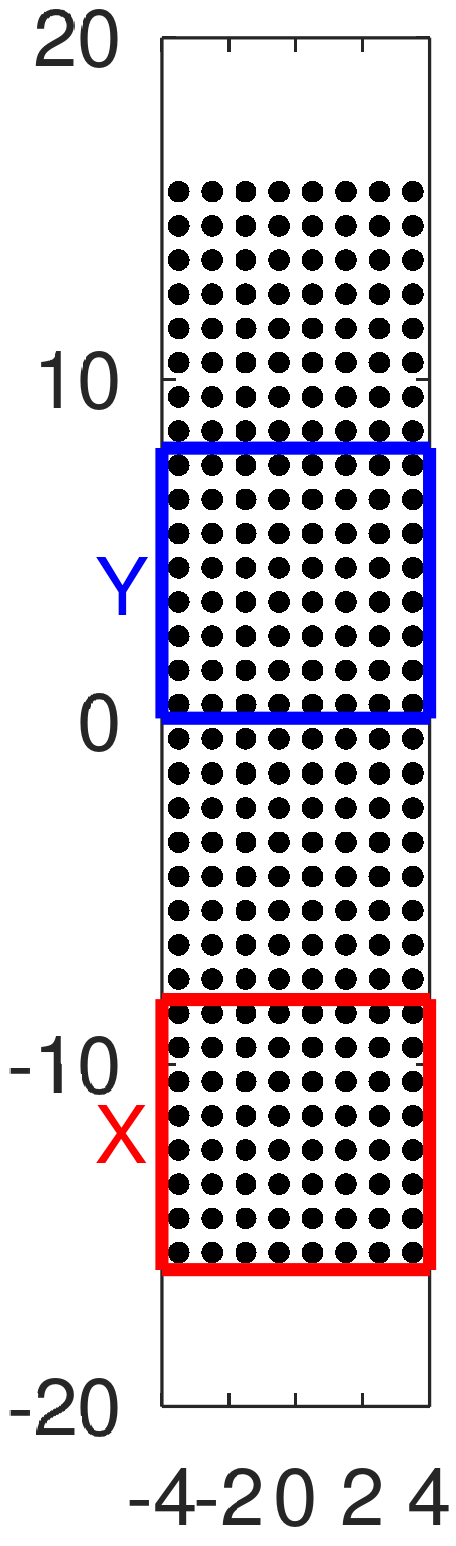}} & \resizebox{0.35\textwidth}{!}{\includegraphics{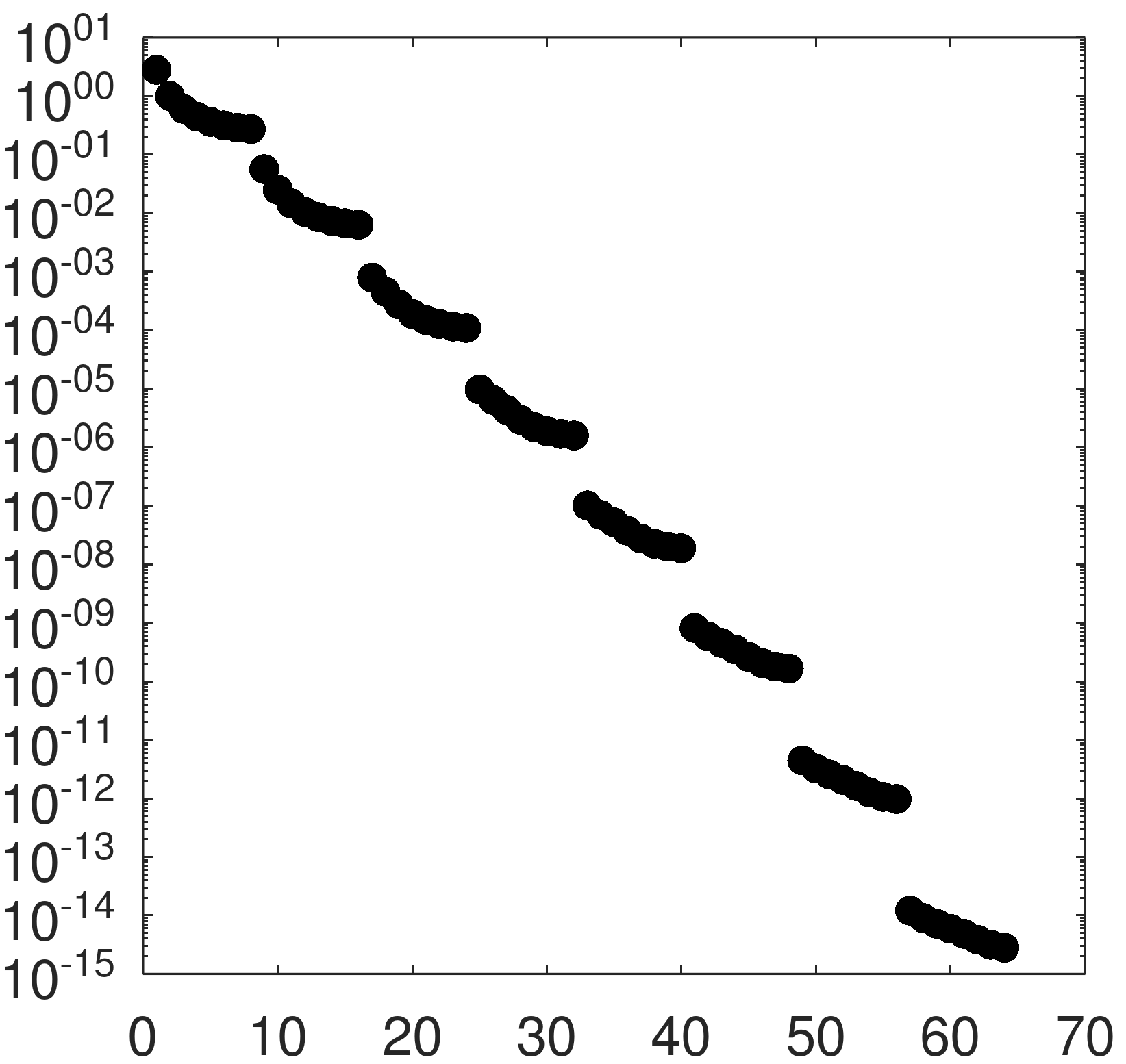}} \\
            \multicolumn{2}{c}{(b)} \\
			\resizebox{0.1\textwidth}{!}{\includegraphics{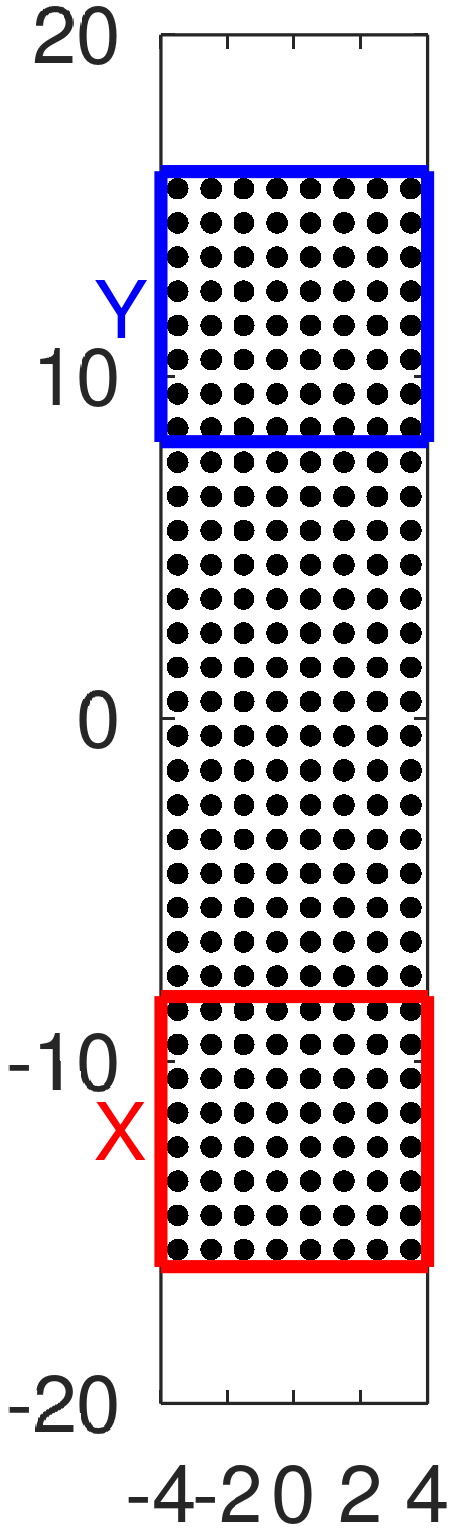}} & \resizebox{0.35\textwidth}{!}{\includegraphics{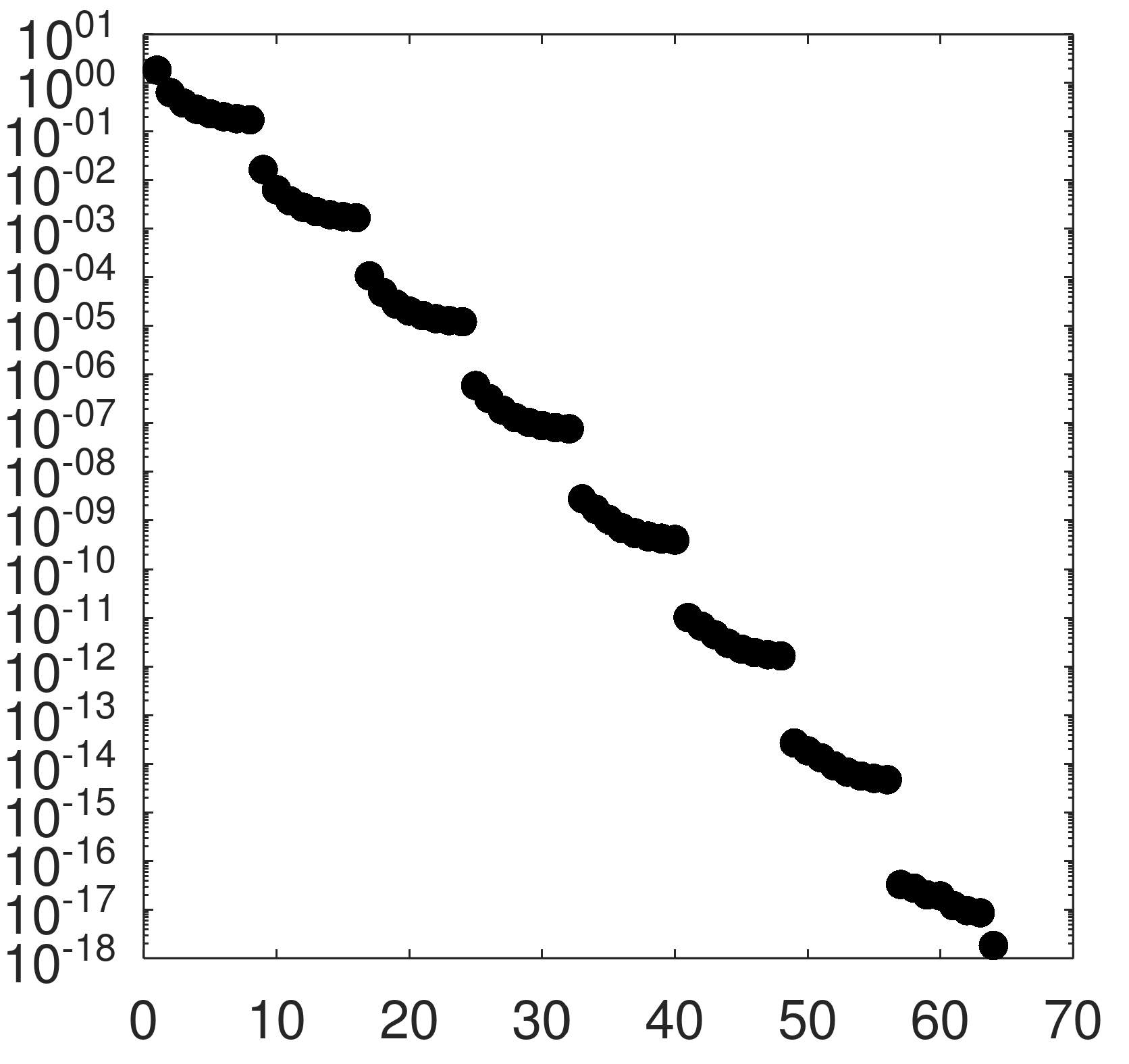}} \\
            \multicolumn{2}{c}{(c)} \\
		\end{tabular}
		\caption{Singular values (unit: a.u.)~of three off-diagonal blocks of the Coulomb matrix of an 8$\times$32 grid of unit point charges, with a grid mesh size of 1 a.u. The corresponding $X$ and $Y$ sets are visualized in real space.}
		\label{fig:coul_grid2}
	\end{figure}

	By comparison, the analogous expansion over the LKE always converges since the LKE is a finite set, and even more, the conclusion that high-energy LKE functions give small matrix elements only requires the smoothness of $K(\mathbf{r},\mathbf{r}')$, not its analyticity, while the Taylor expansion requires both. Therefore, the ODLR property can be observed for regions that do not satisfy the admissibility criterion, e.g.~two rectangular regions stacked face-to-face (Figure~\ref{fig:coul_grid}(a)). Nevertheless, the requirement of a smooth, small-curvature boundary between $X$ and $Y$ still remains: replacing the straight boundary by a zigzag boundary gives a notably slower decay of the singular values (Figure~\ref{fig:coul_grid}(b)), and if the regions $X$ and $Y$ are thoroughly interspersed with each other, the singular values appear to decay only polynomially, and almost all singular values are now non-negligible (Figure~\ref{fig:coul_grid}(c)).
	It is thus clear that the ODLR property of the matrix representations of operators is a direct consequence of the smoothness of both the convolution kernels underlying these operators, and the boundary that separates the spatial regions of interest.
	It should be noted, however, that if $X$ and $Y$ are admissible, $\mathbf{O}^{XY}$ does have a lower numerical rank than if they are not (Figure~\ref{fig:coul_grid2}), i.e.~the admissibility condition is beneficial though not required for achieving a low numerical rank of far off-diagonal blocks.
	
	The ODLR property is intimately related to many popular concepts in computational chemistry. When $K(\mathbf{r},\mathbf{r}') = 1/|\mathbf{r}-\mathbf{r}'|$ is the Coulomb kernel, and the basis functions $\mu$ are sufficiently localized, it is well-known that $\mathbf{O}^{XY}$ can be approximated by the multipolar expansion, provided that $X$ and $Y$ are admissible\cite{FMM1987,FMM1988,FMM1994,xing2020JCP,xing2020SIAM,H2Pack,borm2003Hmatrix} (note that some of the works used the word ``well-separated'' to mean ``admissible''). In the multipolar expansion, one approximates the point charges in $X$ and $Y$ by a few low-order multipoles at the centers of $X$ and $Y$ (hereafter called multipolar expansion centers), respectively, which we assume are fewer in number than the point charges themselves. Then
	\begin{equation}
	\mathbf{O}^{XY} \approx \mathbf{M}^X \mathbf{T}^{XY}(\mathbf{M}^Y)^T,
	\end{equation}
	where $M_{\mu i}^X$ is the contribution of a unit point charge at position $\mu$ to the $i$-th multipole in $X$, and $T_{ij}^{XY}$ is the Coulomb interaction between multipole $i$ in $X$ and multipole $j$ in $Y$. We now immediately see that Eq.~\eqref{eq:pOp} is recovered by setting $\mathbf{p}^X_{\rm L} = (\mathbf{M}^X)^{+,T}$, $\mathbf{p}^Y_{\rm L} = (\mathbf{M}^Y)^{+,T}$ and $\tilde{\mathbf{O}}^{XY}_{\rm LL} = \mathbf{T}^{XY}$ (where ``+'' denotes the Moore-Penrose pseudoinverse). Therefore, the success of the multipolar expansion technique can be traced to the ODLR property of the Coulomb kernel. Indeed, the neglect of high-order multipoles in the multipolar expansion is equivalent to neglecting high-energy LKE functions in Eq.~\eqref{eq:pOp}, as the eigenfunctions of the Laplacian operator in spherical regions have spherical harmonics as their angular parts, and high-order multipoles correspond to spherical harmonics with high angular momenta, which have high kinetic energies.
	
	The admissibility criterion in the multipolar expansion technique arises, because the multipolar expansion method is derived from Eq.~\eqref{eq:Taylor} instead of Eq.~\eqref{eq:pOp}, where $\phi_i(\mathbf{r}'-\mathbf{r}_0)$ is the contribution of a unit point charge at point $\mathbf{r}'$ to the $i$-th multipole centered at $\mathbf{r}_0$ (recall that the contribution of a point charge to a multipole is always given by a polynomial of the position of the point charge), and $c_i(\mathbf{r}-\mathbf{r}_0)$ is the Coulomb potential generated by the $i$-th multipole, evaluated at point $\mathbf{r}$. Specifically, we have the familiar expression\cite{xing2020SIAM}
	\begin{eqnarray}
	\frac{1}{|\mathbf{r}-\mathbf{r}'|} = \sum_{l=0}^{\infty}\sum_{m=-l}^{l} \left( |\mathbf{r}-\mathbf{r}_0|^{-l-1} Y_l^m\left(\frac{\mathbf{r}-\mathbf{r}_0}{|\mathbf{r}-\mathbf{r}_0|}\right) \right) \nonumber\\
	\times \left( |\mathbf{r}'-\mathbf{r}_0|^l Y_l^{-m}\left(\frac{\mathbf{r}'-\mathbf{r}_0}{|\mathbf{r}'-\mathbf{r}_0|}\right) \right), \label{eq:multipole}
	\end{eqnarray}
	but which only converges when $|\mathbf{r}'-\mathbf{r}_0| < |\mathbf{r}-\mathbf{r}_0|$. Here $Y_l^m$ are the spherical harmonic functions. Eq.~\eqref{eq:multipole} has the same form as Eq.~\eqref{eq:Taylor} once the expansion is truncated to a finite angular momentum $l$ (recall that $|\mathbf{r}'-\mathbf{r}_0|^l Y_l^{-m}\left(\frac{\mathbf{r}'-\mathbf{r}_0}{|\mathbf{r}'-\mathbf{r}_0|}\right)$ is a polynomial in $\mathbf{r}'$). From this example, we see that although Eq.~\eqref{eq:Taylor} requires the admissibility condition, while the ODLR property of the Coulomb operator does not, Eq.~\eqref{eq:Taylor} can yield the low-rank decomposition of $\mathbf{O}^{XY}$ without knowing its matrix elements in advance. Conversely, if all elements of $\mathbf{O}^{XY}$ can be calculated in advance, its SVD may reveal a lower numerical rank than the low-rank decomposition constructed by a Taylor expansion, especially when the admissibility condition is not satisfied. We will see in Section~\ref{sec:ODLR_coul} that this insight results in a much faster Coulomb algorithm than the renowned fast multipole moment (FMM)\cite{FMM1987,FMM1988,FMM1994} method.
	
\begin{figure}[htbp]
    \resizebox{0.48\textwidth}{!}{\includegraphics{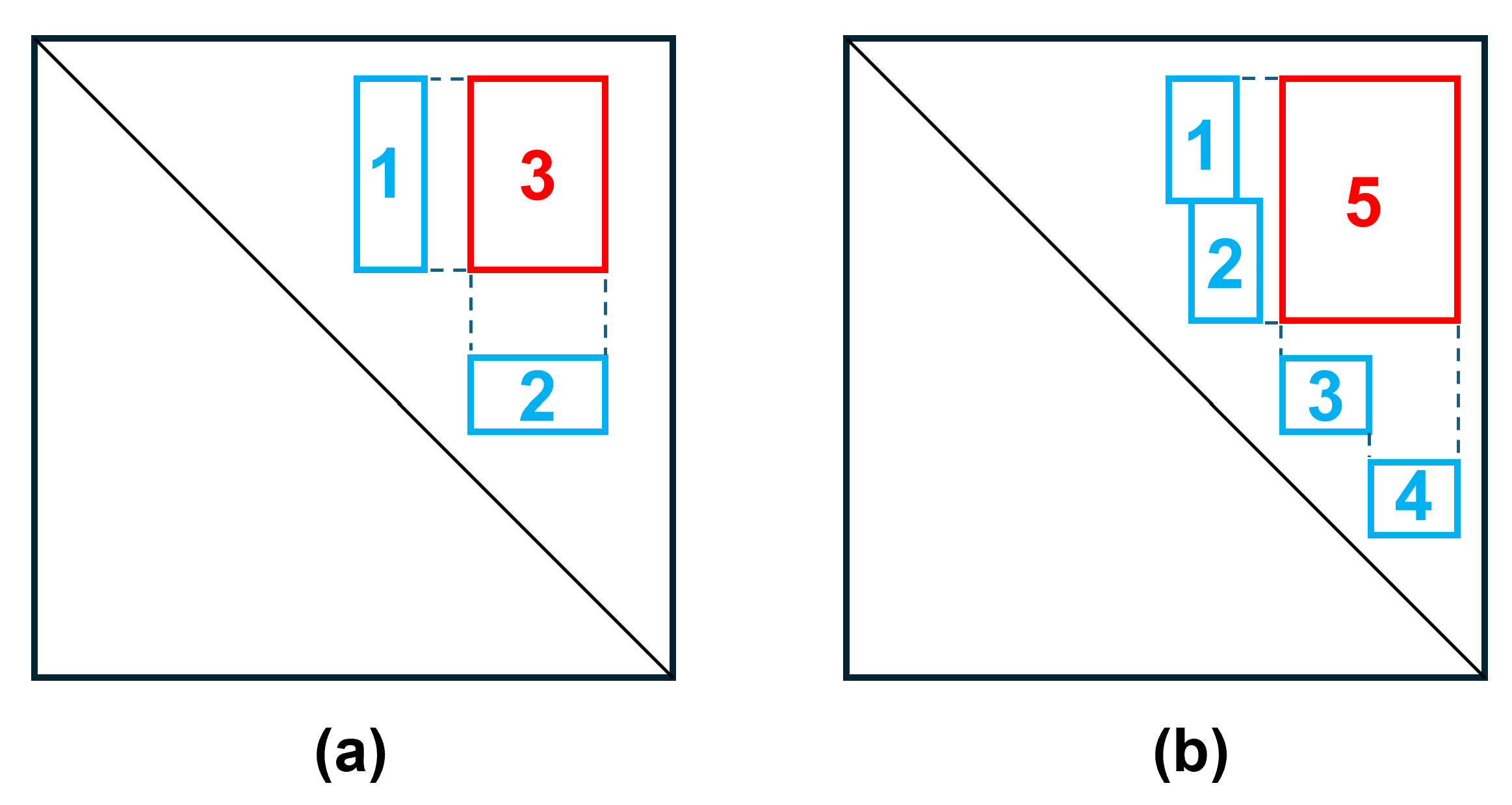}}
		\caption{Reuse of the singular vectors of the off-diagonal blocks of ODLR matrices. (a) Knowledge of the singular vectors of blocks 1 and 2 can speedup the SVD of block 3. (b) Knowledge of the singular vectors of blocks 1-4 can speedup the SVD of block 5.}
		\label{fig:reuse_singular_vectors}
	\end{figure}

	Meanwhile, Eq.~\eqref{eq:pUspV} implies that although the left singular vectors of $\mathbf{O}^{XY}$, $\mathbf{p}^X_{\rm L}\mathbf{U}_{\rm L}^{XY}$, depend on $Y$, the column space of $\mathbf{O}^{XY}$ (the space spanned by the left singular vectors that have non-negligible numerical ranks) is identical to the space spanned by $\mathbf{p}^X_{\rm L}$, which is independent of $Y$ and only dependent on $X$ (unless, of course, in the degenerate case where $\tilde{\mathbf{O}}^{XY}_{\rm LL}$ is accidentally rank-deficient; for brevity, we will not consider such cases unless when necessary); similarly, the row space of $\mathbf{O}^{XY}$ is independent of $X$.
	This observation has two implications. First, given the low-rank decompositions of two far off-diagonal blocks $\mathbf{O}^{XY}$ and $\mathbf{O}^{ZW}$, if $\mathbf{O}^{XW}$ is also a far off-diagonal block, its column and row spaces can be found without any new low-rank decomposition (Figure~\ref{fig:reuse_singular_vectors}(a)). The left and right singular vectors of $\mathbf{O}^{XW}$ can then be found by the SVD of
	\begin{equation}
	\bar{\mathbf{O}}^{XW} = \mathbf{U}^{XY} \mathbf{O}^{XW} (\mathbf{V}^{ZW})^T,
	\end{equation}
	which has much lower dimensions than $\mathbf{O}^{XW}$ itself. Specifically, if
	\begin{equation}
	\bar{\mathbf{O}}^{XW} = \bar{\mathbf{U}}^{XW}\bar{\mathbf{\sigma}}^{XW} (\bar{\mathbf{V}}^{XW})^T, \label{eq:SVD_XW}
	\end{equation}
	then
	\begin{equation}
	\mathbf{O}^{XW} = \left[(\mathbf{V}^{XY})^T\bar{\mathbf{U}}^{XW}\right] \bar{\mathbf{\sigma}}^{XW} \left[(\mathbf{U}^{ZW})^T\bar{\mathbf{V}}^{XW}\right]^T. \label{eq:SVD_XW2}
	\end{equation}
	Thus, the singular vectors of an off-diagonal block can be \emph{reused} to factor the off-diagonal blocks that share the same row and/or column indices as this block.
	Importantly, the singular vectors $(\mathbf{V}^{XY})^T\bar{\mathbf{U}}^{XW}$ and $(\mathbf{U}^{ZW})^T\bar{\mathbf{V}}^{XW}$ can be stored and used in factored form, which can lead to further savings. Taking the former as an example, $\mathbf{V}^{XY}$ is dependent on $X$ but not $W$, and therefore can be reused for different $W$'s; meanwhile, although $\bar{\mathbf{U}}^{XW}$ depends on both $X$ and $W$, it has much smaller dimensions than the singular vector matrix $(\mathbf{V}^{XY})^T\bar{\mathbf{U}}^{XW}$. Therefore, storing $\mathbf{V}^{XY}$ and $\bar{\mathbf{U}}^{XW}$ separately gives more efficient compression than storing their product.
	Alternatively, one can store $\mathbf{V}^{XY}$, $\mathbf{U}^{ZW}$ and $\bar{\mathbf{O}}^{XW}$, without storing the SVD decomposition of the latter, yielding even better compression ratios. The original matrix block $\mathbf{O}^{XW}$ can then be recovered as
	\begin{equation}
	\mathbf{O}^{XW} = (\mathbf{V}^{XY})^T \bar{\mathbf{O}}^{XW} \mathbf{U}^{ZW}.
	\end{equation}

	Another corollary is that the column and row spaces of large far off-diagonal blocks can be assembled from those of smaller blocks (Figure~\ref{fig:reuse_singular_vectors}(b)). Consider a $2\times 2$ far off-diagonal block
	\begin{equation}
	\mathbf{O}^{XY} \equiv \left( \begin{array}{cc} \mathbf{O}^{X_1Y_1} & \mathbf{O}^{X_1Y_2} \\ \mathbf{O}^{X_2Y_1} & \mathbf{O}^{X_2Y_2} \end{array} \right),
	\end{equation}
	where $X=X_1\cup X_2$ and $Y=Y_1\cup Y_2$. The four blocks themselves can be decomposed to give their singular vectors and singular values (here it is assumed that all near-zero singular values and their accompanying singular vectors are removed, i.e. the singular vector matrices are tall rectangular matrices):
	\begin{equation}
	\mathbf{O}^{XY} = \left( \begin{array}{cc} \mathbf{U}^{X_1Y_1}\mathbf{\sigma}^{X_1Y_1} (\mathbf{V}^{X_1Y_1})^T & \mathbf{U}^{X_1Y_2}\mathbf{\sigma}^{X_1Y_2} (\mathbf{V}^{X_1Y_2})^T \\ \mathbf{U}^{X_2Y_1}\mathbf{\sigma}^{X_2Y_1} (\mathbf{V}^{X_2Y_1})^T & \mathbf{U}^{X_2Y_2}\mathbf{\sigma}^{X_2Y_2} (\mathbf{V}^{X_2Y_2})^T \end{array} \right).
	\end{equation}
	We have shown that the column spaces of $\mathbf{O}^{X_1Y_1}$ and $\mathbf{O}^{X_1Y_2}$ are the same. Therefore, it must be possible to express the left singular vectors of both of them as a unitary rotation of those of a third far off-diagonal block, $\mathbf{O}^{X_1Z_1}$:
	\begin{eqnarray}
	\mathbf{U}^{X_1Y_1} & = & \mathbf{U}^{X_1Z_1}\mathbf{Q}^{Z_1Y_1(X_1)}, \\
	\mathbf{U}^{X_1Y_2} & = & \mathbf{U}^{X_1Z_1}\mathbf{Q}^{Z_1Y_2(X_1)},
	\end{eqnarray}
	with $\mathbf{Q}^{Z_1Y_1(X_1)}$ and $\mathbf{Q}^{Z_1Y_2(X_1)}$ being small unitary matrices, whose dimensions are equal to the rank of $\mathbf{O}^{X_1Z_1}$.
	Similarly, we have
	\begin{eqnarray}
	\mathbf{U}^{X_2Y_1} & = & \mathbf{U}^{X_2Z_2}\mathbf{Q}^{Z_2Y_1(X_2)}, \\
	\mathbf{U}^{X_2Y_2} & = & \mathbf{U}^{X_2Z_2}\mathbf{Q}^{Z_2Y_2(X_2)}, \\
	\mathbf{V}^{X_1Y_1} & = & \mathbf{V}^{W_1Y_1}\mathbf{Q}^{W_1X_1(Y_1)}, \\
	\mathbf{V}^{X_2Y_1} & = & \mathbf{V}^{W_1Y_1}\mathbf{Q}^{W_1X_2(Y_1)}, \\
	\mathbf{V}^{X_1Y_2} & = & \mathbf{V}^{W_2Y_2}\mathbf{Q}^{W_2X_1(Y_2)}, \\
	\mathbf{V}^{X_2Y_2} & = & \mathbf{V}^{W_2Y_2}\mathbf{Q}^{W_2X_2(Y_2)},
	\end{eqnarray}
	which involve the singular vectors of three more far off-diagonal blocks, $\mathbf{O}^{X_2Z_2}$, $\mathbf{O}^{W_1Y_1}$, and $\mathbf{O}^{W_2Y_2}$.
	Now we can factorize $\mathbf{O}^{XY}$ as
	\begin{equation}
	\mathbf{O}^{XY} = 
	\left( \begin{array}{cc} \mathbf{U}^{X_1Z_1} & \mathbf{0} \\ \mathbf{0} & \mathbf{U}^{X_2Z_2} \end{array} \right)
	\mathbf{\Sigma}^{XY}
	\left( \begin{array}{cc} \mathbf{V}^{W_1Y_1} & \mathbf{0} \\ \mathbf{0} & \mathbf{V}^{W_2Y_2} \end{array} \right)^T, \label{eq:block_SVD}
	\end{equation}
	\begin{equation}
	\mathbf{\Sigma}^{XY} =
	\left( \begin{array}{cc}
	\mathbf{\Sigma}^{X_1Y_1} & \mathbf{\Sigma}^{X_1Y_2} \\ \mathbf{\Sigma}^{X_2Y_1} & \mathbf{\Sigma}^{X_2Y_2}
	\end{array} \right),
	\end{equation}
    \begin{eqnarray}
    \mathbf{\Sigma}^{X_1Y_1} & = & \mathbf{Q}^{Z_1Y_1(X_1)}\mathbf{\sigma}^{X_1Y_1} (\mathbf{Q}^{W_1X_1(Y_1)})^T, \\
    \mathbf{\Sigma}^{X_1Y_2} & = & \mathbf{Q}^{Z_1Y_2(X_1)}\mathbf{\sigma}^{X_1Y_2} (\mathbf{Q}^{W_2X_1(Y_2)})^T, \\
    \mathbf{\Sigma}^{X_2Y_1} & = & \mathbf{Q}^{Z_2Y_1(X_2)}\mathbf{\sigma}^{X_2Y_1} (\mathbf{Q}^{W_1X_2(Y_1)})^T, \\
    \mathbf{\Sigma}^{X_2Y_2} & = & \mathbf{Q}^{Z_2Y_2(X_2)}\mathbf{\sigma}^{X_2Y_2} (\mathbf{Q}^{W_2X_2(Y_2)})^T.
    \end{eqnarray}
	Thus, assuming $\mathbf{\Sigma}^{XY}$ is full-rank, the column (row) space of $\mathbf{O}^{XY}$ is the direct sum of the column (row) spaces of $\mathbf{U}^{X_1Z_1}$ and $\mathbf{V}^{X_2Z_2}$ ($\mathbf{U}^{W_1Y_1}$ and $\mathbf{V}^{W_2Y_2}$). The left singular vectors of $\mathbf{O}^{XY}$ are thus linear combinations of the left singular vectors of $\mathbf{O}^{X_1Z_1}$ (padded with zeroes in the $X_2$ block), and the left singular vectors of $\mathbf{O}^{X_2Z_2}$ (padded with zeroes in the $X_1$ block); similarly for the right singular vectors. In other words, the singular vectors of a large block can be constructed from those of small blocks outside the large block (Figure~\ref{fig:reuse_singular_vectors}(b)):
	\begin{eqnarray}
	\mathbf{U}^{XY} & = & \left( \begin{array}{cc} \mathbf{U}^{X_1Z_1} & \mathbf{0} \\ \mathbf{0} & \mathbf{U}^{X_2Z_2} \end{array} \right) \tilde{\mathbf{U}}^{XY}, \label{eq:tildeUXY} \\
	\mathbf{V}^{XY} & = & \left( \begin{array}{cc} \mathbf{V}^{W_1Y_1} & \mathbf{0} \\ \mathbf{0} & \mathbf{U}^{W_2Y_2} \end{array} \right) \tilde{\mathbf{V}}^{XY}, \label{eq:tildeVXY}
	\end{eqnarray}
	where $\tilde{\mathbf{U}}^{XY}$ and $\tilde{\mathbf{V}}^{XY}$ are the left and right singular vectors of $\mathbf{\Sigma}^{XY}$:
	\begin{equation}
	\mathbf{\Sigma}^{XY} = \tilde{\mathbf{U}}^{XY} \mathbf{\sigma}^{XY} (\tilde{\mathbf{V}}^{XY})^T,
	\end{equation}
	and $\mathbf{U}^{XY}$, $\mathbf{\sigma}^{XY}$ and $\mathbf{V}^{XY}$ are defined in Eq.~\eqref{eq:pUspV}.
	In the literature, this is sometimes called the \emph{nested basis structure} of ODLR matrices\cite{amestoy2019SIAM,xing2020SIAM}.
	It is also of note that $\mathbf{\Sigma}^{XY}$ may be rank-deficient, in which case the rank of $\mathbf{O}^{XY}$ may be smaller than $\mathrm{rank}(\mathbf{O}^{X_1Z_1}) + \mathrm{rank}(\mathbf{O}^{X_2Z_2})$ (or $\mathrm{rank}(\mathbf{O}^{W_1Y_1}) + \mathrm{rank}(\mathbf{O}^{W_2Y_2})$), instead of being equal to them.
	
	The nested basis structure can be applied recursively to factorize $\mathbf{U}^{XY}$ and $\mathbf{V}^{XY}$ into a product of many matrices, noting that $\mathbf{U}^{X_1Z_1}, \mathbf{U}^{X_2Z_2}, \mathbf{V}^{W_1Y_1}$ and $\mathbf{V}^{W_2Y_2}$ can themselves be decomposed akin to Eq.~\eqref{eq:tildeUXY} and Eq.~\eqref{eq:tildeVXY}; the resulting block-diagonal matrix with four blocks can then be decomposed to separate out a block-diagonal matrix with eight blocks, etc. This results in a product of matrices with an exponentially increasing number of diagonal blocks, i.e.
	\begin{eqnarray}
	\mathbf{U}^{XY} & = & \cdots \left( \begin{array}{cccc} \tilde{\mathbf{U}}^{X_{1,1}Z_{1,1}} & \mathbf{0}  & \mathbf{0}  & \mathbf{0} \\ \mathbf{0} & \tilde{\mathbf{U}}^{X_{1,2}Z_{1,2}}  & \mathbf{0} & \mathbf{0} \\ \mathbf{0}  & \mathbf{0} & \tilde{\mathbf{U}}^{X_{2,1}Z_{2,1}} & \mathbf{0}   \\ \mathbf{0} & \mathbf{0} & \mathbf{0} & \tilde{\mathbf{U}}^{X_{2,2}Z_{2,2}} \end{array} \right) \nonumber\\
	& & \times \left( \begin{array}{cc} \tilde{\mathbf{U}}^{X_1Z_1} & \mathbf{0} \\ \mathbf{0} & \tilde{\mathbf{U}}^{X_2Z_2} \end{array} \right) \tilde{\mathbf{U}}^{XY} \nonumber\\
	& \equiv & \mathbf{U}^{(X,1)}\mathbf{U}^{(X,2)}\cdots \mathbf{U}^{(X,N_{\rm level}^X)} \tilde{\mathbf{U}}^{XY}, \label{eq:block_U}
	\end{eqnarray}
	where $X_{1,1}\cup X_{1,2} = X_1$, $X_{2,1}\cup X_{2,2} = X_2$, $Z_{1,1}\cup Z_{1,2} = Z_1$, and $Z_{2,1}\cup Z_{2,2} = Z_2$. The hierarchical decomposition ends when the remaining blocks have size $O(1)$.
	Matrices to the left ($\mathbf{U}^{(X,i)}$ where $i$ is small) describe small-scale, short-range details of the interaction underlying $\mathbf{O}$ (as they only couple spatially proximal subsets of $X$), while those to the right ($\mathbf{U}^{(X,i)}$ where $i$ is large, as well as $\tilde{\mathbf{U}}^{XY}$ itself) describe large-scale, long-range features of the interaction.
	This approach will be used in Section~\ref{sec:HSS} to remove some logarithmic factors in the storage and computational time scalings related to ODLR matrices.
	
	If the kernel $K(\mathbf{r},\mathbf{r}')$ underlying $\mathbf{O}$ is ``asymptotically smooth'', in the sense that there are constants $C_1, C_2>0$ such that
	\begin{eqnarray}
	& \left|\left(\prod_{i=1}^D \frac{\partial^{\alpha_i}}{\partial r_i^{\alpha_i}} \frac{\partial^{\beta_i}}{\partial r_i'^{\beta_i}}\right) K(\mathbf{r},\mathbf{r}')\right| & \nonumber\\
	& \le C_1(C_2\|\mathbf{r}-\mathbf{r}'\|)^{-\sum_{i=1}^D(\alpha_i+\beta_i)} |K(\mathbf{r},\mathbf{r}')|, \forall \alpha_i, \beta_i \ge 0 & \label{eq:asymp_smooth}
	\end{eqnarray}
	(where $D$ is the number of spatial dimensions, usually 3 in computational chemistry, and $r_i$ is the $i$-th component of the spatial coordinate $\mathbf{r}$),
	and if $X$ and $Y$ satisfy the admissibility condition, the rank of $\mathbf{O}^{XY}$ can be even smaller and close to $\max(\textrm{rank}(\mathbf{O}^{X_1Y_1}),\textrm{rank}(\mathbf{O}^{X_2Y_2}))$, instead of $\textrm{rank}(\mathbf{O}^{X_1Y_1}) + \textrm{rank}(\mathbf{O}^{X_2Y_2})$; as a result, the ranks of admissible blocks do not depend on the sizes of the blocks\cite{hackbusch1999Hmatrix,borm2003Hmatrix,hackbusch2002H2matrix}. This conclusion can be intuitively understood as follows: if the kernel varies sufficiently slowly in the long-range limit, and if the admissibility condition applies, the Taylor expansion Eq.~\eqref{eq:Taylor} is always well within its convergence radius, such that as the sizes of $X$ and $Y$ increase, and their distance increases at the same rate, only an asymptotically constant number of terms $N_{\rm max}$ is needed to converge it to a given accuracy, thereby giving a constant rank.
	We will see in Section~\ref{sec:ODLR_compress} how this fact, combined with the RG structure of ODLR matrices, allows the linear-scaling storage and use of ODLR matrices.
	
	The aforementioned property that the singular vectors of large blocks can be expressed by those of smaller ones, can be interpreted with the language of renormalization group (RG)\cite{wilson1975RG,wilson1983RG}. The RG method treats a collection of spatially proximal particles or sites as a single, coarse-grained particle or site, so that the coarse-grained model captures all the long-range properties of the original interaction, while integrating out some of the short-range details.
	For ODLR matrices, this amounts to using only the dominant singular vectors of small blocks to construct the singular vectors of large blocks; the singular vectors that have negligible singular values correspond to the neglected details.
	One key idea of RG is that the coarse-grained model itself can be again subjected to coarse-graining, \emph{ad infinitum}. This results in an exponential increase of the length scale that can be studied, and eventually gives important emergent quantities in the thermodynamic limit, such as critical exponents.
	It is therefore clear that Eq.~\eqref{eq:block_U}, which performs a scale separation of the singular vectors to yield matrices that interact on a variety of scales, is a manifestation of the philosophy of RG.
	In Section~\ref{sec:ODLR_compress}, we will show how this RG philosophy can be used to devise linear and near-linear scaling algorithms involving ODLR matrices, and Section~\ref{sec:ODLR_coul} will demonstrate how the RG structure of the Coulomb matrix forms the basis of the FMM method. We will also revisit the relation of ODLR and RG in Section~\ref{sec:ODLR_DF}, where an RG-based algorithm for solving density matrices can be reinterpreted under the ODLR framework.
	
	\subsection{Eigenbasis perspective: ODLR property of the matrix functions of local matrices} \label{sec:foundations_matfunc}
	
	Another often cited class of matrices that have the ODLR property are the Green's functions, or inverses, of elliptic linear differential operators\cite{bebendorf2008elliptic,beylkin1991fast,martinsson2009elliptic,lorasp}.
	The aforementioned example of the Coulomb operator belongs to this category, since it is, up to a factor of $-4\pi$, the inverse of the Laplacian operator:
	\begin{equation}
	\nabla^2 V(\mathbf{r}) = -4\pi\rho(\mathbf{r}) \iff V(\mathbf{r}) = \int \frac{\rho(\mathbf{r}')}{|\mathbf{r}-\mathbf{r}'|}d\mathbf{r}'.
	\end{equation}
	Since the non-relativistic Hamiltonian of a multi-particle system
	\begin{equation}
	\mathbf{H} = \sum_i\frac{1}{2m_i} \nabla_i^2 + V(\{\hat{\mathbf{r}}_i\})
	\end{equation}
	is an elliptic linear differential operator, it follows that $\mathbf{H}^{-1}$, or more generally the shifted inverse $(\mathbf{H}-E^*)^{-1}$ (where $E^*$ is an arbitrary energy), has the ODLR property.
	As a special case, the Green's function of a non-interacting Hamiltonian (where $\psi_n$ and $\epsilon_n$ are the $n$-th eigenvector and eigenvalue of $\mathbf{H}$, respectively, and $\eta$ is an infinitesimal shift)
	\begin{equation}
	G_0(\mathbf{r},\mathbf{r}',\omega) = \sum_n \frac{\psi_n(\mathbf{r})\psi_n^*(\mathbf{r}')}{\omega-\epsilon_n+i\eta\,\mathrm{sgn}(\epsilon_n-\mu)}
	\end{equation}
	has the ODLR property for a fixed $\omega$, since it is formally the negative shifted inverse of $\mathbf{H}$, expressed under the eigenstates of position:
	\begin{equation}
	\mathbf{G}_0(\omega) = (\omega-\mathbf{H})^{-1}.
	\end{equation}
	We will see in Section~\ref{sec:ODLR_Hessian} that the ODLR property of the shifted inverse of $\mathbf{H}$ is the basis of the ODLR property of nuclear Hessians.
	
	Actually, we can prove a much stronger fact: any matrix function $f(\mathbf{O})$ of a local matrix $\mathbf{O}$ satisfies the ODLR property, if \emph{most} of the eigenvalues of $\mathbf{O}$ are far from $f(x)$'s discontinuous points.
	Here, a matrix $\mathbf{O}$ is said to be a ``local matrix'', if $O_{\mu\nu}$ decays at least exponentially with respect to the distance between the basis functions $\mu$ and $\nu$, $R_{\mu\nu}$. Note that $f(x)$ can be discontinuous or even singular within the spectrum of $f(\mathbf{O})$. For example, the above examples belong to the special case where $f(x)=1/(x-a)$ for some constant $a$.
	
	Assume that $f(x)$ is continuous in a finite union of closed intervals $I=[a_1,b_1]\cup\ldots\cup[a_n,b_n]$, such that all eigenvalues of $\mathbf{O}$ fall within $[a_1,b_n]$, and all but a small portion of the eigenvalues of $\mathbf{O}$ reside in $I$. According to the Weierstrass approximation theorem, it is possible to uniformly approximate $f(x)$ on $I$ to arbitrary precision using a polynomial
	\begin{equation}
	f(x) \approx \tilde{f}(x) = \sum_{i=0}^{N_{\rm order}} a_i x^i.
	\end{equation}
	Since $\mathbf{O}^i$ is local for any finite positive integer $i$, $\tilde{f}(\mathbf{O})$ is local. As $f(x) \approx \tilde{f}(x)$ when $x\in I$, only those eigenvalues of $\mathbf{O}$ outside $I$ can make significant contributions to the far off-diagonal blocks of $f(\mathbf{O})$. Since there are only a small number of such eigenvalues, and each eigenvalue makes a rank-1 contribution to $f(\mathbf{O})$, the far off-diagonal blocks of $f(\mathbf{O})$ have low rank.
	
	The implications of this fact (which to our best knowledge has not been reported before, except for the case where $f(x)$ can be well-approximated by a low-order rational function\cite{HSS}) are profound. For example, taking $f(x)=\Theta(x-\mu)$, where
	\begin{equation}
	\Theta(x) = \left\{ \begin{array}{cc} 1, & x\le 0 \\ 0, & x>0 \end{array}, \right.
	\end{equation}
	we have that the zero-temperature density matrix $\mathbf{D}=\Theta(\mathbf{F}-\mu)$ of any local Fock matrix $\mathbf{F}$ has the ODLR property, even if some orbitals are very close to the chemical potential $\mu$ (in which case $\mathbf{D}$ would be dense).
	Meanwhile, taking $f(x)=\sqrt{x}$ or $f(x)=1/\sqrt{x}$, we can prove that the square roots and inverse square roots of AO overlap matrices $\mathbf{S}$ are ODLR, since $\mathbf{S}$ is local. The latter fact is especially important since $\mathbf{S}^{-1/2}$ can be quite non-local when the basis set is nearly linear dependent (i.e.~when $\mathbf{S}$ has very small eigenvalues).
	
	The RG structure of ODLR matrices can also be derived for matrix functions of local matrices, although the RG operation is now conducted in eigenvalue (``energy'') space instead of real space\cite{ERG,ERG2,ERG3}. Without loss of generality, consider a function $f(x)$ with a single discontinuity at $x=a$. One can construct a series of intervals
	\begin{eqnarray}
	I_i & = & [a-\delta_i, a-\delta_{i+1}) \cup (a+\delta_{i+1}, a+\delta_i], \nonumber\\
    & & \quad i=0,1,\ldots, \\
	\delta_0 & = & \max (a-\epsilon_{\rm min}, \epsilon_{\rm max}-a),
	\end{eqnarray}
	where $\epsilon_{\rm min}$ and $\epsilon_{\rm max}$ are the smallest and largest eigenvalues of $\mathbf{O}$, respectively, and $\delta_i=k^i\delta_0$ approaches zero monotonically from above (with $0<k<1$ being a constant). We set $f_0(x)\equiv f(x)$ and $\mathbf{O}_0\equiv \mathbf{O}$. For each $i=0,1,\ldots$, one constructs a polynomial approximation to $f_i(x)$, $\tilde{f}_i(x)$, on $I_i$, such that $f_i(\mathbf{O}_i)-\tilde{f}_i(\mathbf{O}_i)$ is a numerically low-rank matrix, and its column/row space is spanned by the eigenvectors of $\mathbf{O}_i$ whose eigenvalues are within $[a-\delta_{i+1},a+\delta_{i+1}]$. Denote the matrix composed of all such eigenvectors as $\mathbf{U}_i$. We then have
	\begin{eqnarray}
	f_i(\mathbf{O}_i)-\tilde{f}_i(\mathbf{O}_i) & \approx & \mathbf{U}_i (f(\mathbf{O}_i)-\tilde{f}_i(\mathbf{O}_i)) \mathbf{U}_i^T \nonumber\\
	& \equiv & \mathbf{U}_i f_{i+1}(\mathbf{O}_i) \mathbf{U}_i^T \nonumber\\
	& = & \mathbf{U}_i f_{i+1}(\mathbf{O}_{i+1}) \mathbf{U}_i^T, \\
	f_{i+1}(x) & = & f_i(x) - \tilde{f}_i(x), \\
	\mathbf{O}_{i+1} & = & \mathbf{U}_i^T\mathbf{O}_i\mathbf{U}_i.
	\end{eqnarray}
	$f_{i+1}(x)$ inherits the discontinuity of $f_i(x)$ at $x=a$. However, while the eigenvalues of $\mathbf{O}_i$ are contained within the interval $[a-\delta_i, a+\delta_i]$, those of $\mathbf{O}_{i+1}$ are contained within the smaller interval $[a-\delta_{i+1}, a+\delta_{i+1}]$. Therefore, we can obtain a polynomial approximation to $f_{i+1}(x)$, i.e.~$\tilde{f}_{i+1}(x)$, on $I_{i+1}$, with comparable expansion order and accuracy with the approximation of $f(x)$ on $I_i$; the proportionally shorter length of $I_{i+1}$ than $I_i$ makes an accurate polynomial approximation easier to achieve, which compensates for the closer distance of $I_{i+1}$ to the discontinuity (which makes a polynomial approximation more difficult). We thus obtain an approximation of $f(\mathbf{O})$ as a nested series:
	\begin{equation}
	f(\mathbf{O}) \approx \tilde{f_0}(\mathbf{O}_0) + \mathbf{U}_0 \left( \tilde{f_1}(\mathbf{O}_1) + \mathbf{U}_1 \left(  \tilde{f_2}(\mathbf{O}_2) + \ldots\right)  \mathbf{U}_1^T \right) \mathbf{U}_0^T. \label{eq:nested}
	\end{equation}
	Now, we argue that the possibility of expanding $f(\mathbf{O})$ as Eq.~\eqref{eq:nested} is equivalent to the ``nested basis structure'' property defined in the previous subsection. To see this, we rewrite Eq.~\eqref{eq:nested} as
	\begin{equation}
	f(\mathbf{O}) \approx \tilde{f_0}(\mathbf{O}_0) + \mathbf{U}_0 \tilde{f_1}(\mathbf{O}_1) \mathbf{U}_0^T + \mathbf{U}_0 \mathbf{U}_1 \tilde{f_2}(\mathbf{O}_2) \mathbf{U}_1^T \mathbf{U}_0^T + \ldots. \label{eq:nested2}
	\end{equation}
	The terms in Eq.~\eqref{eq:nested2} have lower and lower ranks, due to the decreasing sizes of the matrices $\mathbf{O}_i$. Meanwhile, they are less and less local, because they involve eigenvalues of $\mathbf{O}$ that are closer and closer to the discontinuity of $f(x)$. Therefore, the first few terms of Eq.~\eqref{eq:nested2} will contribute negligibly to any off-diagonal block of $f(\mathbf{O})$, $(f(\mathbf{O}))^{XY}$, where $X$ and $Y$ are spatially sufficiently far apart. Suppose that only the term $\mathbf{U}_0 \mathbf{U}_1 \ldots \mathbf{U}_i \tilde{f}_{i+1}(\mathbf{O}_{i+1}) \mathbf{U}_i^T \ldots \mathbf{U}_1^T \mathbf{U}_0^T$ and all subsequent terms contribute significantly to $(f(\mathbf{O}))^{XY}$. Then $(f(\mathbf{O}))^{XY}$ has the form
	\begin{equation}
	(f(\mathbf{O}))^{XY} \approx (\mathbf{U}_0 \mathbf{U}_1 \ldots \mathbf{U}_i)^{X*} \tilde{\mathbf{O}}_i^{XY} ((\mathbf{U}_0 \mathbf{U}_1 \ldots \mathbf{U}_i)^{Y*} )^T,
	\end{equation}
	for some small-dimensional matrix $\tilde{\mathbf{O}}_i^{XY}$. Here the notation $\mathbf{U}^{X*}$ denotes a horizontal section $\{U_{\mu i}, \mu\in X\}$ of the matrix $\mathbf{U}$. Given the SVD of $\tilde{\mathbf{O}}_i^{XY}$,
	\begin{equation}
	\tilde{\mathbf{O}}_i^{XY} = \tilde{\mathbf{U}}^{XY}_i\tilde{\mathbf{\sigma}}^{XY}_i(\tilde{\mathbf{V}}^{XY}_i)^T,
	\end{equation}
	we immediately obtain the SVD of $(f(\mathbf{O}))^{XY}$,
	\begin{eqnarray}
	(f(\mathbf{O}))^{XY} & \approx & \left((\mathbf{U}_0 \mathbf{U}_1 \ldots \mathbf{U}_i)^{X*} \tilde{\mathbf{U}}^{XY}_i\right) \tilde{\mathbf{\sigma}}^{XY}_i \nonumber\\
    & & \times \left((\mathbf{U}_0 \mathbf{U}_1 \ldots \mathbf{U}_i)^{Y*} \tilde{\mathbf{V}}^{XY}_i\right)^T. \label{eq:prod_U}
	\end{eqnarray}
	Thus, although the left singular vectors of $(f(\mathbf{O}))^{XY}$ do depend on $Y$, the column space of $(f(\mathbf{O}))^{XY}$ is independent of $Y$ and is the space spanned by $(\mathbf{U}_0 \mathbf{U}_1 \ldots \mathbf{U}_i)^{X*}$. Similarly, the row space of $(f(\mathbf{O}))^{XY}$ is independent of $X$. We have seen in the last subsection that such independence implies the nested basis structure.
	Moreover, the product of matrices with decreasing locality and dimensions bears close resemblance with Eq.~\eqref{eq:block_U}, except that here $\mathbf{U}_0,\mathbf{U}_1,\ldots$ are merely local but not block-diagonal.
	
	In sum, we have shown that the matrix representations of sufficiently smooth operators in a local basis, as well as many important matrix functions of local matrices, possess the following properties (denoting the matrix by $\mathbf{O}$):
	\begin{itemize}
		\item Property (a): off-diagonal blocks $\mathbf{O}^{XY}$ have low numerical rank, if the regions $X$ and $Y$ can be separated in real space by a smooth boundary (i.e.~they are far off-diagonal blocks).
		\item Property (b): the numerical rank of $\mathbf{O}^{XY}$ decreases as the distance between $X$ and $Y$ increases, if the sizes of the two regions are held fixed.
		\item Property (c): the column/row space of a far off-diagonal block $\mathbf{O}^{XY}$ is approximately independent of $Y$/$X$, respectively.
		\item Property (d): the column space of a far off-diagonal block $\mathbf{O}^{XY}$ is approximately equal to, or is an approximate subspace of, the direct sum of the column spaces of $\mathbf{O}^{X_1Z_1}$ and $\mathbf{O}^{X_2Z_2}$, where $X_1\cup X_2 = X$, and $Z_1, Z_2$ are arbitrary regions chosen to make $\mathbf{O}^{X_1Z_1}$ and $\mathbf{O}^{X_2Z_2}$ far off-diagonal blocks. Similarly for row spaces.
		\item Property (e): applying property (d) recursively, one can factor the left and right singular vectors of a far off-diagonal block $\mathbf{O}^{XY}$ as the product of many matrices with smaller and smaller sizes, as well as poorer and poorer locality.
		\item Property (f): if $X$ and $Y$ additionally satisfy the admissibility condition, the numerical rank of $\mathbf{O}^{XY}$ does not grow asymptotically with the sizes of $X$ and $Y$, and approaches a constant.
	\end{itemize}
	In the next section, we will discuss the use of these properties in the storage, computation and usage of ODLR matrices.
	
	\section{Compact representations of ODLR matrices} \label{sec:ODLR_compress}
	
\begin{figure}[htbp]
        \resizebox{0.48\textwidth}{!}{\includegraphics{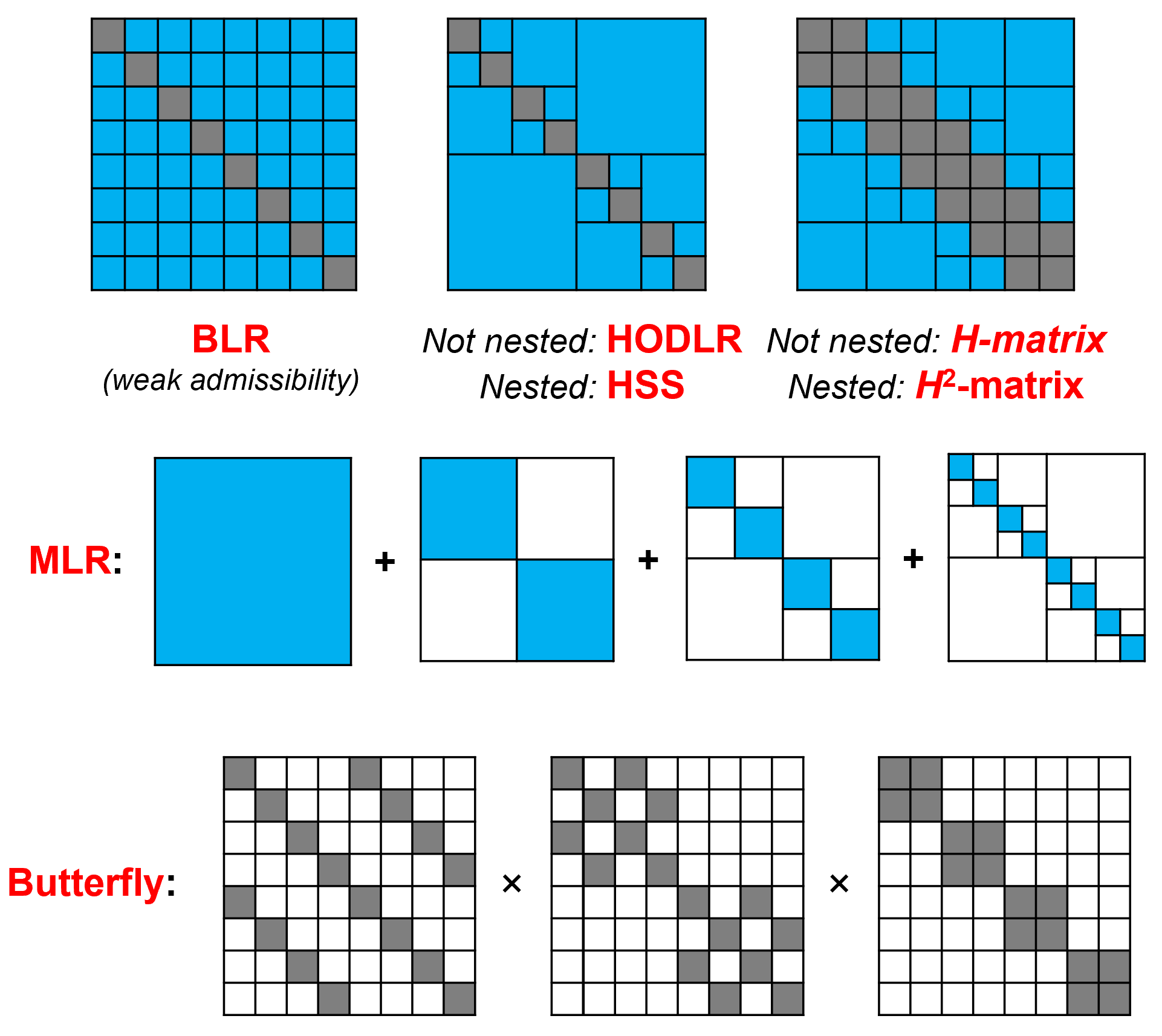}}
		\caption{Schematic representations of different storage formats of ODLR matrices. Grey, blue and white blocks are stored directly, stored as low-rank decomposition, and set as zero, respectively.}
		\label{fig:storage_formats}
	\end{figure}

	Given a matrix that possesses the ODLR property, it remains to find the optimum way to compactly store the matrix, which is a prerequisite for low-scaling calculations involving the matrix. For example, an $N\times N$ dense matrix requires $O(N^2)$ memory to store, if no compression is applied. Computing the matrix-vector product of such a matrix with a vector of length $N$ thus necessarily takes $O(N^2)$ time, since every matrix element must be accessed at least once. Similarly, generating an $N\times N$ dense matrix in uncompressed form takes at least $O(N^2)$ time, as at least $O(1)$ effort is required for generating each matrix element. In general, to devise an $O(N^\alpha)$ ($1\le\alpha<2$) scaling algorithm with an $N\times N$ matrix as input or output, one must compress the matrix such that it is expressible using at most $O(N^\alpha)$ variables, unless the matrix is used as input and not all of its elements are accessed.
	In this section, we review some common ways to compactly store and efficiently compute/use ODLR matrices (Figure~\ref{fig:storage_formats}).
	
	\subsection{Block Low Rank (BLR) format}
	
	The most intuitive way to compress an ODLR matrix is to partition the matrix into many blocks (Figure~\ref{fig:storage_formats}), perform SVD on each off-diagonal block, and store the resulting singular values and singular vectors; the diagonal blocks are stored directly.
	Formally, for an $N\times N$ matrix $\mathbf{O}$, one partitions the $N$ indices into $N_{\rm block}<N$ disjoint sets, $\{X_k\}$:
	\begin{equation}
	\bigcup_{k=1}^{N_{\rm block}} X_k = \{1,2,\ldots,N\},
	\end{equation}
	\begin{equation}
	X_k \cap X_l = \varnothing, \quad \forall k\neq l.
	\end{equation}
	Then, all off-diagonal blocks $\mathbf{O}^{X_kX_l}~(k\neq l)$ are stored as their SVD decomposition
	\begin{equation}
	\mathbf{O}^{X_kX_l} \approx \tilde{\mathbf{U}}^{kl}\mathbf{\sigma}^{kl}(\tilde{\mathbf{V}}^{kl})^T,
	\end{equation}
	where $\tilde{\mathbf{U}}^{kl}$ and $\tilde{\mathbf{V}}^{kl}$ are the left and right singular vectors, respectively, and $\mathbf{\sigma}^{kl}$ is the diagonal matrix of the singular values. Here the negligible singular values (e.g.~those smaller than a threshold $\epsilon$) are truncated off, so that $\tilde{\mathbf{U}}^{kl}$, $\mathbf{\sigma}^{kl}$ and $\tilde{\mathbf{V}}^{kl}$ have sizes $|X_k|\times r_{kl}$, $r_{kl}\times r_{kl}$ and $|X_l|\times r_{kl}$, respectively, where $r_{kl}$ is the numerical rank of $\mathbf{O}^{X_kX_l}$.
	Some further savings can be made by absorbing $\mathbf{\sigma}^{kl}$ into $\tilde{\mathbf{U}}^{kl}$ and $\tilde{\mathbf{V}}^{kl}$, e.g.
	\begin{equation}
	\mathbf{O}^{X_kX_l} \approx \mathbf{U}^{kl}(\mathbf{V}^{kl})^T, \label{eq:OUV}
	\end{equation}
    \begin{equation}
    \mathbf{U}^{kl} = \tilde{\mathbf{U}}^{kl}(\mathbf{\sigma}^{kl})^{1/2}, \quad \mathbf{V}^{kl} = \tilde{\mathbf{V}}^{kl}(\mathbf{\sigma}^{kl})^{1/2}. 
    \end{equation}
	Certain techniques other than SVD, e.g.~truncated QR factorization\cite{borm2003Hmatrix,amestoy2019SIAM}, can directly lead to low-rank decompositions of the form Eq.~\eqref{eq:OUV}, using less computational resources than SVD.
	
	The above compression scheme is called the weak-admissibility BLR format\cite{BLR,BLR2}. Storing some of the ``near'' off-diagonal blocks as dense blocks, one obtains the strong-admissibility BLR format\cite{BLR2,BLR}; for simplicity, in the rest of the manuscript we assume that all BLR matrices have weak-admissibility, and the benefits of strong-admissibility (i.e.~storing some off-diagonal blocks in dense form) will be discussed in Section~\ref{sec:Hmatrix}.
	The compression efficiency depends on the block size: large block sizes imply higher storage overheads for the diagonal blocks, while small block sizes give rise to a large number of blocks and are therefore inefficient for compressing off-diagonal blocks.
	
	Once a matrix is stored in the BLR format, its multiplication with vectors or another BLR matrix with the same index partition scheme can be done in a block-wise fashion\cite{amestoy2019SIAM}. For example, the $X_k$ block of the matrix-vector product $\mathbf{O}\mathbf{v}$ can be evaluated as
	\begin{eqnarray}
	(\mathbf{O}\mathbf{v})^{X_k} & = & \sum_{l=1}^{N_{\rm block}}\mathbf{O}^{X_kX_l}\mathbf{v}^{X_l} \nonumber\\
	& \approx & \sum_{l=1, l\neq k}^{N_{\rm block}} \mathbf{U}^{kl}\left((\mathbf{V}^{kl})^T\mathbf{v}^{X_l}\right) + \mathbf{O}^{X_kX_k}\mathbf{v}^{X_k}. \label{eq:BLR_MVP}
	\end{eqnarray}
	The extra parentheses emphasize that the product $\mathbf{U}^{kl}(\mathbf{V}^{kl})^T\mathbf{v}^{X_l}$ is evaluated from right to left, so that only $r_{kl}(|X_k|+|X_l|)$ multiplications are needed, compared to $|X_k||X_l|$ if $\mathbf{O}^{X_kX_l}$ is stored in dense form\cite{pichon2018}. Note in passing that rank deficiency ($r_{kl}<\min(|X_k|,|X_l|)$) is not sufficient for the BLR format to become advantageous over the dense format, and the stronger criterion $r_{kl}<\frac{|X_k||X_l|}{|X_k|+|X_l|}$ must be satisfied. As large matrices benefit more from cache optimizations, the actual requirement on $r_{kl}$ for observing a speedup is even more stringent.
	
	The product of two BLR matrices with the same index partitioning scheme can similarly become cheaper than the corresponding dense matrix products, if the off-diagonal blocks have sufficiently low rank. On one hand, the product of a compressed off-diagonal block and an uncompressed block can be accelerated similar to Eq.~\eqref{eq:BLR_MVP}; on the other hand, the product of a rank-$r_{kl}$ block $\mathbf{O}_1^{X_kX_l}=\mathbf{U}_1^{kl}(\mathbf{V}_1^{kl})^T$ and a rank-$r_{lm}$ block $\mathbf{O}_2^{X_lX_m}=\mathbf{U}_2^{lm}(\mathbf{V}_2^{lm})^T$ can be evaluated as
	\begin{equation}
	\mathbf{O}_1^{X_kX_l}\mathbf{O}_2^{X_lX_m} = \mathbf{U}_1^{kl} \left((\mathbf{V}_1^{kl})^T\mathbf{U}_2^{lm}\right) (\mathbf{V}_2^{lm})^T,
	\end{equation}
	which requires only $r_{kl}r_{lm}|X_l| + \min(r_{kl}|X_m|(r_{lm}+|X_k|), r_{lm}|X_k|(r_{kl}+|X_m|))$ multiplications, as opposed to $|X_k||X_l||X_m|$ multiplications if both blocks are densely stored. Usually, the result is stored as its low-rank decomposition, so that the resulting matrix is also a BLR matrix, and the number of multiplications can be reduced to $r_{kl}r_{lm}(|X_l| + \min(|X_m|, |X_k|))$.
	
	Moreover, most other linear algebra operations can be performed more efficiently with the BLR format, too. For example, matrix diagonalization, the solution of linear systems and the evaluation of matrix functions can be performed using Krylov subspace methods\cite{cortinovis2022}, which involve the repeated formation of matrix-vector products. Matrix inverses\cite{matrix_inverse} and (inverse) square roots\cite{matrix_invsqrt,ERG2} can be found efficiently by iterative algorithms involving repeated matrix-matrix multiplications. These operations thus inherit the speedups of matrix-vector and matrix-matrix products.
    Meanwhile, dedicated algorithms also exist that can potentially solve linear systems involving BLR matrices faster than Krylov subspace methods\cite{BLR,BLR2}. In fact, efficient solution of linear systems was the primary motive for the design of the BLR format, as well as many of the formats that will be mentioned in the following subsections.
	
	We finalize the subsection by the best-case space complexity of storing BLR matrices, as well as the best-case time complexity of generating and using BLR matrices\cite{amestoy2019SIAM} (the worst-case complexities are trivially equal to dense storage, corresponding to the case where all blocks are full-rank). Assuming that none of the off-diagonal blocks are negligible (so that all blocks have ranks at least one), the best possible compression is obtained when all off-diagonal blocks have rank $O(1)$. It is then easy to show that the optimal number of blocks $N_{\rm block}$ scales as $O(N^{1/2})$, so that $O(N^{3/2})$ storage is required for both the diagonal and off-diagonal blocks, compared to $O(N^2)$ for dense storage. In this case, generating the BLR representation of a dense matrix requires $O(N^2)$ operations. Matrix-vector and matrix-matrix products respectively cost $O(N^{3/2})$ and $O(N^2)$ operations, leading to $O(N^{1/2})$ and $O(N)$ speedups compared to dense linear algebra ($O(N^2)$ and $O(N^3)$, respectively).
	
	\subsection{Hierarchical Off-Diagonal Low Rank (HODLR) format}
	
	While the BLR format exploits the low-rank property of the off-diagonal blocks (property (a) in Section~\ref{sec:foundations}), it fails to use the fact that the numerical rank of $\mathbf{O}^{XY}$ becomes lower as the distance between $X$ and $Y$ increases\cite{BLR} (property (b); Figure~\ref{fig:coul_grid2}). For even more efficient compression, larger blocks should be used for those pairs of basis functions $\mu,\nu$ that are far from each other. The HODLR format\cite{HODLR,hackbusch2004} was thus designed by hierarchically performing BLR compressions on the diagonal blocks of a BLR matrix, yielding a finer and finer partition along the diagonal (Figure~\ref{fig:storage_formats}); the procedure stops when the smallest diagonal blocks have size $O(1)$. If a block is introduced in the $k$-th BLR compression step, the block is said to have level $k$; the total number of levels is denoted as $N_{\rm level}$. Typically, at each step the matrix is divided into $2\times2$ blocks (or more generally $O(1)\times O(1)$ blocks), with roughly equal sizes. Under this setting, we have $N_{\rm level} \sim O(\log N)$, and with the optimistic assumption that all off-diagonal blocks have rank $O(1)$, it requires only $O(N\log N)$ space to store an HODLR matrix (since each level-$i$ block has size $\sim N/2^i$, such that the low-rank decomposition of a level-$i$ block requires $O(N/2^i)$ storage, and there are $2^i$ level-$i$ blocks), which is a significant improvement over BLR. It is straightforward to show that generation of the HODLR representation of a dense matrix, under the same assumptions, costs $O(N^2)$ time\cite{kong2011}.
	
	Unlike the BLR format, the HODLR format is a scale-free compression format, containing blocks with sizes from $O(1)$ to $O(N)$, with no ``characteristic block size''. A matrix element $O_{\mu\nu}$ tends to belong to a small block if $\mu$ and $\nu$ are spatially close to each other, and a large block if they are far from each other. This is consistent with the RG philosophy, where long-range interactions are studied at coarser scales.
	Note that to obtain the $O(N\log N)$ space complexity, it is important to have a variable $N_{\rm level}\sim O(\log N)$ that increases with $N$.
	If $N_{\rm level}\sim O(1)$, the resulting scheme is called multilevel BLR (MBLR)\cite{amestoy2019SIAM}; assuming all off-diagonal blocks have non-zero, $O(1)$ numerical ranks, the optimal storage complexity of MBLR is $O(N^{(N_{\rm level}+2)/(N_{\rm level}+1)})$\cite{amestoy2019SIAM}, which can be arbitrarily close to $O(N\log N)$, but always remains higher than $O(N\log N)$.
	
	HODLR matrices can be easily used in matrix-vector and matrix-matrix product calculations by recursively calling the respective subroutines designed for BLR matrices, remembering that HODLR matrices are just BLR matrices whose dense blocks are also BLR matrices. When evaluating the product of an HODLR matrix with a vector, the level 1 off-diagonal blocks can be multiplied with the corresponding block of the vector, just like the off-diagonal blocks of normal BLR matrices; the level 1 diagonal blocks are themselves BLR matrices composed of level 2 blocks, so that their contributions to the matrix-vector product can also be efficiently calculated, etc.
	Products two HODLR matrices can similarly be performed in a block-wise manner, which involve the products of (1) two BLR blocks, (2) one BLR block and a low-rank block, and (3) two low-rank blocks. Here the efficient algorithms of (1) and (3) are already known and discussed above. To efficiently perform (2), say between an $N\times N$-sized BLR block $\mathbf{O}$ and a rank-$r$, $N\times N$-sized block decomposed as $\mathbf{U}\mathbf{V}^T$, first multiply $\mathbf{O}\mathbf{U}$ (which is equivalent to $r$ matrix-vector products), followed by multiplying the resulting $N\times r$-sized matrix with $\mathbf{V}^T$.
	Again assuming all off-diagonal blocks have rank $O(1)$ and are non-zero, matrix-vector products of HODLR matrices cost $O(N\log N)$ time, and multiplying two HODLR matrices costs $O(N\log^2 N)$ time\cite{amestoy2019SIAM}, which represent almost linear and quadratic speedups compared to dense matrix-vector and matrix-matrix products, respectively. This highlights the benefits of using a hierarchical format over the plain BLR format.
	
	\subsection{$\mathscr{H}$-matrix format} \label{sec:Hmatrix}
	
\begin{figure}[htbp]
			\resizebox{0.48\textwidth}{!}{\includegraphics{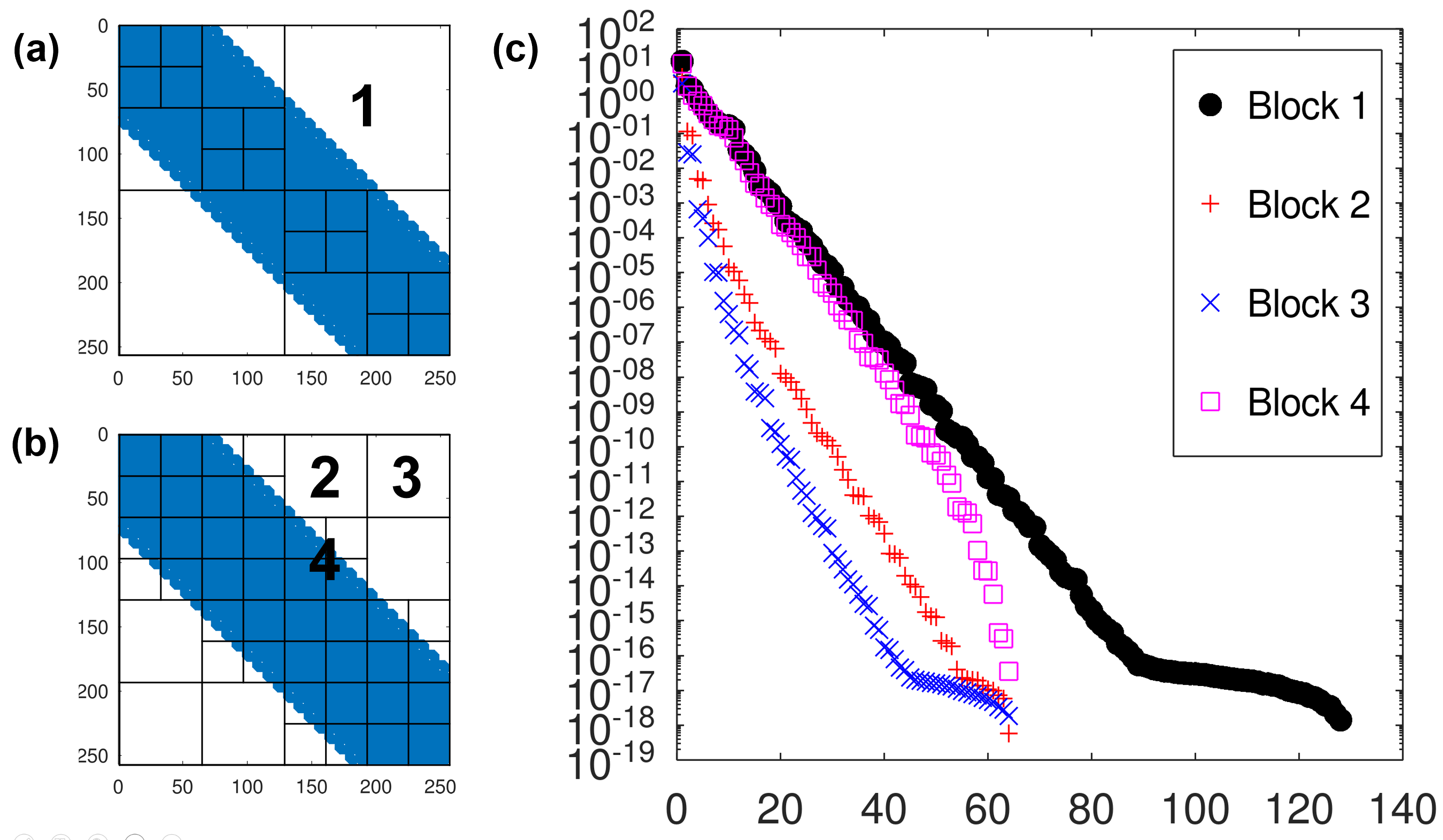}}
		\caption{(a) HODLR and (b) $\mathscr{H}$-partitions of the Coulomb matrix $\mathbf{J}$ of an 8$\times$32 grid of point charges, where matrix elements of $\mathbf{J}$ with absolute value larger than 0.1 are shown in blue; (c) the singular values of selected blocks of $\mathbf{J}$.}
		\label{fig:coul_grid3}
	\end{figure}

	One pitfall of the HODLR format is that the extra logarithmic factors in the $O(N\log N)$ and $O(N\log^2 N)$ time and space complexities remain even if the matrix is local, where standard sparse linear algebra easily leads to $O(N)$ scalings in terms of both time and space complexities, suggesting that the HODLR format is not optimal in the limit of local matrices. Even worse, for two and higher dimensional systems, the off-diagonal blocks of an HODLR matrix generally have numerical ranks higher than $O(1)$, making the scalings even higher\cite{pichon2018}. To see the reason, assume that the basis functions form a connected set, in the sense that it is impossible to separate the basis functions into two admissible regions. Then, the singular values of the level 1 off-diagonal block decay slowly due to inadmissibility. One can easily see this in Figure~\ref{fig:coul_grid3}(a,c), where the highlighted (level 1) off-diagonal block is characterized by a slow singular value decay, but once the block is further partitioned into four smaller blocks, all blocks except the lower left block have a much faster singular value decay (Figure~\ref{fig:coul_grid3}(b,c)).
	
	The $\mathscr{H}$-matrix format\cite{hackbusch1999Hmatrix,hackbusch2000Hmatrix,hackbusch2004,borm2003Hmatrix} is therefore designed by recursively partitioning all inadmissible blocks into smaller blocks (usually, a $2\times 2$ partitioning is used), until all blocks are either small enough (and its contents densely stored), or admissible (and its contents stored as a low-rank decomposition; Figure~\ref{fig:storage_formats}).
    This results in a format with a finer tesselation and a larger number of densely stored entries than the HODLR format.
	For sufficiently large matrices, the $\mathscr{H}$-matrix format often provides better compression efficiencies than the HODLR format, owing to property (f). For example, if the basis functions form a three-dimensional grid in real space, the level 1 off-diagonal block in the HODLR format has rank $O(N^{1/3})$, and therefore the whole matrix requires $O(N^{4/3})$ storage\cite{pichon2018}; while in the $\mathscr{H}$-matrix format, the same block is converted to numerous smaller blocks, so that all low-rank blocks have rank $O(1)$, and the storage requirement is thus reduced to $O(N\log^2 N)$. Note that this is the same as the best-case storage requirement of HODLR matrices, when all off-diagonal blocks have rank $O(1)$; the benefit of $\mathscr{H}$-matrices is that all off-diagonal blocks of an $\mathscr{H}$-matrix do have rank $O(1)$ (owing to property (f) mentioned above), a condition frequently violated for HODLR matrices.
	Similarly, the best-case scalings of generating a $\mathscr{H}$-matrix representation of a dense matrix, as well as performing $\mathscr{H}$-matrix-vector and $\mathscr{H}$-matrix-$\mathscr{H}$-matrix products, are $O(N^2)$, $O(N\log N)$ and $O(N\log^2 N)$, respectively\cite{borm2003Hmatrix}.

    For high-dimensional problems, the ranks of the low-rank blocks, though constant with respect to the matrix size, increase exponentially with respect to the dimension. One possible remedy is to replace low-rank decomposition in the construction of $\mathscr{H}$-matrices by Tucker decompositions\cite{HTLR}, yielding better compression ratios.
	
	\subsection{Hierarchical semi-separable (HSS) and $\mathscr{H}^2$-matrix formats} \label{sec:HSS}
	
	Although the compression ratios and speedup factors of the HODLR and $\mathscr{H}$-matrix formats are remarkable, they are not optimal, as they fail to use property (c) and therefore property (d). The HSS\cite{HSS} and $\mathscr{H}^2$-matrix\cite{hackbusch2002H2matrix} formats were designed to use properties (c) and (d) to further reduce the storage and computation costs of ODLR matrices. They have the same block partition scheme as the HODLR and $\mathscr{H}$-matrix formats, respectively (Figure~\ref{fig:storage_formats}), except that (1) any two low-rank blocks $\mathbf{O}^{X_1Y}$ and $\mathbf{O}^{X_2Y}$ share the same column space, and any two low-rank blocks $\mathbf{O}^{XY_1}$ and $\mathbf{O}^{XY_2}$ share the same row space; (2) if $X=X_1\cup X_2$, then the column space of $\mathbf{O}^{XY}$ is a subspace of the direct sum of the column spaces of $\mathbf{O}^{X_1Z_1}$ and $\mathbf{O}^{X_2Z_2}$, where $\mathbf{O}^{X_1Z_1}$ and $\mathbf{O}^{X_2Z_2}$ are two low-rank blocks; similarly for row spaces.
	
	It can be shown that this ``reuse'' of singular vector information across blocks can reduce the best-case storage cost to $O(N)$. Assuming again that each low-rank block has rank $O(1)$, the HODLR format requires $O(N)$ storage space for blocks belonging to each level, and since there are $O(\log N)$ levels, the whole matrix requires $O(N\log N)$ storage\cite{HODLR}. By contrast, in the HSS format, level $i$ requires $O(2^i)$ storage, which is independent of $N$. To see this, assume that the singular vectors of all level-$(i+1)$ blocks are known. To linearly combine them to yield the singular vectors of a level-$i$ block, one only requires a linear transformation matrix of size $O(1)\times O(1)$, since both the level-$(i+1)$ blocks and the level $i$ blocks have rank $O(1)$. Since there are $2^i$ level-$i$ blocks, the total storage complexity of level-$i$ blocks is $O(2^i)$. The complexity of storing all blocks is thus\cite{HSS} (recalling $N_{\rm level}\sim \log_2N$ for hierarchical $2\times 2$ partitions)
	\begin{equation}
	\underbrace{O(N)}_{\textrm{dense blocks}} + \underbrace{O(\sum_{i=1}^{N_{\rm level}}2^i)}_{\textrm{low-rank blocks}} \sim O(N).
	\end{equation}
	Similar estimates can easily be obtained for $\mathscr{H}^2$-matrices\cite{hackbusch2002H2matrix}. It can be shown, though somewhat tediously, that the best-case complexity of generating an HSS\cite{HSS,HSS_complexity} or $\mathscr{H}^2$-matrix\cite{hackbusch2002H2matrix} representation of a dense matrix is $O(N^2)$.
	Meanwhile, the best-case scalings of the matrix-vector and matrix-matrix products involving HSS\cite{HSS,HSS_complexity} and $\mathscr{H}^2$-matrices\cite{hackbusch2002H2matrix} are $O(N)$, yielding speedup factors of $O(\log N)$ and $O(\log^2 N)$, respectively, compared to $\mathscr{H}$-matrices. This follows from the fact that the singular vectors of the off-diagonal blocks of an $\mathscr{H}^2$-matrix can be expressed as a product of matrices, where all matrices but the last one are block-diagonal (Eq.~\eqref{eq:block_U}). The $X$ block of the matrix-vector product $\mathbf{O}\mathbf{v}$ can then be expressed as
	\begin{eqnarray}
	& & (\mathbf{O}\mathbf{v})^X = \sum_{Y~\textrm{s.t.}~ \mathbf{O}^{XY}~\textrm{dense}}\mathbf{O}^{XY}\mathbf{v}^Y + \sum_{Y~\textrm{s.t.}~ \mathbf{O}^{XY}~\textrm{low rank}}\mathbf{O}^{XY}\mathbf{v}^Y \nonumber\\
	& = & \sum_{Y~\textrm{s.t.}~ \mathbf{O}^{XY}~\textrm{dense}}\mathbf{O}^{XY}\mathbf{v}^Y \nonumber\\
    & & + \sum_{Y~\textrm{s.t.}~ \mathbf{O}^{XY}~\textrm{low rank}}\mathbf{U}^{XY}\mathbf{\sigma}^{XY}(\mathbf{V}^{XY})^T\mathbf{v}^Y \nonumber\\
	& = & \sum_{Y~\textrm{s.t.}~ \mathbf{O}^{XY}~\textrm{dense}}\mathbf{O}^{XY}\mathbf{v}^Y  + \mathbf{U}^{(X,1)}\mathbf{U}^{(X,2)}\cdots \mathbf{U}^{(X,N_{\rm level}^X)} \nonumber\\
	& & \times\sum_{Y~\textrm{s.t.}~ \mathbf{O}^{XY}~\textrm{low rank}} \tilde{\mathbf{U}}^{XY}\mathbf{\sigma}^{XY}(\tilde{\mathbf{V}}^{XY})^T \nonumber\\
	& & \times (\mathbf{V}^{(Y,N_{\rm level}^Y)})^T\cdots (\mathbf{V}^{(Y,2)})^T(\mathbf{V}^{(Y,1)})^T\mathbf{v}^Y. \label{eq:Ov_H2}
	\end{eqnarray}
	The product in the last term is then evaluated from right to left.
	The crucial point here is that different regions $X$ may have different sizes, and thus may yield different numbers of partition levels $N_{\rm level}^X$. For example, if $X$ is partitioned into two regions of roughly the same size, $X_1$ and $X_2$, such that $X=X_1\cup X_2$, then in general $N_{\rm level}^{X_1}=N_{\rm level}^{X_2}=N_{\rm level}^{X}-1$. Nevertheless, the matrix factors of $\mathbf{U}^{XY}$ can be reused to factorize $\mathbf{U}^{X_1Y}$ and $\mathbf{U}^{X_2Y}$, noting that Eq.~\eqref{eq:block_U} implies
	\begin{eqnarray}
	\mathbf{U}^{XY} & = & \left( \begin{array}{cc} \mathbf{U}^{(X_1, 1)} & \mathbf{0} \\ \mathbf{0} & \mathbf{U}^{(X_2, 1)} \end{array} \right) \left( \begin{array}{cc} \mathbf{U}^{(X_1, 2)} & \mathbf{0} \\ \mathbf{0} & \mathbf{U}^{(X_2, 2)} \end{array} \right) \cdots \nonumber\\
	& & \times\left( \begin{array}{cc} \mathbf{U}^{(X_1, N_{\rm level}-1)} & \mathbf{0} \\ \mathbf{0} & \mathbf{U}^{(X_2, N_{\rm level}-1)} \end{array} \right)  \nonumber\\
	& & \times \left( \begin{array}{cc} \tilde{\mathbf{U}}^{X_1Z_1} & \mathbf{0} \\ \mathbf{0} & \tilde{\mathbf{U}}^{X_2Z_2} \end{array} \right) \tilde{\mathbf{U}}^{XY}.
	\end{eqnarray}
	Therefore, intermediate results while evaluating Eq.~\eqref{eq:Ov_H2} can be reused to assemble the intermediates needed for smaller blocks, e.g.~to evaluate $(\mathbf{V}^{(Y,N_{\rm level}^Y)})^T\cdots (\mathbf{V}^{(Y,2)})^T(\mathbf{V}^{(Y,1)})^T\mathbf{v}^Y$, one would first need to evaluate $(\mathbf{V}^{(Y,N_{\rm level}^Y-1)})^T\cdots (\mathbf{V}^{(Y,2)})^T(\mathbf{V}^{(Y,1)})^T\mathbf{v}^Y$, which is nothing but
	\begin{equation}
	\left( \begin{array}{c} (\mathbf{V}^{(Y_1,N_{\rm level}^{Y_1})})^T\cdots (\mathbf{V}^{(Y_1,2)})^T(\mathbf{V}^{(Y_1,1)})^T\mathbf{v}^{Y_1} \\ (\mathbf{V}^{(Y_2,N_{\rm level}^{Y_2})})^T\cdots (\mathbf{V}^{(Y_2,2)})^T(\mathbf{V}^{(Y_2,1)})^T\mathbf{v}^{Y_2} \end{array} \right),
	\end{equation}
	i.e.~a concatenation of the corresponding intermediates of the regions $Y_1$ and $Y_2$. It is this reuse that is responsible for the logarithmic speedups compared to HODLR and $\mathscr{H}$-matrices. Analogous arguments can be used to show that matrix-matrix products between HSS or $\mathscr{H}^2$-matrices are also linear scaling, if all off-diagonal blocks have rank $O(1)$\cite{HSS,hackbusch2002H2matrix}.

    Finally, we briefly mention that all HSS matrices admit a ``telescopic decomposition''\cite{telescopic,telescopic2}
    \begin{equation}
	\mathbf{O} = \mathbf{D}_0 + \mathbf{U}_0 \left( \mathbf{D}_1 + \mathbf{U}_1 \left(  \mathbf{D}_2 + \ldots\right)  \mathbf{U}_1^T \right) \mathbf{U}_0^T,
	\end{equation}
    where $\mathbf{D}_n$ is a block-diagonal matrix with block size $2^n$. Such decompositions clearly resemble Eq.~\eqref{eq:nested}, and has found use in evaluating matrix functions of $\mathbf{O}$\cite{telescopic,telescopic2}.
	
	\subsection{Multilevel Low Rank (MLR) format} \label{sec:MLR}
	
	Instead of directly approximating its blocks, another way to compactly store ODLR matrices is to express it as a sum of matrices with decreasing locality and rank. Then, since only the low-rank matrices but not the local matrices contribute to the far off-diagonal blocks, the far off-diagonal blocks automatically have low rank.
	Such a representation arises naturally from Eq.~\eqref{eq:nested}.
	In the MLR format\cite{MLR,MLR2}, it is customary to further require the local matrices to be block-diagonal. The matrices therefore have fewer and fewer (but larger and larger) blocks, and except possibly for the smallest blocks, all blocks have low rank (Fig.~\ref{fig:storage_formats}). The number of summed matrices is called the \emph{level} of the MLR representation.
    A special case of MLR format is the low rank plus diagonal (LRPD) format\cite{LRPD,LRPD2}, which is a level-2 MLR format where the local matrix is diagonal (i.e.~block-diagonal with a block size of 1).
	
	Assuming that each block has rank $O(1)$, the scalings of storage cost, as well as computational time of matrix-vector and matrix-matrix products, are the same as HODLR matrices ($O(N \log N)$, $O(N \log N)$ and $O(N \log^2 N)$, respectively)\cite{MLR}. This is understandable given that an HODLR matrix
	\begin{equation}
	\mathbf{O} = \left( \begin{array}{cc} \mathbf{O}^{XX} & \mathbf{U}^{X}(\mathbf{V}^{Y})^T \\ \mathbf{V}^{Y}(\mathbf{U}^{X})^T & \mathbf{O}^{YY} \end{array} \right)
	\end{equation}
	can be trivially converted to a level-2 MLR representation:
	\begin{eqnarray}
	\mathbf{O} & = & \left( \begin{array}{cc} \mathbf{O}^{XX} - \mathbf{U}^{X}(\mathbf{U}^{X})^T & \mathbf{0} \\ \mathbf{0} & \mathbf{O}^{YY} - \mathbf{V}^{Y}(\mathbf{V}^{Y})^T \end{array} \right) \nonumber\\
	& & + \left( \begin{array}{c} \mathbf{U}^{X}\\ \mathbf{V}^{Y} \end{array} \right) \left( \begin{array}{cc} (\mathbf{U}^{X})^T & (\mathbf{V}^{Y})^T \end{array} \right). \label{eq:HODLR2MLR}
	\end{eqnarray}
	The diagonal blocks of the first term, $\mathbf{O}^{XX} - \mathbf{U}^{X}(\mathbf{U}^{X})^T$ and $\mathbf{O}^{YY} - \mathbf{V}^{Y}(\mathbf{V}^{Y})^T$, are themselves ODLR matrices, since the terms $\mathbf{U}^{X}(\mathbf{U}^{X})^T$ and $\mathbf{V}^{Y}(\mathbf{V}^{Y})^T$ have low rank, and therefore do not destroy the low rank character of the off-diagonal blocks of $\mathbf{O}^{XX}$ and $\mathbf{O}^{YY}$. The first term of Eq.~\eqref{eq:HODLR2MLR} can thus be recursively decomposed into the sum of matrices with more and more blocks, eventually yielding a multilevel MLR description of $\mathbf{O}$.
	
	While the MLR format is intimately related to the HODLR format, its expressive power can be improved dramatically if the involved matrices are merely required to be local, but not block-diagonal. In this case, the low-rank matrices in the MLR representation only needs to represent far off-diagonal matrix elements well, and all matrix elements $O_{\mu\nu}$ with small inter-site distances $R_{\mu\nu}$ can be taken care of by the local matrices in the MLR representation. This is in accord with the improvement of expressive power from HODLR matrices to $\mathscr{H}$-matrices, or from HSS matrices to $\mathscr{H}^2$-matrices. We will see in Sec.~\ref{sec:ODLR_Hessian} how a level-2 MLR representation of the nuclear Hessian enables the calculation of seminumerical Hessians using only $O(1)$ gradients, while conventional algorithms require $O(N_{\rm atom})$ gradients.
	
	\subsection{Butterfly, BP and BPBP and related decompositions} \label{sec:butterfly}
	
	The success of the MLR format raises a natural question: is it possible to express ODLR matrices as \emph{products} of simpler matrices? Eq.~\eqref{eq:prod_U} shows that this is possible for the far off-diagonal blocks of ODLR matrices (property (e)), but is an analogous decomposition possible for the whole matrix?
	For translationally invariant convolutional operators with the form Eq.~\eqref{eq:convolution}:
	\begin{equation}
	\left(\hat{O}f\right)(\mathbf{r}) = \int K(\mathbf{r}-\mathbf{r}')f(\mathbf{r}')d\mathbf{r}', \label{eq:convolution2}
	\end{equation}
	the answer is affirmative. A standard technique for evaluating convolutions like Eq.~\eqref{eq:convolution2} is to perform Fourier transform $\hat{\mathcal{F}}$ on the input, which converts convolutions with $K(\mathbf{r}-\mathbf{r}')$ to multiplications with its reciprocal space representation $(\hat{\mathcal{F}}K)(\mathbf{k})$. The final result is then found by the inverse Fourier transform $\hat{\mathcal{F}}^{-1}$:
	\begin{equation}
	(\hat{O}f)(\mathbf{r}) = (\hat{\mathcal{F}}^{-1}(\hat{\mathcal{F}}K)(\hat{\mathcal{F}}f))(\mathbf{r}).
	\end{equation}
	This expression can be discretized on a regular grid, yielding a matrix representation. For example, for one-dimensional convolution operators we can write
	\begin{equation}
	\mathbf{O} = (\mathbf{\mathcal{F}}^{(N)})^{-1}\mathbf{\mathcal{O}}\mathbf{\mathcal{F}}^{(N)},
	\end{equation}
	where $\mathbf{\mathcal{F}}^{(N)}$ is the discrete Fourier transform matrix on a one-dimensional grid with $N$ points, and is given by (here and in the rest of this subsection, $i$ denotes the imaginary unit, and unlike in many literatures, the indices start from 1 instead of 0, in accord with the remaining parts of the article)
	\begin{equation}
	\mathcal{F}_{kl}^{(N)} = e^{-2\pi i(k-1)(l-1)/N}. \label{eq:Fourier}
	\end{equation}
	Crucially, $\mathbf{\mathcal{O}}$ is very local, and becomes diagonal as the grid approaches completeness, since it represents a purely multiplicative operation that multiplies the input function value at each reciprocal point $\mathbf{k}$ by $(\hat{\mathcal{F}}K)(\mathbf{k})$, and does not mix the function values at different reciprocal points.
	Thus, in this specific case, $\mathbf{O}$, though neither local nor low-rank, becomes local after a basis transformation to the reciprocal space, and the transformation matrix $\mathbf{\mathcal{F}}^{(N)}$ is independent of the details of $\mathbf{O}$, only the fact that the latter represents the real-space discretization of a convolution operator with a sufficiently smooth kernel.
	
	One special property of $\mathbf{\mathcal{F}}^{(N)}$ is that multiplications of $\mathbf{\mathcal{F}}^{(N)}$ with a vector can be done in $O(N\log N)$ time, when $N = 2^M$ is a power of 2. In this case, $\mathbf{\mathcal{F}}^{(N)}$ can be decomposed as\cite{dao2019butterfly}
	\begin{eqnarray}
	\mathbf{\mathcal{F}}^{(N)} & = & \mathbf{B}^{(N)} \left( \begin{array}{cc} \mathbf{\mathcal{F}}^{(N/2)} & \mathbf{0} \\ \mathbf{0} & \mathbf{\mathcal{F}}^{(N/2)} \end{array} \right) \mathbf{P}^{(N)}, \label{eq:FFT_recursion} \\
	\mathbf{B}^{(N)} & = & \left( \begin{array}{cc} \mathbf{I} & \mathbf{\Omega}^{(N/2)} \\ \mathbf{I} & -\mathbf{\Omega}^{(N/2)} \end{array} \right), \\
	\Omega_{kl}^{(N/2)} & = & \delta_{kl} e^{2\pi i(k-1)/N}, \\
	P_{kl}^{(N)} & = & \left\{ \begin{array}{cc} \delta_{k(2l-1)}, & l \le N/2 \\ \delta_{k(2l-N)}, & l>N/2 \end{array} \right. .
	\end{eqnarray}
	We see that $\mathbf{B}^{(N)}$ is a sparse matrix with three stripes, one along the diagonal $k=l$ and two along the lines $k=l\pm N/2$, while $\mathbf{P}^{(N)}$ is a permutation matrix. Eq.~\eqref{eq:FFT_recursion} relates $\mathbf{\mathcal{F}}^{(N)}$ to the FFT transformation matrix with halved dimensions, $\mathbf{\mathcal{F}}^{(N/2)}$, which immediately suggests that $\mathbf{\mathcal{F}}^{(N/2)}$ can be further expressed using $\mathbf{\mathcal{F}}^{(N/4)}$, etc., until one reaches $\mathbf{\mathcal{F}}^{(1)}$, which equals the scalar 1. This yields the following decomposition
	\begin{eqnarray}
	\mathbf{\mathcal{F}}^{(N)} & = & \mathbf{\mathcal{B}}^{(N)} \mathbf{\mathcal{P}}^{(N)}, \label{eq:BP} \\
	\mathbf{\mathcal{B}}^{(N)} & = & \mathbf{B}^{(N)} \left( \begin{array}{cc} \mathbf{B}^{(N/2)} & \mathbf{0} \\ \mathbf{0} & \mathbf{B}^{(N/2)} \end{array} \right) \nonumber\\
	& & \times \left( \begin{array}{cccc} \mathbf{B}^{(N/4)} & \mathbf{0}  & \mathbf{0}  & \mathbf{0} \\ \mathbf{0} & \mathbf{B}^{(N/4)}  & \mathbf{0} & \mathbf{0} \\ \mathbf{0}  & \mathbf{0} & \mathbf{B}^{(N/4)} & \mathbf{0}   \\ \mathbf{0} & \mathbf{0} & \mathbf{0} & \mathbf{B}^{(N/4)} \end{array} \right) \nonumber\\
	& & \times \cdots, \label{eq:butterfly_B} \\
	\mathbf{\mathcal{P}}^{(N)} & = & \cdots \left( \begin{array}{cccc} \mathbf{P}^{(N/4)} & \mathbf{0}  & \mathbf{0}  & \mathbf{0} \\ \mathbf{0} & \mathbf{P}^{(N/4)}  & \mathbf{0} & \mathbf{0} \\ \mathbf{0}  & \mathbf{0} & \mathbf{P}^{(N/4)} & \mathbf{0}   \\ \mathbf{0} & \mathbf{0} & \mathbf{0} & \mathbf{P}^{(N/4)} \end{array} \right) \nonumber\\
	& & \times \left( \begin{array}{cc} \mathbf{P}^{(N/2)} & \mathbf{0} \\ \mathbf{0} & \mathbf{P}^{(N/2)} \end{array} \right) \mathbf{P}^{(N)}.
	\end{eqnarray}
	Eq.~\eqref{eq:butterfly_B} is called a \emph{butterfly decomposition}\cite{dao2019butterfly,dao2020kaleidoscope,li2015butterfly,li2017butterfly} of $\mathbf{\mathcal{B}}^{(N)}$, and the factors on its right hand side are called the \emph{butterfly factors} of $\mathbf{\mathcal{B}}^{(N)}$; the name comes from the fact that each factor is the direct sum of many $2\times 2$ transformations (or Jacobi rotations), which are typically depicted in a way that resembles butterflies\cite{oppenheim1972butterfly}. Matrices that have butterfly decompositions are called butterfly matrices.
	Since each factor of a butterfly matrix has only $O(N)$ non-zero entries, and there are $O(\log N)$ factors, the product of $\mathbf{\mathcal{F}}^{(N)}$ with a vector can be evaluated in $O(N\log N)$ time. In fact, this is the basis of the famous Cooley-Tukey fast Fourier transform (FFT) algorithm\cite{FFT}.
	It is also useful to call Eq.~\eqref{eq:BP} a BP decomposition\cite{dao2019butterfly}, defined as a butterfly decomposition times a permutation matrix.
	
	The above derivation assumes that the grid is one-dimensional; for $D$-dimensional grids, the Fourier transformation matrix has to be labeled by the number of grid points along each of the Cartesian dimensions, i.e.~$\mathbf{\mathcal{F}}^{(N_1,N_2,\ldots,N_m)}$. If each of the grid point numbers $N_1,N_2,\ldots,N_m$ are powers of two, then it follows that $\mathbf{\mathcal{F}}^{(N_1,N_2,\ldots,N_m)}$ also admits a BP decomposition, since similar to expressing $\mathbf{\mathcal{F}}^{(N)}$ by $\mathbf{\mathcal{F}}^{(N/2)}$, one can express $\mathbf{\mathcal{F}}^{(N_1,N_2,\ldots,N_m)}$ by $\mathbf{\mathcal{F}}^{(N_1/2,N_2,\ldots,N_m)}$, which can in turn be expressed by $\mathbf{\mathcal{F}}^{(N_1/2,N_2/2,\ldots,N_m)}$, etc., until $\mathbf{\mathcal{F}}^{(N_1/2,N_2/2,\ldots,N_m/2)}$; the latter can further be expressed by $\mathbf{\mathcal{F}}^{(N_1/4,N_2/2,\ldots,N_m/2)}$, etc. Thus, again, $\mathbf{\mathcal{F}}^{(N_1,N_2,\ldots,N_m)}$ can be decomposed into a butterfly matrix (with factors having $1, 2, 4, \ldots$ blocks) times a permutation matrix, analogous to $\mathbf{\mathcal{F}}^{(N)}$.
    When $D$ is large, however, more efficient compressions are possible taking advantage of Tucker decompositions\cite{tensor_butterfly}.
	
	Since the inverse of $(\mathbf{\mathcal{F}}^{(N)})$,
	\begin{equation}
	(\mathcal{F}_{kl}^{(N)})^{-1} = \frac{1}{N}e^{2\pi i(k-1)(l-1)/N} = \frac{1}{N}(\mathcal{F}_{kl}^{(N)})^*,
	\end{equation}
	is proportional to the complex conjugate of $(\mathbf{\mathcal{F}}^{(N)})$, it follows that $(\mathcal{F}_{kl}^{(N)})^{-1}$ also adopts a BP decomposition
	\begin{equation}
	(\mathcal{F}_{kl}^{(N)})^{-1} = \frac{1}{N} (\mathbf{\mathcal{B}}^{(N)})^* \mathbf{\mathcal{P}}^{(N)}.
	\end{equation}
	We can thus express the original convolution operator $\mathbf{O}$ as
	\begin{equation}
	\mathbf{O} = \frac{1}{N} (\mathbf{\mathcal{B}}^{(N)})^* \mathbf{\mathcal{P}}^{(N)} \mathbf{\mathcal{O}} \mathbf{\mathcal{B}}^{(N)} \mathbf{\mathcal{P}}^{(N)}.
	\end{equation}
	Finally, when $\mathbf{\mathcal{O}}$ is diagonal, the matrix $\frac{1}{N}\mathbf{\mathcal{O}}$ can be absorbed into $\mathbf{\mathcal{B}}^{(N)}$ without changing the sparsity patterns of the latter's butterfly factors, giving a decomposition of the form
	\begin{equation}
	\mathbf{O} = (\mathbf{\mathcal{B}}^{(N)})^* \mathbf{\mathcal{P}}^{(N)} \tilde{\mathbf{\mathcal{B}}}^{(N)} \mathbf{\mathcal{P}}^{(N)}. \label{eq:BPBP}
	\end{equation}
	Eq.~\eqref{eq:BPBP} is called a BPBP decomposition\cite{dao2019butterfly}. Note that even if $\mathbf{\mathcal{O}}$ is not diagonal, $\mathbf{O}$ may still have (or can be well approximated by) a BPBP decomposition, since one can use arbitrary factors in place of $\mathbf{\mathcal{B}}^{(N)}$/$\tilde{\mathbf{\mathcal{B}}}^{(N)}$ and arbitrary permutation matrices $\mathbf{\mathcal{P}}^{(N)}$ (the two permutation matrices in Eq.~\eqref{eq:BPBP} can also be different if necessary), as long as $\mathbf{\mathcal{B}}^{(N)}$/$\tilde{\mathbf{\mathcal{B}}}^{(N)}$ still admits butterfly decompositions.

\begin{figure}[htbp]
			\resizebox{0.48\textwidth}{!}{\includegraphics{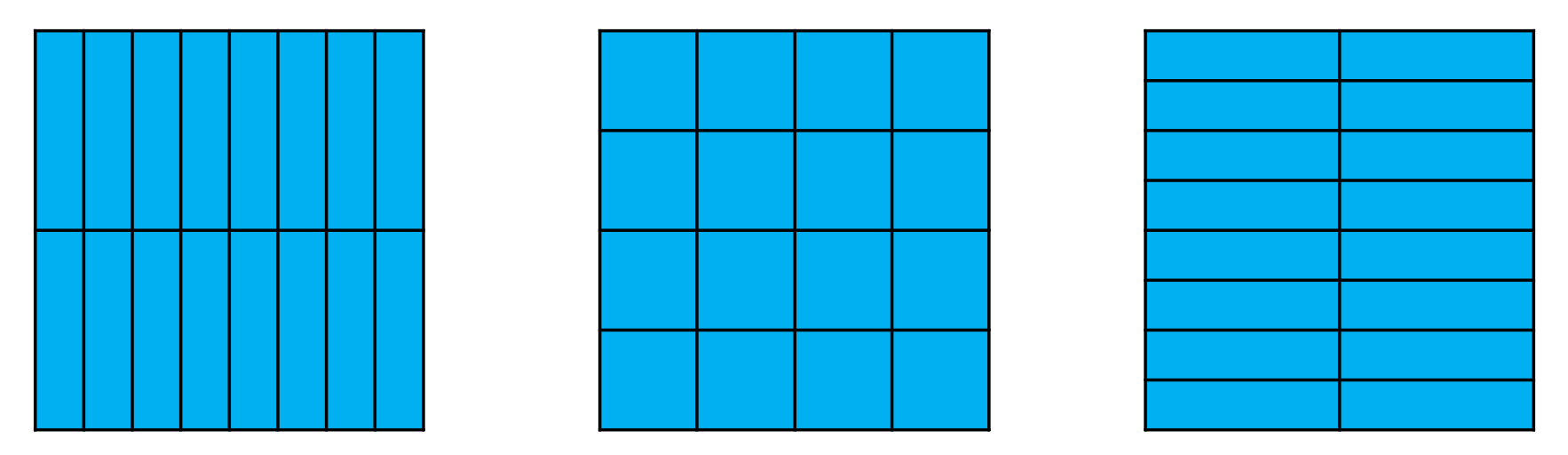}}
		\caption{Three tesselation schemes of a $16\times 16$ complementary low-rank matrix, where all blocks are required to be low-rank.}
		\label{fig:CLR}
	\end{figure}

	However, the above approach does not work for other ODLR matrices, not even those that represent convolutions with non-translationally-invariant kernels. It also fails in general for matrix functions of local matrices (Sec.~\ref{sec:foundations_matfunc}), as seen from the simple fact that the density matrices of metallic systems, despite being ODLR, are generally not local in reciprocal space. To our best knowledge, it is an open question whether general ODLR matrices possess BPBP-like decompositions.
    Nevertheless, a closely related family of matrices, \emph{complementary low-rank matrices}, are known to have butterfly decompositions\cite{li2015butterfly}. As its narrowest definition, complementary low-rank matrices are $2^L\times 2^L$ matrices that, when tessellated into $2^l\times 2^{L-l}$ ($l$ is an integer that is neither close to 1 nor close to $L$) rectangular blocks, all blocks have low numerical rank (Figure~\ref{fig:CLR}). This requirement is more stringent than ODLR in the sense that certain diagonal blocks need to be low-rank as well, but less stringent than ODLR in the sense that the largest low-rank blocks only need to have area $2^L$, instead of $O(4^L)$ as in ODLR matrices. Efficient algorithms for finding the butterfly decompositions of these matrices are known\cite{zheng2023butterfly,zheng2023butterfly2,le2022butterfly,le2025butterfly}.

	\section{Selected applications in computational chemistry}
	
	From this section on, we demonstrate the practical use of the ODLR property to speed up calculations in computational chemistry.
	
	\subsection{ODLR property of the Coulomb interaction} \label{sec:ODLR_coul}
	
	As we have already shown in Sec.~\ref{sec:foundations_realspace}, the Coulomb matrix between point charges, $\mathbf{J}$, has the ODLR property; the underlying physics is that the Coulomb interaction between two groups of point charges can be approximated as the interaction between a handful of multipoles.
	Therefore, the most rudimentary use of the ODLR property of $\mathbf{J}$ consists of dividing the system into molecular fragments, calculating the low-order multipoles of each fragment, and calculate the interactions between spatially distant fragments using the multipoles, instead of the point charges themselves. This is equivalent to expressing $\mathbf{J}$ in the BLR format, where each block represents the interaction between two fragments, and the singular vectors of the low-rank blocks are linear combinations of low-order multipoles. Such approaches are widely used in coarse-graining methods\cite{CG1,CG2,CG3,CG4,CG5}, as well as approximate explicit solvation methods such as the effective fragment potential (EFP) method\cite{EFP}. When the fragment sizes do not increase as a function of the system size (which is usually the case), these approaches do not improve the scaling of the interaction energy calculation (i.e.~the scaling remains $O(N^2)$), but only improves the prefactor.
	
	To also improve the computational scaling, one has to compress $\mathbf{J}$ in more sophisticated ways. The FMM method\cite{FMM1987,FMM1988,FMM1994} is equivalent to expressing $\mathbf{J}$ as an  $\mathscr{H}^2$-matrix.
	In FMM, the real space is divided into a number of non-overlapping ``boxes'', so that each point charge belongs to one and only one box. Then, each ``box'' is recursively divided into $2^D$ smaller boxes (for three-dimensional problems, $2^3=8$ boxes). Subsequently, for each box, a set of multipoles is calculated and placed at the center of the box, such that they approximately reproduce the electrostatic field generated by the point charges within the box. Finally, the total interaction energy of a point charge $A$ with all other charges $B\neq A$ is calculated using a multiresolution approach: the interactions of point charges $B$ that are closest to $A$ (the near-field charges) are calculated exactly, while those point charges that are further and further from $A$ (the far-field charges) are accounted for by replacing them with the multipoles associated with larger and larger boxes. By this way, the electrostatic fields of far-field charges are calculated using coarse-grained representations of the point charges, while those of near-field charges are calculated using finer representations or even treated exactly, yielding an optimal cost-accuracy balance.
	
	Since large boxes are used only when the box's distance to $A$ is at least on the same order as the size of the box itself, the aforementioned algorithm only compresses admissible blocks of $\mathbf{J}$, converting $\mathbf{J}$ into an $\mathscr{H}$-matrix instead of an HODLR matrix. As we have discussed before, however, $\mathscr{H}$-matrix representations only lead to $O(N \log N)$ scalings but not $O(N)$ scalings. To get rid of the logarithmic factor, one must express the singular vectors of large blocks as linear combinations of those of smaller blocks. In FMM, this amounts to calculating the multipoles of large boxes from the multipoles of its constituent small boxes, instead of directly from the point charges themselves, which indeed renders FMM linear scaling, as is well known.
	
	We then discuss the generalization of FMM to the ERI tensor, which is commonly termed continuous FMM (CFMM)\cite{CFMM1,CFMM2,CFMM3,CFMM4}. First of all, an AO ERI $(\mu\nu|\kappa\lambda)$ is negligible if either $\mu$ is distant from $\nu$, or $\kappa$ is distant from $\lambda$. Therefore, we only need to consider the Coulomb matrix between ``significant AO pairs'' $\mu\nu$ and $\kappa\lambda$, i.e.~AO pairs whose overlaps are not negligible; as discussed before, this alone leads to an $O(N^2)$ scaling since there are only $O(N)$ significant AO pairs. Secondly, the overlap densities $\mu\nu$ and $\kappa\lambda$ are continuous density distributions instead of point charges; though they are rather local in the sense that they decay quickly in real space, they do not decay to exactly zero at any finite radius. Thus, the CFMM treatment of the ERI tensor starts from a truncation of the tails of the overlap densities, converting them to charge distributions with finite extents. Only then can the admissibility criterion be used to give convergent multipolar expansions. Without such truncation, the multipolar expansions of either $\mu\nu$ or $\kappa\lambda$ around any point in real space are divergent.
	This however raises a practical issue: $(\mu\nu|\kappa\lambda)$ can only be evaluated with CFMM when the overlap between $\mu\nu$ and $\kappa\lambda$ is comparable to or smaller than the required energy accuracy (typically in the nanohartree range, as the energy error of CFMM is not a smooth function of the nuclear coordinates, so that nanohartree accuracy is required for the potential energy surface to be practically smooth). The ERIs between any closer AO pairs (the near-field ERIs) have to be evaluated exactly. Even though there are only $O(N)$ near-field ERIs, they introduce a large prefactor to the algorithm, and a late crossover with conventional prescreening-based algorithms (for example, the Gaussian program only switches on FMM when the number of atoms is greater than 60, while even larger thresholds are used for highly symmetric molecules).
	
	To remedy this, we recall that the reason why the multipolar expansion requires non-overlapping charge densities to converge, is because it involves a Taylor expansion of the charge densities (Eq.~\eqref{eq:multipole}), which diverges whenever $|\mathbf{r}'-\mathbf{r}_0| > |\mathbf{r}-\mathbf{r}_0|$ for \emph{any} pair of points $\{\mathbf{r}\in X,\mathbf{r}'\in Y\}$ that belong respectively to the two density distributions. The Taylor expansion formulation of the ODLR property of $\mathbf{J}$ was used in CFMM, because it directly gives the singular vectors of far-field off-diagonal blocks of $\mathbf{J}$ (e.g.~$\mathbf{J}^{XY}$), without first having to calculate the elements of $\mathbf{J}$. However, since it is already known that $\mathbf{J}^{XY}$ has low numerical rank $r$, one can always estimate the whole matrix by sampling $O(r\min(|X|,|Y|))$ elements of $\mathbf{J}^{XY}$, regardless of whether the charge densities in $X$ and $Y$ overlap with each other or not, even though the resultant singular vectors of $\mathbf{J}^{XY}$ are no longer necessarily related to charge multipoles.
	Xing, Huang and Chow\cite{xing2020JCP,xing2020SIAM,H2Pack} incorporated the proxy point method into this algorithm to further reduce the cost, giving the $\mathscr{H}^2$-ERI method, which expresses $\mathbf{J}$ as an $\mathscr{H}^2$-matrix, but does not require that the densities $\mu\nu$ and $\kappa\lambda$ do not overlap. In the $\mathscr{H}^2$-ERI method, the singular vectors of $\mathbf{J}^{XY}$ are estimated by nuclear repulsion integrals $(\mu\nu|A)$, where $\mu\nu \in X$, and $A\in Y$ are ``proxy points'' that span the whole region $Y$, where the overlap density $\kappa\lambda$ is mostly distributed. Since it does not rely on Taylor expansions of the electronic density, $\mathscr{H}^2$-ERI significantly reduces the number of near-field ERIs compared to CFMM. Such reduction has yielded speedups and memory reductions of 5-18 folds\cite{xing2020SIAM} compared to FMM, with errors comparable to CFMM.
	From this example, we can see that although CFMM uses the ODLR property of the ERI tensor, it does not exploit the property to the greatest extent possible.
	
	For periodic systems using the plane wave basis, another way of efficiently evaluating the Coulomb contribution to the Fock matrix is via FFT. As the Coulomb kernel is diagonal in the reciprocal space:
	\begin{equation}
	J_{\mathbf{k}\mathbf{k}'} = \frac{4\pi}{|\mathbf{k}|^2}\delta(\mathbf{k}-\mathbf{k}'),
	\end{equation}
	the application of the Coulomb kernel becomes linear scaling in the reciprocal space. The whole Coulomb building process is thus dominated by the FFT and inverse FFT before and after applying the Coulomb kernel, respectively, and thus scales as $O(N\log N)$. As discussed in Section~\ref{sec:butterfly}, the $O(N\log N)$ scaling is due to the availability of butterfly decomposition of the Fourier transform and inverse Fourier transform matrices, which is in turn closely related to the ODLR property of convolution operators.
	However, precisely because of the use of the butterfly decomposition, Coulomb algorithms that rely on Cooley-Tukey FFT cannot achieve scalings better than $O(N\log N)$, unlike CFMM.
	
	We note that alternative linear or near-linear scaling Coulomb algorithms exist. First, instead of calculating the electric multipoles in CFMM boxes, the multiresolution analysis (MRA) method\cite{MADNESS,MADNESS_J,beylkin1991fast} uses Daubechies wavelets to expand the densities, where the wavelets live in CFMM boxes of varying sizes. The resulting coefficients under the wavelet basis however play a similar role as the multipoles in CFMM: the Coulomb interaction between two highly oscillatory wavelets decays quickly with respect to the distance between them, just like the interaction between high-order multipoles decays quickly with their distance. As a result, MRA achieves linear scaling Coulomb build, but one drawback of MRA relative to CFMM is that the same wavelets have to be used as the orbital basis, while any localized orbital basis can be used in CFMM.
    Another example is the bubblepole (BUPO) algorithm developed by the ORCA team\cite{BUPO} for the linear scaling contraction of three-center RI integrals $(\mu\nu|P)$ with density matrices or RI auxiliary densities, which uses ``bubbles'' of arbitrary shapes, in place of the boxes in CFMM; meanwhile, a large ``bubble'' can contain a varying number of lower-level bubbles, in contrast to FMM where a large box always contains 8 small boxes. Regardless, the BUPO method achieves linear scaling also owing to the multipolar expansion of the Coulomb operator, which, as we have repeatedly shown, is related to the ODLR property of the Coulomb matrix. The particle mesh Ewald (PME) method\cite{PME} is an alternative to FFT for periodic systems, which uses direct summation for near-field charges, and FFT for far-field charges. As the near-field contribution scales as $O(N)$, while the far-field contribution scales as $O(N\log N)$, the overall scaling of PME is dominated by the FFT part.
	At this point, reach the interesting conclusion that \emph{all} of the above linear or near-linear scaling Coulomb algorithms can be interpreted as exploiting the ODLR property of the Coulomb matrix. We conjecture that this is true for all the other linear or near-linear scaling Coulomb algorithms, including those that will be developed in the future.
	
	\subsection{ODLR property of density and Fock matrices} \label{sec:ODLR_DF}
	
	As mentioned in Section~\ref{sec:introduction}, the Fock matrices $\mathbf{F}$ of pure, semilocal functionals (or functionals with only short-range exact exchange, e.g.~HSE03\cite{HSE03a,HSE03b,HSE03c,HSE03d} and HSE06\cite{HSE06}) are local, and therefore trivially possess the ODLR property. Owing to their locality, the zero-temperature density matrices $\mathbf{D}$ of such functionals are also ODLR, since they are step functions of $\mathbf{D}$ (i.e.~$\mathbf{D} = \Theta(\mathbf{F}-\mu)$), as discussed in Section~\ref{sec:foundations_matfunc}. In this subsection, we will first discuss the use of $\mathbf{D}$'s ODLR property in the efficient calculation of the density matrix using pure functionals. Then, we prove that the Fock matrices of hybrid functionals are ODLR. We finally discuss the ODLR property of density matrices from hybrid functionals.
	
	To separate $\mathbf{D}$ into a local component plus a low-rank component, we need to replace $\Theta(\epsilon-\mu)$ by a similar, but smooth, function $f(\epsilon)$. A natural choice is the Fermi-Dirac function
	\begin{equation}
	f(\epsilon; \beta,\mu) = \frac{1}{1 + e^{\beta(\epsilon-\mu)}},
	\end{equation}
	for some inverse temperature $0 < \beta < +\infty$, and a chemical potential $\mu$ that gives the correct integrated number of electrons. Then, $\mathbf{D}$ can be decomposed as
	\begin{eqnarray}
	\mathbf{D} & = & \mathbf{D}(\beta) + \Delta\mathbf{D}, \label{eq:D_decomp} \\
	\mathbf{D}(\beta) & = & f(\mathbf{F}; \beta,\mu), \\
	\Delta\mathbf{D} & = & \mathbf{D}-\mathbf{D}(\beta).
	\end{eqnarray}
	The first term on the right hand side of Eq.~\eqref{eq:D_decomp} is nothing but the density matrix at inverse temperature $\beta$, and is therefore local irrespective of whether the system has a gap, as is well-known\cite{kohn1996,goedecker1998}; however it should be stressed that any other smooth approximation of the step function will equally suffice (see the discussion in Section~\ref{sec:foundations_matfunc}), and the only benefit of the Fermi-Dirac function is that the resulting $\mathbf{D}(\beta)$ has a physical meaning.
	The fact that $\Delta\mathbf{D}$ has low rank has a simple physical interpretation: only the occupation numbers of the frontier orbitals will change significantly upon changing the electronic temperature from a finite value to zero, such that only frontier orbitals would contribute significantly to $\Delta\mathbf{D}$, and the contribution therefore has low rank because each orbital makes a rank-1 contribution.
    As a corollary, the singular vectors of $\Delta\mathbf{D}$, and therefore those of the far off-diagonal blocks of $\mathbf{D}$, should resemble the frontier orbitals of the whole system.
	
	The above considerations immediately suggest a low-scaling algorithm, the filter diagonalization method\cite{FD1,FD2,FD3,FD4,FD5}, for finding the density matrix of a local Fock matrix $\mathbf{F}$ (in the rest of this subsection, for simplicity, we will assume the basis set is orthonormal; extension to non-orthogonal AO basis sets is straightforward). One first calculates the density matrix at a finite inverse temperature $\beta$, which is linear scaling since the Fermi-Dirac function can be well-approximated by a Chebyshev polynomial of fixed degree, and it takes only $O(N)$ time to evaluate an $O(1)$-degree polynomial of $\mathbf{F}$.
	Then, one can construct a pseudo-projector
	\begin{equation}
	\mathbf{G} = \frac{\partial}{\partial \mu} f(\mathbf{F}; \beta,\mu)
	\end{equation}
	in linear time, also via its Chebyshev polynomial approximation. $\mathbf{G}$ is numerically highly rank-deficient, since only the $M$ frontier orbitals closest to the chemical potential $\mu$ make non-zero contributions to it (where $M$ is a small integer). To find the space spanned by these $M$ orbitals, one can calculate $M$ columns of $\mathbf{G}$, followed by orthonormalization to yield a projector $\bar{\mathbf{G}}$ onto that space. One can then project $\mathbf{F}$ onto this reduced space:
	\begin{equation}
	\mathbf{F}^{(1)} = \bar{\mathbf{G}}^T\mathbf{F}\bar{\mathbf{G}},
	\end{equation}
	and diagonalize $\mathbf{F}^{(1)}$ to yield the canonical orbitals $\mathbf{C}^{(1)}$ within this reduced space. Finally, $\Delta\mathbf{D}$ is assembled by
	\begin{equation}
	\Delta\mathbf{D} = \bar{\mathbf{G}}^T \left(\sum_p (\mathbf{C}^{(1)}_p)^T (\Theta(\epsilon_p-\mu) - f(\epsilon_p; \beta,\mu)) \mathbf{C}^{(1)}_p \right) \bar{\mathbf{G}},
	\end{equation}
	where $\epsilon_p$ is the orbital energy of canonical orbital $p$, and $\mathbf{C}^{(1)}_p$ is the vector of the coefficients of orbital $p$ in the reduced space.
	
	However, the above algorithm cannot achieve near-linear scaling: if $\beta$ is independent of system size, $\mathbf{F}^{(1)}$ will have dimensions $O(N)\times O(N)$, just like $\mathbf{F}$ itself. On the other hand, if $\beta$ increases with the system size in such a way that $\mathbf{F}^{(1)}$ has dimensions $O(1)\times O(1)$, the calculation of $\mathbf{D}(\beta)$ will take more than $O(1)$ iterations, again leading to superlinear scaling. This is similar to the reason why the BLR format cannot achieve near-linear scaling.
	In the energy renormalization group (ERG) method\cite{ERG,ERG2}, Roi and Baer chose $\beta=\beta_0$ as independent of the system size, but used filter diagonalization to diagonalize $\mathbf{F}^{(1)}$, where the inverse temperature $\beta_1 = q\beta (q>1)$ used in this new filter diagonalization calculation is chosen to be higher than $\beta$.
	This leads to a series of effective Fock matrices with smaller and smaller dimensions, where the filter diagonalization of $\mathbf{F}^{(n)}$ at inverse temperature $\beta_n = q^n \beta$ yields the reduced space Fock matrix $\mathbf{F}^{(n+1)}$. Such a procedure eventually gives $\mathbf{D}$ as a telescopic series:
	\begin{eqnarray}
	\mathbf{D} & = & \mathbf{D}(\beta) + \sum_{i=1}^{N_{\rm max}} \mathbf{D}^{(i)} + (\mathbf{D} - \mathbf{D}(\beta_{N_{\rm max}})), \label{eq:telescopic} \\
	\mathbf{D}^{(i)} & = & \mathbf{D}(\beta_i) - \mathbf{D}(\beta_{i-1}).
	\end{eqnarray}
    Here the number of terms in the summation $N_{\rm max}$ is chosen such that the last term of Eq.~\eqref{eq:telescopic} has rank $O(1)$, allowing the final small-dimensional Fock matrix to be exactly diagonalized in $O(1)$ time.
	As $\mathbf{D}^{(i)}$ has progressively lower ranks as $i$ increases, this series is an MLR representation of $\mathbf{D}$.
	The ERG method can achieve an $O(N \log^2 N)$ scaling, where the logarithmic factor can be understood as due to using the MLR format (see Section~\ref{sec:MLR}). Unfortunately, the prefactor of the ERG method is large: in tight-binding models of one-dimensional metals, a crossover with conventional Fock diagonalization is only seen after the system size reaches 4000 sites\cite{ERG}. As a result, the ERG method has, to our best knowledge, never been implemented in any electronic structure program. It is unclear whether expressing $\mathbf{D}$ using, e.g.~the $\mathscr{H}^2$-matrix format, can lead to a lower scaling or reduce the prefactor.
	
	Although this has not been demonstrated yet, the ERG method can in principle also be applied to dense, ODLR Fock matrices (provided that the Fock matrix has already been compressed as an HODLR, $\mathscr{H}$, HSS or $\mathscr{H}^2$-matrix), while retaining linear or near-linear scaling. This is because the evaluation of $O(1)$-order Chebyshev polynomials of these matrices is linear/near-linear scaling, just like that of local matrices, thanks to the linear/near-linear scaling algorithms for multiplying ODLR matrices (Section~\ref{sec:ODLR_compress}). In this case, the series Eq.~\eqref{eq:telescopic} is a sum of ODLR matrices with lower and lower ranks. Therefore, the resulting density matrix $\mathbf{D}$ is also an ODLR matrix.
	
	
	Now, we will show that the exact exchange Fock matrix $\mathbf{K}=\mathbf{K}[\mathbf{D}]$ from ODLR density matrices $\mathbf{D}$ (dropping the spin index for convenience):
	\begin{equation}
	K_{\mu\nu} = \sum_{\kappa\lambda} (\mu\kappa|\nu\lambda) D_{\kappa\lambda}, \label{eq:K}
	\end{equation}
	though being dense, is also ODLR. Consider a far off-diagonal block, $\mathbf{K}^{XY}$. As $(\mu\kappa|\nu\lambda)$ is only non-negligible when $\mu$ and $\kappa$ are spatially close, and $\nu$ and $\lambda$ are spatially close, the summation in Eq.~\eqref{eq:K} can be restricted to $\kappa\in X'$ and $\lambda\in Y'$, where the regions $X'\supset X$ and $Y'\supset Y$ are only slightly larger than $X$ and $Y$, respectively. As $\mathbf{D}^{X'Y'}$ has low rank:
	\begin{equation}
	\mathbf{D}^{X'Y'} = \mathbf{U}^{X'Y'}_{\rm D}\mathbf{\sigma}_{\rm D}^{X'Y'}(\mathbf{V}^{X'Y'}_{\rm D})^T,
	\end{equation}
	and the $XX',YY'$ block of the ERI tensor $I_{\mu\kappa,\nu\lambda}=(\mu\kappa|\nu\lambda)$ also has low rank (here $\mu\kappa$ and $\nu\lambda$ are understood as compound indices, so that the tensor can be shaped into a matrix):
	\begin{equation}
	\mathbf{I}^{XX',YY'} = \mathbf{U}^{XX',YY'}_{\rm I}\mathbf{\sigma}^{XX',YY'}_{\rm I}(\mathbf{V}^{XX',YY'}_{\rm I})^T,
	\end{equation}
	one can build $\mathbf{K}^{XY}$ as
	\begin{eqnarray}
	\mathbf{K}^{XY} & = & \mathbf{U}^{XY}_{\rm K} \mathbf{\sigma}_{\rm K}^{XY} (\mathbf{V}^{XY}_{\rm K})^T, \label{eq:K_SVD} \\
	(\mathbf{U}^{XY}_{\rm K})_{\mu,PQ} & = & \sum_{\kappa} (\mathbf{U}^{XX',YY'}_{\rm I})_{\mu\kappa,P} (\mathbf{U}^{X'Y'}_{\rm D})_{\kappa Q}, \label{eq:K_SVD_U} \\
	(\mathbf{V}^{XY}_{\rm K})_{\nu,PQ} & = & \sum_{\lambda} (\mathbf{V}^{XX',YY'}_{\rm I})_{\nu\lambda,P} (\mathbf{V}^{X'Y'}_{\rm D})_{\lambda Q},\label{eq:K_SVD_V} \\
	(\mathbf{\sigma}_{\rm K}^{XY})_{PQ,RS} & = & \delta_{PR}\delta_{QS} (\mathbf{\sigma}^{XX',YY'}_{\rm I})_{PP} (\mathbf{\sigma}_{\rm D}^{X'Y'})_{QQ}, \label{eq:K_SVD_sigma}
	\end{eqnarray}
	where the indices $P$ and $Q$ run over the singular vectors of $\mathbf{I}^{XX',YY'}$ and $\mathbf{D}^{X'Y'}$, respectively. Eqs.~\eqref{eq:K_SVD}-\eqref{eq:K_SVD_sigma} show that the numerical rank of $\mathbf{K}^{XY}$ is no more than the product of the numerical ranks of $\mathbf{I}^{XX',YY'}$ and $\mathbf{D}^{X'Y'}$, and is therefore $O(1)$, finishing our proof that $\mathbf{K}$ is ODLR.
    The singular vectors of $\mathbf{K}^{XY}$, Eqs.~\eqref{eq:K_SVD_U}-\eqref{eq:K_SVD_V}, are given by the contractions of the singular vectors of $\mathbf{I}^{XX',YY'}$ (which are related to three-center RI integrals, Eq.~\eqref{eq:RI}), and the singular vectors of $\mathbf{D}^{X'Y'}$ (which are approximately equal to the frontier orbitals).
    The complete Fock matrix is therefore also ODLR, since all other components (one-electron Hamiltonian, Coulomb potential, XC potential) of the Fock matrix are local.
	
	One final question is whether the density matrices of hybrid functional calculations are also ODLR. One can show that this is true if the initial guess density matrix is ODLR, and the SCF calculation converges in $O(1)$ iterations. This is because (1) the Fock matrix built from an ODLR density matrix is an ODLR matrix, and (2) the density matrix obtained from diagonalizing an ODLR Fock matrix is an ODLR matrix. However, the SCF convergence of gapless systems is often difficult, and the number of iterations may increase with the system size\cite{iOI}, in which case the off-diagonal blocks of the Fock and density matrices may in principle have higher and higher ranks as the SCF iterations proceed, and eventually become (almost) full rank. Thus, despite favorable numerical evidence (see, e.g.~Fig.~\ref{fig:C32}(e)), it is still unknown whether the density matrices of gapless systems always possess the ODLR property if hybrid functionals are used.
	
	\subsection{ODLR property of LMO coefficient matrices}  \label{sec:ODLR_LMO}
	
	LMOs can be generated by a variety of approaches, for example the Foster-Boys\cite{FBLoc1,FBLoc2}, Pipek-Mezey\cite{PMLoc} and Edminston-Ruedenberg\cite{ERLoc1,ERLoc2} methods, and the resulting orbitals may have different extents of locality or other mathematical structures.
	In principle, even if a system has local LMOs, or LMOs whose coefficient matrices are ODLR, a specific localization algorithm may still fail to find it, and instead give much more delocalized or unstructured LMOs.
	It is therefore hard to show whether the LMO coefficient matrices from all available localization approaches are ODLR matrices, especially for gapless systems.
    In particular, the Foster-Boys, Pipek-Mezey and Edminston-Ruedenberg methods all rely on the extremization of some localization criterion, which are non-convex optimization problems; it is a formidable task to decide if all minima/maxima of the localization criterion give ODLR LMO coefficient matrices.
    Rather, in this section we will only show that a few selected localization algorithms give ODLR LMO coefficient matrices.
	
	When reasonable, orthonormal guesses of the occupied and virtual LMOs, $\mathbf{C}_0^{\rm occ}$ and $\mathbf{C}_0^{\rm vir}$, are known, a set of exact LMOs $\mathbf{C}^{\rm occ}$ and $\mathbf{C}^{\rm vir}$ (in the sense that they span the exact occupied and virtual subspaces) can be found by projecting the LMOs onto the occupied and virtual spaces, respectively, followed by orthonormalization\cite{ACR-FLMO}:
	\begin{eqnarray}
	\mathbf{C}^{\rm occ} & = & \mathbf{D} \mathbf{S} \mathbf{C}_0^{\rm occ} (\mathbf{S}^{\rm occ})^{-1/2}, \label{eq:DSC1} \\
	\mathbf{S}^{\rm occ} & = & (\mathbf{C}_0^{\rm occ})^T \mathbf{S} \mathbf{D} \mathbf{S} \mathbf{D} \mathbf{S} \mathbf{C}_0^{\rm occ}, \\
	\mathbf{C}^{\rm vir} & = & (\mathbf{I}-\mathbf{D}\mathbf{S}) \mathbf{C}_0^{\rm vir} (\mathbf{S}^{\rm vir})^{-1/2}, \label{eq:DSC2} \\
	\mathbf{S}^{\rm vir} & = & (\mathbf{C}_0^{\rm vir})^T (\mathbf{I}-\mathbf{S}\mathbf{D}) \mathbf{S} (\mathbf{I}-\mathbf{D}\mathbf{S}) \mathbf{C}_0^{\rm vir}.
	\end{eqnarray}
	Here $\mathbf{D}\mathbf{S}$ and $\mathbf{I}-\mathbf{D}\mathbf{S}$ project MOs onto the occupied and virtual spaces, respectively.
	The initial guess LMOs are often obtained from fragment SCF calculations, followed by localization of the fragments' CMOs using another method (e.g.~the Foster-Boys or Pipek-Mezey localization methods). This gives LMOs $\tilde{\mathbf{C}}_0^{\rm occ}$ and $\tilde{\mathbf{C}}_0^{\rm vir}$ which are orthogonal within each fragment, but not between fragments.
	Therefore, they are symmetrically orthogonalized:
	\begin{eqnarray}
	\mathbf{C}_0^{\rm occ} & = & \tilde{\mathbf{C}}_0^{\rm occ}((\tilde{\mathbf{C}}_0^{\rm occ})^T\mathbf{S}\tilde{\mathbf{C}}_0^{\rm occ})^{-1/2}, \label{eq:pFLMO_orth1} \\
	\mathbf{C}_0^{\rm vir} & = & \tilde{\mathbf{C}}_0^{\rm vir}((\tilde{\mathbf{C}}_0^{\rm vir})^T\mathbf{S}\tilde{\mathbf{C}}_0^{\rm vir})^{-1/2}. \label{eq:pFLMO_orth2}
	\end{eqnarray}
	The above approach is called the ``top-down, least-change'' fragment LMO (FLMO) localization method\cite{FLMO,ACR-FLMO}.
	Note that here we have assumed that the fragments are non-overlapping. For the treatment of overlapping fragments (as would be necessary when the fragmentation method splits covalent bonds), please refer to the original FLMO references\cite{FLMO,ACR-FLMO}.
	The FLMO method has been extended to a hierarchical approach where the fragments are further fragmented to give even smaller fragments, which has been termed the iterative orbital interaction (iOI) method\cite{iOI}.
	
\begin{figure}[htbp]
			\resizebox{0.48\textwidth}{!}{\includegraphics{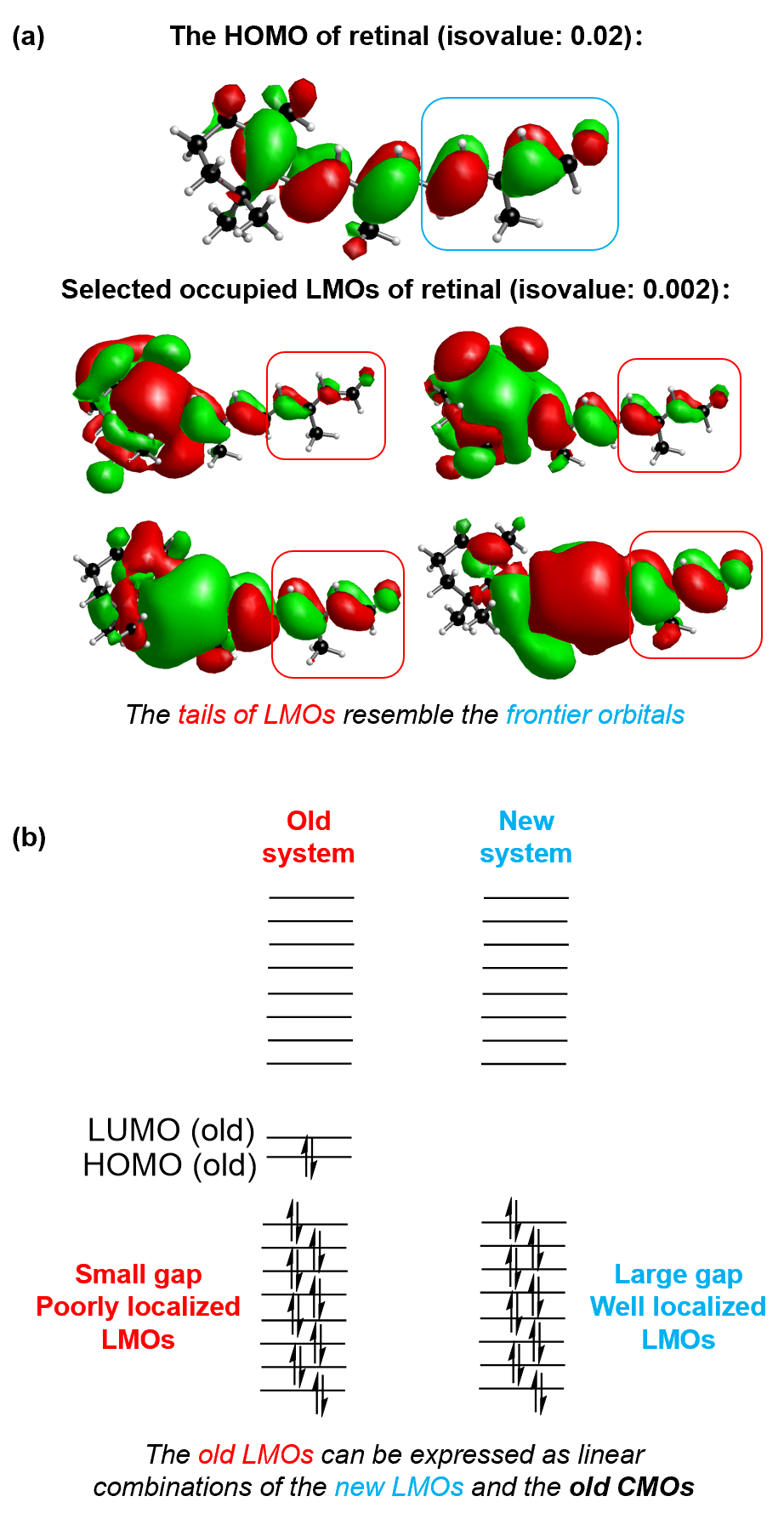}}
		\caption{(a) The HOMO and selected LMOs of retinal (a system with a small but non-zero gap, 3.04 eV) at the B3LYP-D3/def2-SV(P)\cite{DEF2SVP} level of theory, highlighting the tails of the LMOs. The geometry was optimized at the same level. (b) Heuristic explanation of the ODLR property of LMO coefficient matrices.}
		\label{fig:LMOtail}
	\end{figure}

	We are now ready to show that $\mathbf{C}^{\rm occ}$ and $\mathbf{C}^{\rm vir}$ are ODLR.
	The orthogonalization Eqs.~\eqref{eq:pFLMO_orth1}-\eqref{eq:pFLMO_orth2} deteriorates the locality of $\tilde{\mathbf{C}}_0^{\rm occ}$ and $\tilde{\mathbf{C}}_0^{\rm vir}$, as $\mathbf{C}_0^{\rm occ}$ and $\mathbf{C}_0^{\rm vir}$ are no longer strictly localized on one fragment. Nevertheless, they are still exponentially localized since the LMO-basis overlap matrices $(\tilde{\mathbf{C}}_0^{\rm occ})^T\mathbf{S}\tilde{\mathbf{C}}_0^{\rm occ}$ and $(\tilde{\mathbf{C}}_0^{\rm vir})^T\mathbf{S}\tilde{\mathbf{C}}_0^{\rm vir}$ are local, which (combined with the fact that they are well-conditioned, as $\tilde{\mathbf{C}}_0^{\rm occ}$ and $\tilde{\mathbf{C}}_0^{\rm vir}$ are not too far from being orthonormal) means that their inverse square roots are local.
	On the other hand, $\mathbf{S}^{\rm occ}$ and $\mathbf{S}^{\rm vir}$ are ODLR since they are products of ODLR matrices and local matrices, and therefore their inverse square roots are also ODLR.
	Therefore, $\mathbf{C}^{\rm occ}$ and $\mathbf{C}^{\rm vir}$ are ODLR since they are the products of local matrices and ODLR matrices.
    Moreover, the singular vectors of the far off-diagonal blocks of $\mathbf{C}^{\rm occ}$ and $\mathbf{C}^{\rm vir}$ inherit those of $\mathbf{D}$, i.e.~the tails of LMOs resemble the frontier orbitals of the same system. This effect does not only occur in gapless systems, but can be seen in gapped systems as well (Figure~\ref{fig:LMOtail}(a)).

    There is another more intuitive, though not rigorous way to see why LMO coefficient matrices are ODLR. Consider a system with a small HOMO-LUMO gap, but a much larger HOMO-1-LUMO+1 gap (Figure~\ref{fig:LMOtail}(b)). The occupied and virtual spaces cannot be well-localized due to the small gap\cite{kohn1959}. However, if the HOMO and LUMO are removed, the remaining orbitals can be much better localized to give compact LMOs, since the new system has a much larger gap. As the old occupied (virtual) LMOs can be expressed as linear combinations of the new occupied (virtual) LMOs and the HOMO (LUMO) of the old system, it follows that the poor locality of the old occupied (virtual) LMOs must be due to admixtures of the old HOMO (LUMO), i.e.~the tails of occupied (virtual) LMOs resemble the HOMO (LUMO). An off-diagonal block consisting of the tails of many occupied or virtual LMOs therefore has numerical rank one. This argument is not rigorous because the relation between localizability and gap only applies to local Hamiltonians\cite{kohn1959}, and the Hamiltonian with HOMO and LUMO projected out is no longer local. Nevertheless, it provides a more physically intuitive picture than the proof given earlier.
	
	The ``top down, least change'' method requires diagonalization of the Fock matrix, in order to assemble the density matrix $\mathbf{D}$. While the diagonalization can of course be circumvented by e.g.~the ERG method mentioned in the previous subsection, a more efficient method is the ``bottom-up, least-change'' method\cite{FLMO,ACR-FLMO,SMH}, which generates the LMOs of the current SCF iteration from those of the previous SCF iteration, without recourse to the density matrix.
	The Fock matrix $\mathbf{F}$ is first transformed to the LMO basis, using the LMO coefficients of the previous SCF iteration, $\mathbf{C}^{\rm old}$:
	\begin{equation}
	\mathbf{F}^{\rm LMO} \equiv \left( \begin{array}{cc} \mathbf{F}^{\rm oo} & \mathbf{F}^{\rm ov} \\ \mathbf{F}^{\rm vo} & \mathbf{F}^{\rm vv} \end{array} \right) = (\mathbf{C}^{\rm old})^T \mathbf{F} \mathbf{C}^{\rm old}.
	\end{equation}
	To obtain the correct occupied and virtual subspaces, one only needs to apply a unitary transformation $\mathbf{U}$ that zeros out the occupied-virtual and virtual-occupied blocks, $\mathbf{F}^{\rm ov}$ and $\mathbf{F}^{\rm vo}$, resulting in a block-diagonalization of $\mathbf{F}^{\rm LMO}$ as opposed to a full diagonalization.
	The resulting LMOs are identical to the ``top-down, least-change'' LMOs\cite{ACR-FLMO}.
	
	Unlike the ODLR property of density matrices, the ODLR property of LMO coefficient matrices has apparently never been used in electronic structure calculations, in either solving the LMOs (i.e.~orbital localization) or using the LMOs (in, e.g.~post-SCF methods and TDDFT).
	One possible usage in solving the LMOs is to recast the ``bottom-up, least-change'' method into the solution of a Riccatti equation\cite{ACR-FLMO}:
	\begin{equation}
	\mathbf{F}^{\rm vo} + \mathbf{F}^{\rm vv}\mathbf{X} - \mathbf{X}\mathbf{F}^{\rm oo} + \mathbf{X}\mathbf{F}^{\rm oo}\mathbf{X} = \mathbf{0},
	\end{equation}
	such that $\mathbf{U}$ can be assembled by
	\begin{equation}
	\mathbf{U} = \left( \begin{array}{cc} \mathbf{I} & -\mathbf{X}^T \\ \mathbf{X} & \mathbf{I} \end{array} \right) \left( \begin{array}{cc} (\mathbf{I}+\mathbf{X}^T\mathbf{X})^{-1/2} & \mathbf{0} \\ \mathbf{0} & (\mathbf{I}+\mathbf{X}\mathbf{X}^T)^{-1/2} \end{array} \right).
	\end{equation}
	As linear-scaling solution algorithms of the Riccatti equation for ODLR matrices $\mathbf{F}^{\rm LMO}$ are known\cite{borm2003Hmatrix}, this provides a possible way of obtaining LMOs of gapless systems with linear cost.
    Another possibility is to express $\mathbf{U}$ as a butterfly/BP/BPBP decomposition (or other similar decompositions), taking advantage of the fact that $\mathbf{U}$ is often expressed as the product of many Jacobi rotations\cite{FLMO,ACR-FLMO,iOI,SMH}, just like butterfly matrices.
	
	\subsection{ODLR property of Hessians} \label{sec:ODLR_Hessian}
	Finally, we discuss the ODLR property of nuclear Hessians. While one can certainly construct artificial, approximate potential energy surfaces whose Hessians are not ODLR, we will show that the exact Hessian of any molecular system is ODLR\cite{O1NumHess_note}. Thus, in some sense, any ``sufficiently good'' approximate potential energy surface will give ODLR Hessians, although it is beyond the scope of the present manuscript to make the wording ``sufficiently good'' precise.
	
	The electronic energy $E_I$ of the $I$-th state of a system is given by the Schr\"o{}dinger equation:
	\begin{equation}
	\hat{H} |\Psi_I\rangle = E_I |\Psi_I\rangle.
	\end{equation}
    Here the full electronic Hamiltonian $\hat{H}$ should not be confused with the Hessian $\mathbf{H}$ (where the indices $A,B$ and $i,j$ denote atoms and electrons, respectively, $Z_A$ and $Z_B$ are nuclear charges, and $R$/$r$ represent interparticle distances):
    \begin{equation}
	\hat{H} = \sum_{A<B} \frac{Z_A Z_B}{R_{AB}} -\frac{1}{2}\sum_i \nabla_i^2 - \sum_{iA} \frac{Z_A}{r_{iA}} + \sum_{i<j} \frac{1}{r_{ij}}. \label{eq:hatH}
	\end{equation}
	Suppose that the wavefunction $|\Psi_I\rangle$ does not explicitly depend on the nuclear coordinates, which is true in the basis set limit. Then the Hellmann-Feynman theorem holds:
	\begin{equation}
	\frac{\partial E_I}{\partial \xi_j} = \langle\Psi_I| \frac{\partial \hat{H}}{\partial \xi_j} |\Psi_I\rangle, \label{eq:HFey}
	\end{equation}
	where $\xi_j$ is the $j$-th nuclear coordinate.
	Differentiation of Eq.~\eqref{eq:HFey} with respect to another nuclear coordinate $\xi_i$ yields
	\begin{eqnarray}
	& & H_{ij} \equiv \frac{\partial^2 E_I}{\partial \xi_i \partial \xi_j} \nonumber\\
	& = & \langle\Psi_I| \frac{\partial^2 \hat{H}}{\partial \xi_i \partial \xi_j} |\Psi_I\rangle + \left(\langle\Psi_I| \frac{\partial \hat{H}}{\partial \xi_j} |\frac{\partial \Psi_I}{\partial \xi_i}\rangle + \mathrm{c.c.}\right) \label{eq:2ndderiv0} \\
	& = & \langle\Psi_I| \frac{\partial^2 \hat{H}}{\partial \xi_i \partial \xi_j} |\Psi_I\rangle + \left(\langle\Psi_I| \frac{\partial \hat{H}}{\partial \xi_j} \hat{R} \frac{\partial \hat{H}}{\partial \xi_i} |\Psi_I\rangle + \mathrm{c.c.}\right) \label{eq:2ndderiv1} \\
	& \equiv & H_{ij}^{(0)} + (H_{ij}^{(1)} + \mathrm{c.c.}), \label{eq:2ndderiv}
	\end{eqnarray}
	where the resolvent $\hat{R}$ is defined by
	\begin{eqnarray}
	\hat{R} & = & \hat{Q} (\hat{Q}\hat{H}\hat{Q}-E_I)^{-1} \hat{Q}, \label{eq:R} \\
	\hat{Q} & = & 1 - |\Psi_I\rangle \langle\Psi_I|. \label{eq:Q}
	\end{eqnarray}
	The reduction of Eq.~\eqref{eq:2ndderiv0} to Eq.~\eqref{eq:2ndderiv1} is elementary and follows from expanding
	\begin{equation}
	\frac{\partial}{\partial \xi_i}\left((\hat{H}-E_I)|\Psi_I\rangle\right)=0,
	\end{equation}
	and using $E_I = \langle\Psi_I|\hat{H}|\Psi_I\rangle$ and $\hat{Q}\hat{H} = \hat{H}\hat{Q}$.
	
	The term $H_{ij}^{(0)} \equiv \langle\Psi_I| \frac{\partial^2 \hat{H}}{\partial \xi_i \partial \xi_j} |\Psi_I\rangle$ is ODLR, since the only term of $\hat{H}$ that is neither linear nor constant in the nuclear coordinates is the nuclear-nuclear repulsion term (the first term of Eq.~\eqref{eq:hatH}), and its second order derivative is an asymptotically smooth convolution kernel.
	Specifically, if $\xi_i = \alpha_A$ and $\xi_j = \beta_B$ (where $\alpha,\beta \in \{x,y,z\}$ and $A,B$ $(B\neq A)$ denote atoms), then
	\begin{equation}
	H_{ij}^{(0)} =  Z_A Z_B \left(\frac{\delta_{\alpha\beta}}{R_{AB}^3} - 3\frac{(\alpha_B-\alpha_A)(\beta_B-\beta_A)}{R_{AB}^5} \right). \label{eq:H0ij}
	\end{equation}
	It is easy to verify that Eq.~\eqref{eq:H0ij} satisfies Eq.~\eqref{eq:asymp_smooth}.
	To prove that the second term $H_{ij}^{(1)}$ is also ODLR, note that the matrix representation of $\hat{R}$ in a local basis, $\mathbf{R}$ (which is a matrix whose dimension is equal to the number of the electronic states of the system, $N_{\rm state}$) is ODLR. To see the latter fact, note that since $\hat{Q}$ is ODLR under the same basis (it is the sum of a local matrix and a rank-1 matrix, Eq.~\eqref{eq:Q}), $\hat{Q}\hat{H}\hat{Q}-E_I$ is ODLR, which means that $(\hat{Q}\hat{H}\hat{Q}-E_I)^{-1}$ and therefore $\hat{Q} (\hat{Q}\hat{H}\hat{Q}-E_I)^{-1} \hat{Q}$ is ODLR. Then, noting that $H_{ij}^{(1)}$ can be written as the product of an $N_{\rm atom}\times N_{\rm state}$ ODLR matrix, an $N_{\rm state}\times N_{\rm state}$ ODLR matrix, and an $N_{\rm state}\times N_{\rm atom}$ ODLR matrix:
	\begin{eqnarray}
	\mathbf{H}^{(1)} & = & \mathbf{g}^T\mathbf{R}\mathbf{g}, \\
	g_{Ji} & = & \langle \Psi_J | \frac{\partial \hat{H}}{\partial \xi_i} |\Psi_I\rangle,
	\end{eqnarray}
	it follows that $\mathbf{H}^{(1)}$ is also ODLR. Interestingly, $g_{Ji}$ is nothing but the non-adiabatic coupling matrix elements (NACMEs) between states $I$ and $J$\cite{NAC-TDDFT}. Here the ODLR property of $\mathbf{g}$ follows from the fact that only the nuclear attraction and nuclear repulsion terms (the first and third terms of Eq.~\eqref{eq:hatH}) contribute to $\frac{\partial \hat{H}}{\partial \xi_i}$, and the nuclear derivatives of these two terms have the ODLR property.
	
	Note that the above proof does not assume that the system is in its ground state. Moreover, when state $I$ is not degenerate or near-degenerate (when state $I$ is the ground state, this is equivalent to the condition that the lowest excitation energy is not too small, or the ground state has a large HOMO-LUMO gap), $\mathbf{R}$ is not only ODLR but also local, as can be seen from the fact that $\hat{Q}\hat{H}\hat{Q}-E_I$ in Eq.~\eqref{eq:R} do not have near-zero eigenvalues. Thus, in this case, the Hessian $\mathbf{H}$ is not only ODLR but also local.
    
\begin{figure}[htbp]
		\begin{tabular}{c}
			\resizebox{0.48\textwidth}{!}{\includegraphics{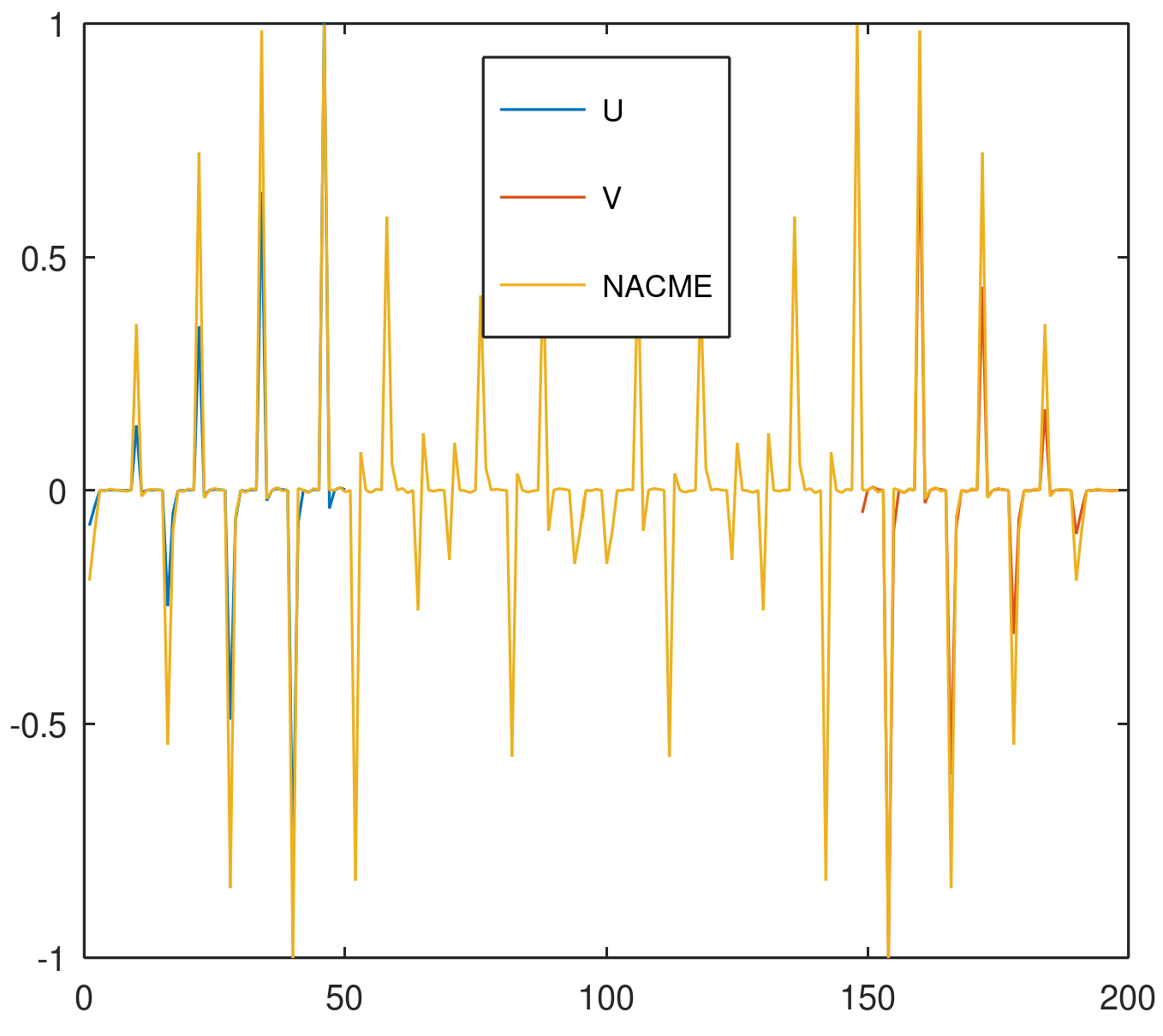}} \\ (a) \\ \resizebox{0.48\textwidth}{!}{\includegraphics{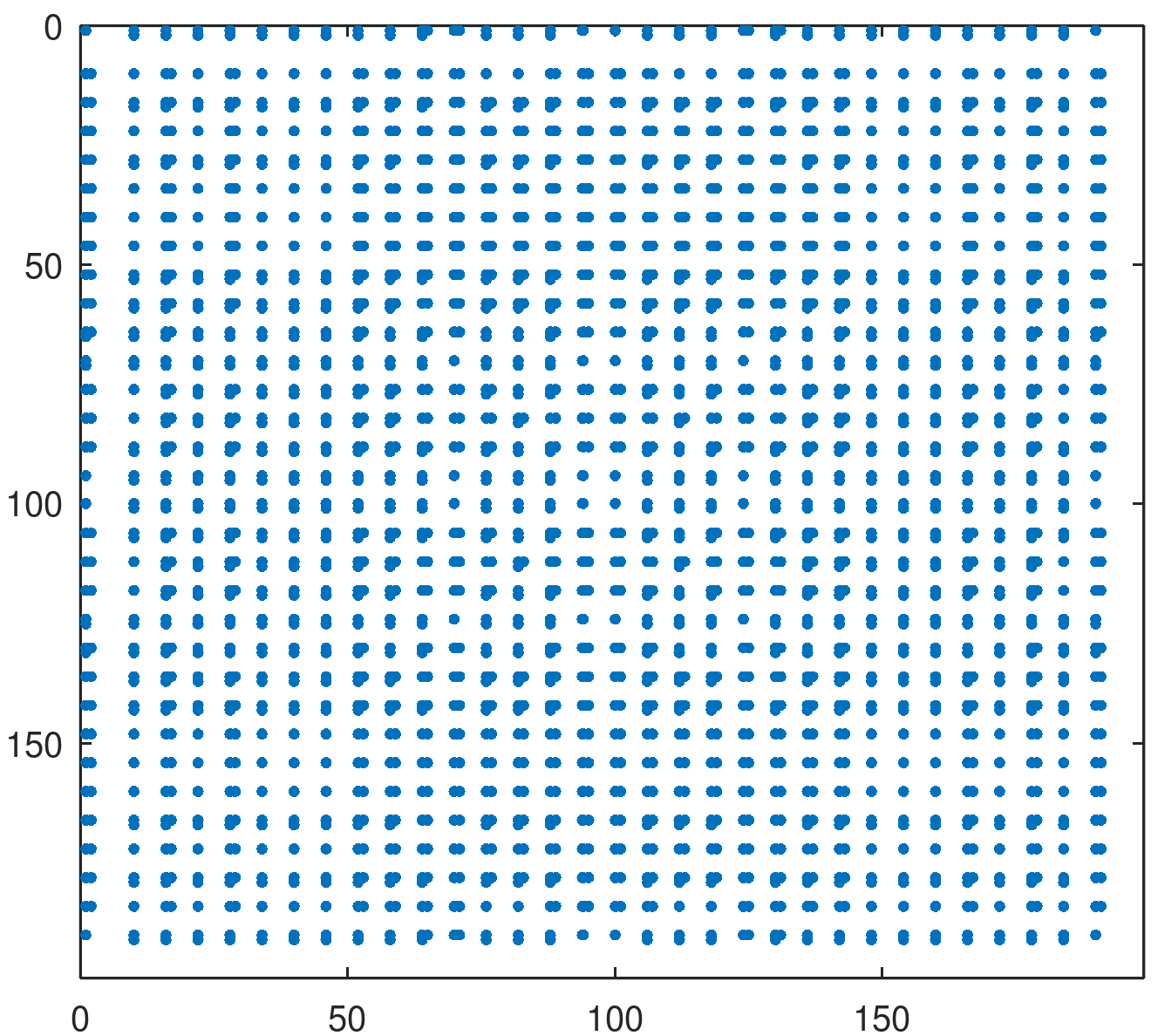}} \\ (b)
		\end{tabular}
		\caption{(a) The left and right singular vectors of the upper right $50\times 50$ block of the Hessian of the $T_1$ state of \ce{C32H34}, at the B3LYP-D3/def2-SVP level of theory, compared against the U-TDDFT $T_1$-$T_2$ NACME at the same level. All vectors are normalized so that the largest absolute value is 1. (b) Elements of the outer product of the $T_1$-$T_2$ NACME with itself, showing elements whose absolute values are greater than $10^{-2}$.}
		\label{fig:NACME}
	\end{figure}

    If the energy of state $J$ is close to state $I$, but there is no other state with similar energy as state $I$, then the off-diagonal blocks of $\mathbf{H}$ will be dominated by $\mathbf{g}^T_JR_{JJ}\mathbf{g}_J$ (where $\mathbf{g}_J$ is the column vector formed by $\{g_{Ji}\}, i=1,\ldots,N_{\rm atom}$; contrary to earlier in this subsection, here both $\mathbf{g}$ and $\mathbf{R}$ are expressed in the eigenbasis of $\hat{H}$), as $R_{JJ}=1/(E_J-E_I)$ is large. In this case, the singular vectors of the far off-diagonal blocks of $\mathbf{H}$ have a physical meaning: they are proportional to the NACMEs between states $I$ and $J$.
    We demonstrate this numerically by comparing the dominant left and right singular vectors of the upper right $50\times 50$ block of the Hessian of triplet ($T_1$ state) \ce{C32H34} (Figure~\ref{fig:off_diagonal_SVD}(e)), with the $T_1$-$T_2$ NACME of the same molecule (Figure~\ref{fig:NACME}(a)). The studied block is very close to rank-1, with the largest singular value (0.021) two orders of magnitude larger than the second largest singular value (0.00057). On the other hand, the $T_2$ state is moderately closer to the $T_1$ state than higher lying triplet states are to $T_1$ (the $T_1$-$T_2$, $T_1$-$T_3$ and $T_1$-$T_4$ energy differences are 0.88, 1.44 and 1.72 eV, respectively; note that despite the higher accuracy of spin-adapted TDDFT compared to U-TDDFT\cite{XTDDFTgrad,XTDDFTNACME}, here U-TDDFT was used to be consistent with the unrestricted Kohn-Sham treatment when calculating the Hessian itself). We thus see that the singular vectors coincide well with the $T_1$-$T_2$ NACME; in fact, the checkerboard pattern of the non-negligible elements of the Hessian (Figure~\ref{fig:C32}) can be well reproduced by the outer product of the $T_1$-$T_2$ NACME with itself (Figure~\ref{fig:NACME}(b)). In a broader context, the possibility of extracting NACMEs from the off-diagonal blocks of Hessians reminds one of the popular approach of estimating NACMEs or hopping probabilities from the curvatures of potential energy surfaces, such as in Landau-Zener\cite{LandauZener1,LandauZener2}, Zhu-Nakamura\cite{ZhuNakamura} and Baeck-An\cite{BaeckAn,BaeckAnDynamics} dynamics, in the nonadiabatic molecular dynamics community.
    For molecules with sufficiently large gaps, however, the singular vectors of the off-diagonal blocks of $\mathbf{H}$ can no longer be interpreted as NACMEs. On one hand, the NACMEs between state $I$ and numerous other states may have similar contributions to the singular vectors, such that the singular vectors are linear combinations of many NACMEs; on the other hand, $\mathbf{H}^{(0)}$ may now contribute substantially to the off-diagonal blocks of $\mathbf{H}$, further complicating the assignment of their singular vectors.

	The ODLR property of Hessians has been used to devise an algorithm, O1NumHess\cite{O1NumHess}, that calculates the seminumerical Hessian of an $N_{\rm atom}$-atom molecule using only $O(1)$ gradients, instead of the $O(N_{\rm atom})$ gradients required by conventional seminumerical Hessian algorithms (more specifically, $3N_{\rm atom}+1$ for single-sided numerical differentiation of analytic gradients, and $6N_{\rm atom}$ for double-sided differentiation).
	The Hessian is first expressed as an MLR matrix. In fact, for molecules up to about 200 atoms, a two-level MLR matrix suffices:
	\begin{equation}
	\mathbf{H} = \mathbf{H}^{\mathrm{local}} + \mathbf{H}^{\mathrm{lowrank}}.\label{eq:HlHlr}
	\end{equation}
	The algorithm first calculates the gradients for a few displaced geometries, and finds a quadratic approximation to the potential energy surface that reproduces the gradients as closely as possible, under the constraint that the Hessian of the potential energy surface, $\mathbf{H}^{\mathrm{local}}$, is local (but not necessarily block-diagonal). $\mathbf{H}^{\mathrm{local}}$ is often a good approximation to $\mathbf{H}$ for large-gap systems, but a poor approximation for small-gap systems. In either case, a second correction, $\mathbf{H}^{\mathrm{lowrank}}$, is sought for, so that the corrected Hessian reproduces the gradients even better; the correction can be non-local, but is constrained to be low-rank. The final $\mathbf{H}$ reproduces the correct structure of the exact Hessian far away from the diagonal, which is important for the correct description of low vibrational frequencies and thermochemical quantities.
	
	\section{Summary and outlook}
	
	In this Perspective, we reviewed the theoretical basis of the ODLR property of a wide range of matrices involved in computational chemistry, as well as techniques to efficiently store and use ODLR matrices, with special emphasis on Coulomb, density, Fock, LMO coefficient and Hessian matrices (Table~\ref{table:summary}). Despite the vast body of literature on ODLR matrices, many of the cited references are from the mathematics community, and it is fair to say that the utility of the ODLR property is severely underrated in computational chemistry, especially compared to the existing literature that uses sparsity and/or low-rank properties of matrices. In particular, the ODLR properties of the Fock and LMO coefficient matrices have not been discussed in previous works, and were only first mentioned in this work. It is thus hoped that the present contribution will stimulate novel use of the ODLR property in computational chemistry, leading to memory and computational time savings for quantities that are neither local nor low-rank, where low-scaling algorithms were often thought to be impossible to devise.

	\begin{table*}[htbp]
		\centering
		\caption{ODLR matrices used in computational chemistry, and ways to exploit their ODLR property. For the meanings of the abbreviations, see text.}
        \setlength{\tabcolsep}{2pt}
		\begin{tabular}{ccccc}
			\hline\hline
			Matrix & Method & Storage Format & Singular vectors$^a$ & Scaling \\
            \hline
			Coulomb matrix & FMM & $\mathscr{H}^2$-matrix & Multipoles & $O(N)$ \\
			ERI tensor & FMM, BUPO, MRA, $\mathscr{H}^2$-ERI & $\mathscr{H}^2$-matrix & Multipoles & $O(N)$ \\
			ERI tensor & FFT, PME & BPBP decomposition & Multipoles & $O(N\log N)$ \\
			Density matrix & ERG$^b$, unknown$^{c}$ & MLR$^b$, unknown$^{c}$ & Frontier orbitals & $O(N\log^2 N)$ $^b$, unknown$^{c}$ \\
			Fock matrix$^d$ & unknown & unknown & $\mathbf{K}$(front.~orb.)$^e$ & unknown \\
			LMO coefficient matrix$^d$ & unknown & unknown & Frontier orbitals & unknown \\
			Nuclear Hessian matrix & O1NumHess & MLR & NACMEs & $O(N)^f$ \\
			\hline\hline
		\end{tabular}\label{table:summary}
        \\
		\footnotesize{$^a$ Physical interpretations of the singular vectors of far off-diagonal blocks.}
		\\
		\footnotesize{$^b$ Pure functionals.}
		\\
		\footnotesize{$^c$ Hybrid functionals.}
        \\
		\footnotesize{$^d$ The ODLR property was unknown until this work.}
        \\
		\footnotesize{$^e$ Exact exchange contribution from frontier orbitals.}
        \\
		\footnotesize{$^f$ Assuming that each gradient calculation takes $O(N)$ time, as is the case for linear scaling electronic structure methods.}
	\end{table*}
	
	Especially interesting is the possibility of linear scaling SCF calculations for gapless systems, at zero electronic temperature. Such calculations have generally been believed to be impossible due to a lack of exploitable sparsity, and while the ERG method achieves $O(N\log^2 N)$ scaling for pure functionals, its large prefactor has severely hindered practical use, and existing implementations can only achieve $O(N^2)$ or worse scalings for hybrid functionals, where the Fock matrix is dense.
	However, the present work demonstrates, for the first time, the possibility of solving the electronic structure of gapless systems at zero electronic temperature in $O(N)$ time, either without (i.e.~direct solution of the density matrix from the Fock matrix) or with recourse to LMOs, and either with our without long-range exact exchange.
	Work is ongoing in our laboratory to devise and implement such an algorithm.
	
	\begin{acknowledgments}
		The author acknowledges support by the Qilu Young Scholar project of Shandong University, as well as helpful discussions with Prof.~Wenjian Liu (Shandong University).
	\end{acknowledgments}
	
	\begin{Notes}
		The author declares no competing financial interest.
	\end{Notes}
	
	\bibliographystyle{cicc}
	\bibliography{ODLR}
	\label{lastpage}
\end{document}